\newif \ifonecol
\onecoltrue 

\ifonecol
       \documentclass[journal,onecolumn, draftcls, 12pt]{IEEEtran} 
\else
       \documentclass[journal,letterpaper, 12pt]{IEEEtran}  
\fi

\usepackage[utf8]{inputenc}
\usepackage[T1]{fontenc}
\usepackage[hidelinks]{hyperref}
\usepackage{url}
\usepackage{ifthen}
\usepackage{cite}
\usepackage[cmex10]{amsmath} 
\usepackage{amsfonts,amssymb}
\usepackage{graphicx}
\usepackage{color}
\usepackage{ifthen}
\usepackage{comment}
\usepackage[arxiv]{optional} 

\usepackage[ruled]{algorithm2e}
\usepackage{float}
\usepackage{enumitem}

\allowdisplaybreaks[1]

\usepackage{tikz}
\usetikzlibrary{calc,shapes,arrows,positioning}
\tikzset{
       solid node/.style={circle,draw,inner sep=1.5,fill=black},
       hollow node/.style={circle,draw,inner sep=1.5}
}

\usepackage{mathtools}

\usepackage{empheq}

\begin{document}
\title{Sequential Bayesian Learning with A Self-Interested Coordinator}

 \author{%
   \IEEEauthorblockN{Xupeng Wei, Achilleas Anastasopoulos\\}
   \IEEEauthorblockA{University of Michigan\\
                     Ann Arbor, MI 48109, USA\\
                     Email: \texttt{\{xupwei,anastas\}@umich.edu}}
 }

\maketitle




\optv{2col}{
}

\def\cA{\mathcal{A}}
\def\cB{\mathcal{B}}
\def\cC{\mathcal{C}}
\def\cD{\mathcal{D}}
\def\cE{\mathcal{E}}
\def\cF{\mathcal{F}}
\def\cG{\mathcal{G}}
\def\cH{\mathcal{H}}
\def\cN{\mathcal{N}}
\def\cP{\mathcal{P}}
\def\cR{\mathcal{R}}
\def\cS{\mathcal{S}}
\def\cT{\mathcal{T}}
\def\cU{\mathcal{U}}
\def\cV{\mathcal{V}}
\def\cW{\mathcal{W}}
\def\cX{\mathcal{X}}
\def\cY{\mathcal{Y}}
\def\cZ{\mathcal{Z}}

\def\tw{\tilde{w}}
\def\pih{\hat{\pi}}
\def\Pih{\hat{\Pi}}

\newtheorem{lemma}{Lemma}
\newtheorem{fact}{Fact}
\newtheorem{theorem}{Theorem}
\newtheorem{corollary}{Corollary}
\newtheorem{assumption}{Assumption}
\newtheorem{definition}{Definition}
\newtheorem{proposition}{Proposition}
\newtheorem{mdpmodel}{Decision Process}

\newcommand{\ve}[1]{\boldsymbol{#1}}
\newcommand{\eqdef}{\stackrel{\scriptscriptstyle \triangle}{=}}
\newcommand{\mdef}{\stackrel{\text{\tiny def}}{=}}
\newcommand{\E}{\mathbb{E}}
\def\Real{\mathbb{R}}
\def\Integer{\mathbb{Z}}
\def\Natural{\mathbb{N}_0}
\def\P{\mathbb{P}}

\def\Tr{\mathsf{T}}
\def\cM{\mathcal{M}}
\def\cL{\mathcal{L}}
\def\one{\boldsymbol{1}}

\newcommand{\red}[1]{\textcolor{red}{#1}}
\newcommand{\blue}[1]{\textcolor{blue}{#1}}

\begin{abstract}

       Social learning refers to the process by which networked strategic agents learn an unknown state of the world by observing private state-related signals as well as other agents' actions. In their classic work, Bikhchandani, Hirshleifer and Welch showed that information cascades occur in social learning, in which agents blindly follow others' behavior, and consequently, the actions in a cascade reveal no further information about the state.

       In this paper, we consider the introduction of an information coordinator to mitigate information cascades. The coordinator commits to a mechanism, which is a contract that agents may choose to accept or not. If an agent enters the mechanism, she pays a fee, and sends a message to the coordinator indicating her private signal (not necessarily truthfully). The coordinator, in turn, suggests an action to the agents according to his knowledge and interest. We study a class of mechanisms that possess properties such as individual rationality for agents (i.e., agents are willing to enter), truth telling, and profit maximization for the coordinator. We prove that the coordinator, without loss of optimality, can adopt a summary-based mechanism that depends on the complete observation history through an appropriate sufficient statistic. Furthermore,  we show the existence of a mechanism which strictly improves social welfare, and results in strictly positive profit, so that such a mechanism is acceptable for both agents and the coordinator, and is beneficial to the agent community. Finally, we analyze the performance of this mechanism and show significant gains on both aforementioned metrics.

\end{abstract}



\section{Introduction}
\label{sec:intro}

Learning in general social networks is characterized by the following salient features: there is an unknown state of the world (e.g., the value of a new product or a new technology) that agents want to estimate in order to improve their well being. Agents themselves do not observe this state of the world directly, but only indirectly through some private signals (e.g., recommendations from friends, etc). Since agents are selfish they may not want to share this private information. Nevertheless, they take actions (e.g., buy the product or adopt the new technology) and these actions are publicly observed in the network by all agents. As a result, the action of each agent is based on the public information available (e.g., buying actions of previous agents) and their own private information. Through this process, information about the state of the world, which is beneficial for the entire community, is only partially revealed through the actions of the agents. This partial revelation of the private information may lead to catastrophic behavior in social networks: although the community as a whole has sufficient information to accurately estimate the state of the world, because of the
partial revelation of this information through the agents' actions, the actual information revealed in the network is minimal.
This scenario was studied in the seminal works~\cite{bikhchandani1992,banerjee1992simple,smith2000pathological}, with the key result being the existence of a phenomenon called \emph{information cascade} which is shown to occur almost surely in sequential social learning.
%
%
When a cascade occurs, the agent's belief of the state is dominated by her observation of previous actions so that her action is independent of her private signal, and therefore uninformative for the successive agents. An information cascade can be catastrophic if it starts with an agent's belief supportive of the wrong state, because even if the successors all hold private signals in favor of the correct state, the information cannot be passed through the actions, and consequently all the successors take actions consistent with the wrong state estimate.
The results of these seminal works raise the following question: if strategic agents are not allowed to have direct communication with each other, are there any approaches to disseminating information more efficiently throughout the network?

A number of ideas have been proposed (a detailed literature review is presented in the next subsection) to postpone or avoid information cascades.
They include such ideas as introducing noisy or extra observations, allowing agents to query others, revisit the marketplace, etc.
Most of these approaches manage to somewhat alleviate--but not fully eliminate--the formation of (bad) cascades, at a possible expense of individual welfare. Furthermore, they may require additional assumptions, such as unconditional cooperation from previous agents without the need of extra incentives, which are difficult to guarantee in real-world applications.

In this paper, we approach the problem in a more direct way. Motivated by the fact that information cascades are caused by insufficient information sharing, we introduce an information coordinator to facilitate information dissemination. The coordinator serves as a third party who aggregates agents' information and provides agents with recommendations. The coordinator commits to a \textit{mechanism}, which is a public contract specifying the coordinator's behavior during the social learning process. Each agent may choose to sign the contract with the coordinator or not. If an agent chooses to cooperate, she pays an entrance fee, privately shares her information with the coordinator, and in turn receives a recommendation from the coordinator as to whether she would buy the product or not.
This process resembles a consultation in real life. The client (agent) pays to meet the consultant (coordinator), tells the consultant about her own situation, and asks for advice. The consultant is experienced because he learns from the past consultations. Although he is not allowed to disclose others' private information due to professional ethics, he can still make recommendations based on what he knows.

For this approach to work, a number of concerns have to be taken into account in the design process. On the agents' side, due to privacy concerns, they may not be willing to share their information with the coordinator truthfully. Thus appropriate incentives have to be introduced. Furthermore, for agents to join the mechanism, an improvement in their individual utility has to be guaranteed by design.
From the coordinator's viewpoint, the contract has to be designed in such a way that it results in positive profit in order for him to have incentive to maintain the mechanism.

A second concern is mechanism complexity. A mechanism relying heavily on the exact history details is not practical. For example, even if both the signal space and the action space are finite, the history space grows exponentially with time, so that recommendations that depend on the specific history are unrealistically complex. On the other hand if one designs recommendations based on a summary of the observation history one may incur losses in the profitability of the coordinator.
In this paper we show that without loss of optimality, it is sufficient for the coordinator to design a time-invariant mechanism depending only on a summary of the observation history.

The last concern is efficiency. As a metaphor, in Braess's paradox, a newly-constructed route may worsen the traffic throughput of an area. In social learning with a coordinator, it is possible that a badly designed mechanism leads to worse learning processes that decrease the social welfare. To settle this concern, we show the existence of a specific mechanism that is beneficial from the viewpoints of both the coordinator and the community of agents. Although this specific mechanism is not the optimal for the coordinator, it brings substantial profit to the coordinator while at the same time increases the social utility. Such mechanism may be useful is a scenario when a neutral entity, such as a government, may want to recruit a for-profit organization to serving as the coordinator so as to improve the overall social welfare.

\subsection{Specific Contributions}

We introduce a self-interested information coordinator to the sequential Bayesian learning model. The contributions are summarized as follows:
\begin{itemize}
       \item Adopting perfect Bayesian equilibrium as a solution concept, we introduce a class of candidate mechanisms that we call ``feasible and profit maximizing'' (FPM) with the properties of individual rationality (IR), truth-telling (TT) and profit maximization (PM).
       \item We propose a subset of FPM mechanisms called summary-based mechanisms, with the property that the recommendations are only a function of a sufficient statistic of the observation history with finite domain. We prove that the coordinator can restrict attention to summary-based mechanism without loss of profitability.
           Furthermore, we characterize the optimal summary-based mechanism through backward dynamic programming equations.
       \item We investigate a special type of FPM mechanisms with nonnegative tax (NT-FPM). We show the existence of NT-FPM mechanisms with positive social welfare improvement by explicitly constructing a specific mechanism that we call the ``no switch if indifference'' (NSII) mechanism. The NSII mechanism is proved to be strictly profitable for the coordinator, without any harm to agents' individual welfare.
       \item Delving deeper into the NSII mechanism, we conduct an exact performance analysis for the gross/net average social utility and average profit of the coordinator for that mechanism. We show that these quantities can be evaluated precisely by means of recursive equations. Numerical analysis based on the above results show a substantial improvement on both the net social welfare and the coordinator profit compared to the case without a coordinator.
\end{itemize}

\subsection{Related Literature}

In the literature of social learning, the seminal works~\cite{bikhchandani1992,banerjee1992simple} characterize information cascades in basic social learning models. The authors of~\cite{smith2000pathological} extend the setting to multiple types of agents and continuous signal space. A number of works such as~\cite{sgroi2002optimizing, banerjee2004word, ho2015robust, monzon2017aggregate} remove the assumption of sequentially observational learning. In these works, agents may have imperfect observation of the previous actions, or may not be clear about the order of decision making. Social learning in general networks are investigated in~\cite{lobel2015information, song2016social}. Efforts to resolve the information cascade problem are reported in~\cite{le2017information, su2019social, BiHeAn22}. Specifically, \cite{le2017information} shows that the reviews provided by previous agents do not necessarily decrease the probability of getting into a bad cascade, while \cite{su2019social} proposes a question-based approach, which could eliminate bad cascades with the assumption of truthful responses. Recently, \cite{BiHeAn22} demonstrated that some bad cascades could be avoided with patient, non-myopic agents.

The idea of recommendation mechanism design in this paper borrows ideas from ``information design'' and ``mechanism design''.
In information design, the designer influences agent behaviors by taking advantage of information asymmetry. \cite{kamenica2011bayesian} provides a basic static model in the field of information design. In this model, the designer, known as a sender, is allowed to reveal state-relevant signals to agents (receivers) to influence agents' actions. The designer commits to an information mechanism, which specifies how the signals are generated conditioned on the state. The agents then form a belief with signals and take actions. The paper~\cite{bergemann2019information} reveals a connection between information design and correlated equilibrium in game theory. Some works like~\cite{ely2017beeps, lingenbrink2019optimal, sayin2021bayesian, farhadi2022dynamic} extend information design to dynamic settings.
In mechanism design~\cite{borgers2015introduction}, information flows in a reverse direction.  The designer (as a receiver), commits to a mechanism specifying the allocation and taxes as functions of agents' reported messages.
In some information design works~\cite{kolotilin2017persuasion,heydaribeni2021joint}, the designer implements a mechanism in order to know agents' private types and makes customized recommendations. In these works, the private type is useless for the recommendation to a different agent, because the state is known by the designer, which is not the case in our work.
In contrast, the information gathering in mechanism design in our work serves as a complementary part for information design, which helps the coordinator collect information from agents so as to build information asymmetry.

The remainder of this paper is organized as follows. The model of social learning with a coordinator is described in Section~\ref{sec:model}. Section~\ref{sec:pbe} characterizes the equilibrium behavior of agents and the coordinator. Section~\ref{sec:sb_md} shows the sufficiency of summary-based mechanisms for the coordinator's optimization problem. Switching to the point of view of the agents' community, Section~\ref{sec:suboptimal} presents the existence of a subclass of mechanisms that increase net social welfare and give strictly positive profit to the coordinator by the explicit construction of the NSII mechanism. Numerical results are presented in Section~\ref{sec:num}. Section~\ref{sec:conclusion} concludes the paper.

\section{Model}
\label{sec:model}

The model to be discussed introduces a self-interested coordinator into the basic observational learning model of Bikhchandani, Hirshleifer and  Welch (``BHW'' for short)~\cite{bikhchandani1992}. We first restate the basic learning model. This basic model will then be augmented with a coordinator together with a mechanism which induces a dynamic game. We are interested in the outcome of this dynamic game at equilibrium.

\subsection{Sequential Bayesian Learning without A Coordinator}
In the basic model,
there is an unknown \emph{state of the world} $W \in\cW=\{-1,1\}$, which models for instance the (unobserved) quality of a good. An infinite sequence of agents $t=1,2,\ldots$  are coming to the marketplace, each at time $t$, take an action $A_t\in \cA=\{-1,1\}$ and then leave the marketplace forever.
Every agent's objective is to match her action with the \emph{state of the world}, i.e., agent~$t$'s utility is $u(w,a_t)=1\{a_t=w\}$\footnote{For simplicity of exposition we are assuming a slightly different utility from the one considered in~\cite{bikhchandani1992} which was $u(w,a_t)=w(a_t+1)/2$.}. The prior belief $Q^w(w)=\P(W=w)=1/2$ for $w=\pm 1$ is a common knowledge. In the original setting, each agent~$t$ privately observes $W$ through a binary symmetric channel with a crossover probability $p \in (0,1/2)$, and we denote this private observation by $Y_t\in\cY=\{-1,1\}$. We define $Q^y(y \vert w)=\P(Y=y \vert W=w)$, so that $Q^y(-w \vert w)=p$ (and $\bar{p}=1-p$). At each time $t$, only agent $t$ takes action $a_t$ based on her observation of the action history $a_{1:t-1}$ (here $x_{j:k}$ represents $x_j,\ldots,x_k$, and sometimes we also use the notation $x^k=x_{1:k}$) and her private observation $y_t$. Agents are Bayesian learners, so agent~$t$'s rationality dictates that
\begin{equation}
       a_t \in \arg \max_a \P(W=a \mid a_{1:t-1},y_t).
\end{equation}

\subsection{Sequential Bayesian Learning with A Coordinator}
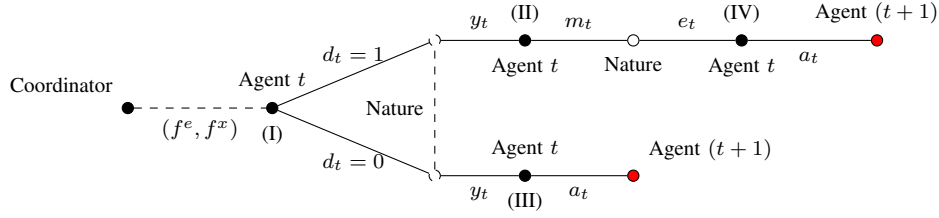
\begin{figure}
       \centering
\ifonecol
       \begin{tikzpicture}[scale=1.2,font=\scriptsize]
\else
       \begin{tikzpicture}[scale=0.75,font=\scriptsize]
\fi
                     \tikzstyle{level 1}=[level distance=16mm,sibling distance=35mm]
                     \tikzstyle{level 2}=[level distance=18mm,sibling distance=15mm]
                     \tikzstyle{level 3}=[level distance=10mm,sibling distance=20mm]
                     \tikzstyle{level 4}=[level distance=12mm,sibling distance=18mm]
                     \tikzstyle{level 5}=[level distance=12mm,sibling distance=15mm]
                     \tikzstyle{level 6}=[level distance=15mm,sibling distance=10mm]
                     \node(0)[solid node,label=above left:{Coordinator}]{}
                     [grow=right]
                            child{node(0-0)[solid node,label=above :{Agent $t$}, label=below:{(I)}]{}
                                   %
                                   child{node(0-0-0)[hollow node]{}
                                          child{node[solid node,label=above:{Agent $t$}, label=below: {(III)}]{}
                                                 child{node[solid node, fill=red,label=above right:{Agent $(t+1)$}]{}
                                                 edge from parent[solid,left] node[below]{$a_t$}}
                                          edge from parent[solid,left] node[below]{$y_t$}}
                                   edge from parent[solid,left] node[below]{$d_t=0$}}
                                   child{node(0-0-1)[hollow node]{}
                                          child{node[solid node,label=below:{Agent $t$}, label=above:{(II)}]{}
                                                 child{node[hollow node,label=below:{Nature}]{}
                                                        child{node[solid node,label=below:{Agent $t$}, label=above:{(IV)}]{}
                                                               child{node[solid node,fill=red,label=above:{Agent $(t+1)$}]{} edge from parent[solid,left] node[below]{$a_t$}}
                                                        edge from parent[solid,left] node[above]{$e_t$}}
                                                 edge from parent[solid,right] node[above]{$m_t$}}
                                          edge from parent[solid,right] node[above]{$y_t$}}
                                   edge from parent[solid,right] node[above]{$d_t=1$}}
                            edge from parent[dashed] node[below]{$(f^e,f^x)$}
                            }
                     ;
                     \draw[dashed](0-0-0)to(0-0-1);
                     \node[left] at ($.5*(0-0-0)+.5*(0-0-1)$){Nature};
       \end{tikzpicture}
       \caption{Part of the game tree for the mechanism-induced game. The dashed line for ``Nature''  indicates that the realization of $y_t$ is independent of $d_t$. The red node represents the next subgame for agent $(t+1)$.}
       \label{fig:game_tree}
\end{figure}

In this augmented model, the coordinator first commits to a \emph{mechanism}. A mechanism is a contract between the coordinator and the agents, the contents of which are known to all. If agents accept this contract, they pay a service fee (tax) to the coordinator, report necessary information, and receive recommended actions.
The overall process is depicted in Fig.~\ref{fig:game_tree} and is detailed below.

First, the coordinator announces the mechanism, which is a family of functions $f=(f^e,f^x)$, the meaning of which will be explained in the following.
Following this revelation, before agent~$t$ takes an action, she decides whether to join the mechanism or not. This decision is publicly observed, and is denoted by $d_t\in\cD=\{0,1\}$, with $d_t=1$ meaning ``join'', and $d_t=0$ meaning ``not join'', based on agent~$t$'s observed history $d^{t-1},a^{t-1}$. Subsequently, agent~$t$ receives her private signal $y_t$. If she chooses not to join the mechanism, she needs to take action $a_t$ based only on $d^{t-1},a^{t-1},y_t$. If she chooses to join, she pays an entrance fee $f^{x}_t(d^{t-1},a^{t-1})$ and reports a message $m_t\in\cM=\cW$ (confidential to other agents) directly to the coordinator. This message is supposed to be the private signal $y_t$ that the agent reports to the coordinator (but the agent may choose to misreport).
After that, agent~$t$ receives (confidentially) an action recommendation $e_t\in \cA$ generated by the conditional distribution $f^e_t(\cdot \vert d^{t-1},a^{t-1},m^t) \in \Delta(\cA)$ based on the public history $d^{t-1},a^{t-1}$, and the private information $m^t$. Then, agent~$t$ takes action $a_t=e_t$\footnote{This type of behavior is called obedience. In our setting, it can be achieved by imposing a large penalty to disobedient agents. This can be achieved since the coordinator knows $e_t$ and also observes $a_t$.} that is publicly observed. The two functions $f^e$ and $f^x$ constitute the contract between the coordinator and the agents and in our setting the coordinator commits to this contract (so that both functions are known to the agents).

The mechanism induces a sequential game where both the coordinator and agents are the players.
As a result, the timeline for the realizations of random variables can be written as
\begin{equation*}
       F,W, \ldots, D_t, Y_t, (M_t, E_t), A_t, D_{t+1}, \ldots,
\end{equation*}
where $F=(F^e, F^x)$ is the mechanism specified by the coordinator, and $M_t,E_t$ are not realized if $D_t=0$.
We emphasize once more that the coordinator commits to the mechanism $F$ and this is exactly the reason we indicate it above as an observed action by the agents and not merely as a strategy that emerges as an equilibrium solution.

With the introduction of the coordinator, if agent~$t$ joins the mechanism, her net utility is $u(w,a_t) - f^x_t(d^{t-1},a^{t-1})$, and the coordinator's profit is a sum of the taxes received from each agent~$t$ (discounted by $\delta^{t-1}$, $\delta \in (0,1)$). Otherwise, the agent's utility will be $u(w,a_t)$ and the coordinator will not extract any profit from her.

In this problem, we are interested in mechanisms $f=(f^x,f^e)$ with the following properties:
\begin{itemize}
       \item \emph{Individual Rationality (IR).} For a Bayesian rational agent, joining the mechanism brings an expected net utility which is no worse than not to join.
       \item \emph{Truth-telling (TT).} When joining the mechanism, agents are willing to report $m_t=y_t$.
\end{itemize}

\begin{definition}
       A mechanism $f$ is said to be feasible if it satisfies the IR and TT conditions.
\end{definition}




\subsection{The Induced Dynamic Game}
Assume rationality in this setting, i.e., all the agents and the coordinator take actions to maximize their expected payoff based on the available information, and this is a common knowledge to all. This means the agents and coordinator are facing a dynamic game, part of which is sketched in Figure~\ref{fig:game_tree} without clarifying at every stage who observes what. To describe precisely the induced dynamic game, we state the \emph{information set} (``infoset'' for short)
of each decision maker at each decision-making node.

In this game, the coordinator takes the first move $f=(f^e,f^x)$ (i.e., committing to a mechanism) before the realizations of all random variables. An agent~$t$ would see the following possible infosets:
 \begin{enumerate}
       \item Infoset $h^d_t = (f,d^{t-1},a^{t-1})$ relating to decision $D_t \sim g^d_t(\cdot \vert h^d_t)$ (Node (I) in Figure~\ref{fig:game_tree});
       \item Infoset $h^m_t = (f,d^{t-1},a^{t-1},d_t=1,y_t)$ relating to decision $M_t \sim g^m_t(\cdot \vert h^m_t)$ (Node (II));
       \item Infoset $h^a_t = (f,d^{t-1},a^{t-1},d_t=0,y_t)$ relating to decision $A_t \sim g^a_t(\cdot \vert h^a_t)$ (Node (III)).
\end{enumerate}
Node (IV) also represents an infoset, but we don't consider it because $a_t=e_t$ is enforced at this node by the problem setting (obedience).
In the following, we denote agents' strategies by $g=(g^d,g^m,g^a)$.

For this game, perfect Bayesian equilibrium (PBE) is an appropriate solution concept. A PBE is described by an \emph{assessment} consisting of a strategy profile~$(f,g)$\footnote{Rigorously, $f$ is an action instead of a strategy. Since the coordinator chooses $f$ based on only the fixed common prior~$Q^w$, the action can represent the strategy itself.} and a \emph{belief system}~$\beta$ for each player in each information set. The belief system~$\beta$ is defined as a mapping from an information set to a probability distribution over all possible histories within this information set. As we will see later, only the unobservable variables $W,M^{t-1}$ are of agent~$t$'s interest, so beliefs of the form $\beta: \cH \to \Delta(\cW\times \cM^{t-1})$ suffice (where $\cH$ denotes the set of all infosets). The details of PBE of the proposed model will be discussed in the next section.

\section{Perfect Bayesian Equilibrium}
\label{sec:pbe}


\begin{definition}
       \label{def:PBE}
       A PBE is an assessment consisting of a strategy profile~$(f,g)$ and a belief system~$\beta$ satisfying
       \begin{enumerate}
              \item \emph{Sequential Rationality.} In every information set, the strategy given by the profile~$(f,g)$ is a best response toward the belief specified by~$\beta$, i.e., for every agent $t$ (resp. the coordinator) and information set $h$ of this player, the strategy $g_t$ (resp. $f$) maximizes her expected payoff under the belief $\beta(\cdot \vert h)$.
              \item \emph{Consistency on Equilibrium.} $\beta$ is consistent with the probability distribution induced by the strategy profile~$(f,g)$ at equilibrium. This consistency requires that the belief update on the equilibrium path satisfies Bayes' rule.
              \item \emph{Plausible Off-Equilibrium Belief.} The conditional belief for the off-equilibrium information set should also be specified by~$\beta$, in a reasonable way.
       \end{enumerate}
\end{definition}

For the third requirement in Definition~\ref{def:PBE}, there are various interpretations on ``a reasonable way'' in the literature (see \cite[Chap. 8]{fudenbergtirole}, \cite{gonzalez2014notion}, and \cite[Chap.~5]{mailath2018modeling}). We elaborate on this in the following section.


\subsection{Consistent Belief System}
\label{subsec:consistent}

In this part, we elaborate on the requirements for the consistent belief system in the definition of PBE, namely, consistency on equilibrium and plausibility off equilibrium, given~$(f,g)$.
The belief system is a map $\beta:\cH \to \Delta(\cW \times \cM^*)$,  where $\cH$ is the collection of infosets, $\cM^* = \bigcup_{n=1}^\infty \cM^{t-1}$ is a collection of private message histories.
Though there are unobservable variables other than $w$ and $m^{t-1}$ for agent~$t$, when discussing her rationality we will see it is sufficient for her to focus on $w, m^{t-1}$.

Denote by~$\P^{g}(\cdot)$ the probability distribution induced by~$g$ (note that we do not use the notation $\P^{f,g}$, because $f$ is a part of infosets $h^d_t, h^m_t, h^a_t$). Intuitively, a consistent belief~$\beta$ on~$w, m^{t-1}$ is evaluated by
\begin{equation}
       \label{eq:belief_onpath}
       \beta(w,m^{t-1} \vert h_t) = \frac{\P^{g}(w,m^{t-1},h_t)}{\P^{g}(h_t)}.
\end{equation}
However, the above is not a good choice for defining a consistent belief system for two reasons: (i) the complexity of probability computation on history $h_t$ increases as time $t$ increases, (ii) the belief also needs to be formed on zero probability histories, but the value of the fraction with zero denominator is undefined. Instead, we construct a consistent belief system recursively using Bayes' rule. Next, the rule of belief update will be stated, for the histories on or off-equilibrium paths.

\subsubsection{Belief update on the equilibrium path}

On the equilibrium path, there are two types of the infosets $h^d_t, h^m_t$ with $d_\tau=1, \tau \leq t$, given the mechanism is feasible. On the equilirbium path, $\P^{g}(h_t)>0$, so one may use Bayes' rule to update the belief without any concerns. Belief update starts from the prior belief
\begin{equation}
       \label{eq:belief_hd0}
       \beta(w \vert h^d_1) = Q^w(w),
\end{equation}
where $h^d_1 = (f)$. The belief at infosets of type $h^d_t$ can be evaluated recursively by Bayes' rule
\ifonecol
\begin{align}
              \beta&(w,m^t \vert h^d_{t+1}) = \frac{\P^{g}(w,m^t,d_t=1,a_t \vert h^d_t)}{\P^{g}(d_t=1,a_t \vert h^d_t)} \nonumber\\
              &= \frac{\sum_{y_t}\beta(w,m^{t-1} \vert h^d_t)Q^y(y_t \vert w)g^d_t(1 \vert h^d_t) g^m_t(m_t \vert h^m_t)f^e_t(a_t \vert d^{t-1},a^{t-1},m^t)  }{\P^{g}(d_t=1,a_t \vert h^d_t)} \nonumber\\
              \label{eq:belief_hd}
              &= \frac{\sum_{y_t}\beta(w,m^{t-1} \vert h^d_t)Q^y(y_t \vert w)g^m_t(m_t \vert h^m_t) f^e_t(a_t \vert d^{t-1},a^{t-1},m^t)}
              {\sum_{w, m^{t-1}, y_t}
              \beta(w,m^{t-1} \vert h^d_t)Q^y(y_t \vert w)g^m_t(m_t \vert h^m_t) f^e_t(a_t \vert d^{t-1},a^{t-1},m^t)},
\end{align}
\else
\begin{align}
       \beta&(w,m^t \vert h^d_{t+1}) = \frac{\P^{g}(w,m^t,d_t=1,a_t \vert h^d_t)}{\P^{g}(d_t=1,a_t \vert h^d_t)} \nonumber\\
       \label{eq:belief_hd}
       &= \frac{\splitfrac{\sum_{y_t}\beta(w,m^{t-1} \vert h^d_t)Q^y(y_t \vert w)}{\cdot g^m_t(m_t \vert h^m_t) f^e_t(a_t \vert d^{t-1},a^{t-1},m^t)}}
       {\splitfrac{\sum_{w, m^{t-1}, y_t}
       \beta(w,m^{t-1} \vert h^d_t)Q^y(y_t \vert w)}{\cdot g^m_t(m_t \vert h^m_t) f^e_t(a_t \vert d^{t-1},a^{t-1},m^t)}}
\end{align}
\fi

where the second equality holds because a feasible $f$ results in $g^d_t(1 \vert h^d_t) = 1$ for rational agent~$t$.

The belief at infosets $h^m_t=(h^d_t,d_t=1,y_t)$ is formed privately by agent~$t$ due to the newly coming private information $y_t$. This belief can be calculated by Bayes' rule as
\ifonecol
\begin{equation}
       \label{eq:belief_hm}
       \beta(w,m^{t-1} \vert h^m_t) = \frac{\P^{g}(w,m^{t-1},d_t=1,y_t \vert h^d_t)}{\P^{g}(d_t=1,y_t \vert h^d_t)} = \frac{Q^y(y_t \vert w)\beta(w,m^{t-1} \vert h^d_t)}{\sum_{\bar{w},\bar{m}^{t-1}} Q^y(y_t \vert \bar{w})\beta(\bar{w},\bar{m}^{t-1} \vert h^d_t) }.
\end{equation}
\else
\begin{align}
       \beta&(w,m^{t-1} \vert h^m_t) = \frac{\P^{g}(w,m^{t-1},d_t=1,y_t \vert h^d_t)}{\P^{g}(d_t=1,y_t \vert h^d_t)} \nonumber\\
       \label{eq:belief_hm}
       =& \frac{Q^y(y_t \vert w)\beta(w,m^{t-1} \vert h^d_t)}{\sum_{\bar{w},\bar{m}^{t-1}} Q^y(y_t \vert \bar{w})\beta(\bar{w},\bar{m}^{t-1} \vert h^d_t) }.
\end{align}
\fi

\subsubsection{Belief update off the equilibrium path}

Since the belief assignment for the off-equilibrium paths is more technical and does not directly affect subsequent analysis, the details are relegated to Appendix~\ref{appx:belief_off_path}.

\subsection{Agents' Rationality}

This subsection describes agents' rationality under a given mechanism $f=(f^x_t,f^e_t)$ and a belief system $\beta$. The analysis uncovers how agents' rationality is related to the constraints for IR and TT of a feasible mechanism. In this subsection, the analysis assumes $f$ is given. Notation $U_t(x \vert h)$ is a shorthand for agent~$t$'s expected payoff for taking action~$x$ at infoset~$h$;
notation $(m^{t-1}, y_t)$ represents the $t$-length sequence $m^t$ with $m_t=y_t$, and similarly, $(m^{t-1},-y_t)$ represents $m^t$ with $m_t = -y_t$.
The analysis is done in a backward recursive manner. First consider the decision $m_t$ at infoset $h^m_t=(h^d_t,d_t=1,y_t)$. The expected utility for choosing a certain $m_t$ can be evaluated by
\ifonecol
\begin{equation}
       U_t(m_t \vert h^m_t) = \P^{g}_\beta(A_t=W\mid h^m_t,m_t) - f^x_t(d^{t-1},a^{t-1}),
\end{equation}
\else
\begin{align}
       &U_t(m_t \vert h^m_t) \nonumber\\
       &= \P^{g}_\beta(A_t=W\mid h^m_t,m_t) - f^x_t(d^{t-1},a^{t-1}),
\end{align}
\fi
where the above probability can be derived by analyzing the corresponding joint probability
\ifonecol
\begin{align}
       \P^{g}_\beta&(a_t,w\mid h^m_t,m_t) = \frac{\sum_{m^{t-1}} \P^{g}_\beta(a_t,m_t, y_t,m^{t-1},w\mid h^d_t)}{\P^{g}_\beta(y_t,m_t \mid h^d_t)} \nonumber\\
       \label{eq:g^m_intermediate}
       &= \frac{\sum_{m^{t-1}} \beta(w,m^{t-1} \vert h^d_t) Q^y(y_t \vert w)  f^e_t(a_t \vert d^{t-1},a^{t-1},m^t)}{\sum_{w,m^{t-1}} \beta(w,m^{t-1} \vert h^d_t) Q^y(y_t \vert w)}.
\end{align}
\else
\begin{align}
       \P^{g}_\beta&(a_t,w\mid h^m_t,m_t) \nonumber\\
       &= \frac{\sum_{m^{t-1}} \P^{g}_\beta(a_t,m_t, y_t,m^{t-1},w\mid h^d_t)}{\P^{g}_\beta(y_t,m_t \mid h^d_t)} \nonumber\\
       \label{eq:g^m_intermediate}
       &= \frac{\splitfrac{\sum_{m^{t-1}} \beta(w,m^{t-1} \vert h^d_t) Q^y(y_t \vert w)  }{\cdot f^e_t(a_t \vert d^{t-1},a^{t-1},m^t)}}{\sum_{w,m^{t-1}} \beta(w,m^{t-1} \vert h^d_t) Q^y(y_t \vert w)}.
\end{align}
\fi
The expected utility is therefore
\ifonecol
\begin{equation}
       \label{eq:val_m}
       U_t(m_t \vert h^m_t) = \frac{\sum_{w,m^{t-1}} \beta(w,m^{t-1} \vert h^d_t) Q^y(y_t \vert w)  f^e_t(w \vert d^{t-1},a^{t-1},m^t)}{\sum_{w,m^{t-1}} \beta(w,m^{t-1} \vert h^d_t) Q^y(y_t \vert w)} - f^x_t(d^{t-1},a^{t-1}).
\end{equation}
\else
\begin{align}
       &U_t(m_t \vert h^m_t) \nonumber \\
       &= \frac{\splitfrac{\sum_{w,m^{t-1}} \beta(w,m^{t-1} \vert h^d_t) Q^y(y_t \vert w)}{\cdot f^e_t(w \vert d^{t-1},a^{t-1},m^t)}  }{\sum_{w,m^{t-1}} \beta(w,m^{t-1} \vert h^d_t) Q^y(y_t \vert w)} \nonumber\\
       \label{eq:val_m}
       &\qquad - f^x_t(d^{t-1},a^{t-1}).
\end{align}
\fi

Sequential rationality implies that for any $m_t$ with $g^m_t(m_t \vert h^m_t) > 0$ we have
\begin{equation}
       \label{eq:g^m_t}
       m_t \in \arg \max_{\bar{m}}\,  U_t(\bar{m} \vert h^m_t).
\end{equation}
The truth-telling requirement of $f$ requires $U_t(y_t \vert h^m_t) \geq U_t(-y_t \vert h^m_t)$. This provides TT constraints for all $h^m_t$:
\ifonecol
\begin{equation}
       \label{eq:truth-telling}
       \sum_{w,m^{t-1}} \beta(w,m^{t-1} \vert h^d_t) Q^y(y_t \vert w)  \left(f^e_t(w \vert d^{t-1},a^{t-1},(m^{t-1},y_t)) - f^e_t(w \vert d^{t-1},a^{t-1},(m^{t-1},-y_t))\right)\geq 0.
\end{equation}
\else
\begin{align}
       \sum_{w,m^{t-1}} & \beta(w,m^{t-1} \vert h^d_t) Q^y(y_t \vert w) \nonumber \\
       \cdot &\left(f^e_t(w \vert d^{t-1},a^{t-1},(m^{t-1},y_t)) \right. \nonumber\\
       \label{eq:truth-telling}
       &\left.- f^e_t(w \vert d^{t-1},a^{t-1},(m^{t-1},-y_t))\right)\geq 0.
\end{align}
\fi

We now consider the infoset at the decision for $a_t$ in the case that the user did not join the mechanism, $h^a_t=(h^d_t,d_t=0,y_t)$. The expected utility for choosing $a_t$ at $h^a_t$ is
\begin{equation}
       \label{eq:val_a}
       U_t(a_t \vert h^a_t) = \frac{\sum\limits_{m^{t-1}} Q^y(y_t \vert a_t) \beta(a_t,m^{t-1} \vert h^d_t)}{\sum\limits_{w,m^{t-1}}Q^y(y_t \vert w)\beta(w,m^{t-1} \vert h^d_t)}.
\end{equation}
Therefore, for any $a_t$ with  $g^a_t(a_t \vert h^a_t)>0$ we have
\begin{equation}
       \label{eq:g^a_t}
       a_t \in \arg \max_{\bar{a}} \, U_t(\bar{a} \vert h^a_t).
\end{equation}
       As a result, $g^a_t$ satisfies the constraint: if $g^a_t(a_t \vert h^a_t) > 0$,
\ifonecol
       \begin{equation}
              \sum_{m^{t-1}} Q^y(y_t \vert a_t) \beta(a_t,m^{t-1} \vert h^d_t)  \geq \sum_{m^{t-1}} Q^y(y_t \vert -a_t) \beta(-a_t,m^{t-1} \vert h^d_t).
       \end{equation}
\else
       \begin{align}
              \sum_{m^{t-1}}& Q^y(y_t \vert a_t) \beta(a_t,m^{t-1} \vert h^d_t) \nonumber \\
              & \geq \sum_{m^{t-1}} Q^y(y_t \vert -a_t) \beta(-a_t,m^{t-1} \vert h^d_t).
       \end{align}
\fi

Now consider the decision to join the mechanism or not. For this decision, agent~$t$ compares the expected payoff for joining $\E^{g} \left\{ U_t(Y_t \vert h^d_t,d_t=1,Y_t) \vert h^d_t \right\}$, with the expected payoff for not joining $\E^{g} \left\{ \sum_{a_t} g^a_t(a_t \vert h^d_t,Y_t) U_t(a_t \vert h^d_t,d_t=0,Y_t) \vert h^d_t \right\}$. The explicit form of the expected utility for choosing $d_t$ at $h^d_t$ is
\ifonecol
       \begin{align}
              &U_t(d_t \vert h^d_t) \nonumber\\
              \label{eq:val_d}
              &= \begin{cases}
                     \sum_{y_t, w, m^{t-1}} \beta(w,m^{t-1} \vert h^d_t)Q^y(y_t \vert w) f^e_t(w \vert d^{t-1},a^{t-1},(m^{t-1}, y_t)) - f^x_t(d^{t-1}, a^{t-1}), & d_t=1, \\
                     \sum_{y_t, w, m^{t-1}} \beta(w,m^{t-1} \vert h^d_t)Q^y(y_t \vert w) g^a_t(w \vert h^d_t,y_t) , & d_t = 0.
              \end{cases}
       \end{align}
\else
       \begin{align}
              &U_t(d_t \vert h^d_t) \nonumber\\
              \label{eq:val_d}
              &= \begin{cases}
                     \sum\limits_{y_t, w, m^{t-1}} \beta(w,m^{t-1} \vert h^d_t)Q^y(y_t \vert w)\\
                     \qquad \cdot  f^e_t(w \vert d^{t-1},a^{t-1},(m^{t-1}, y_t)) \\
                     \qquad \qquad - f^x_t(d^{t-1}, a^{t-1}), \\
                     \qquad \qquad \qquad \qquad \qquad \text{ if }d_t=1, \\
                     \sum\limits_{y_t, w, m^{t-1}} \beta(w,m^{t-1} \vert h^d_t)Q^y(y_t \vert w) \\
                     \qquad \cdot g^a_t(w \vert h^d_t,y_t),  \\
                     \qquad\qquad \qquad \qquad \qquad \text{ if }d_t = 0.
              \end{cases}
       \end{align}
\fi

Therefore, for any $d_t$ with $g^d_t(d_t \vert h^d_t)>0$ we have
\begin{equation}
       \label{eq:g^d_t}
       d_t \in \arg \max_{\bar{d}} U_t(\bar{d}  \vert  h^d_t).
\end{equation}
For a feasible $f$, IR requires $g^d_t(1 \vert h^d_t)>0$. This requires the following for all on-path $h^d_t$,
\ifonecol
\begin{equation}
       \label{eq:IR}
       \sum_{y_t, w, m^{t-1}} \beta(w,m^{t-1} \vert h^d_t)Q^y(y_t \vert w)  \left(f^e_t(w \vert d^{t-1},a^{t-1},(m^{t-1},y_t)) - g^a_t(w \vert h^d_t,y_t)\right) \geq f^x_t(d^{t-1}a^{t-1}).
\end{equation}
\else
\begin{align}
       &\sum_{y_t, w, m^{t-1}} \beta(w,m^{t-1} \vert h^d_t)Q^y(y_t \vert w)  \nonumber\\
       &\left(f^e_t(w \vert d^{t-1},a^{t-1},(m^{t-1},y_t)) - g^a_t(w \vert h^d_t,y_t)\right) \nonumber\\
       \label{eq:IR}
       &\geq f^x_t(d^{t-1}a^{t-1}).
\end{align}
\fi

Given the form of expected payoff functions \eqref{eq:val_m}, \eqref{eq:val_a} at decision nodes (II) and (III) in Figure~\ref{fig:game_tree}, it is sufficient for agent~$t$ to use strategies $g^m_t$ and $g^a_t$ depending only on the belief $\beta_t$, since her expected payoff depends on the complete history only through $\beta_t$. Furthermore, if she adopts a belief-based $g^m_t, g^a_t$, then at decision node (I) her expected payoff depends on the complete history only through $\beta_t$ as well (see~\eqref{eq:val_d}), resulting in the sufficiency of using belief-based $g^d_t$. Therefore, it is reasonable to restrict attention to strategies $g$ determined by the best response correspondence as
\begin{equation}
       g_t \in \text{BR}_t(f_t,\beta), \quad t = 1,2,\ldots,
\end{equation}
where the $\text{BR}_t(\cdot,\cdot)$ is defined by~\eqref{eq:g^d_t}, \eqref{eq:g^a_t} and~\eqref{eq:g^m_t}. We further assume that when $\text{BR}_t(\cdot,\cdot)$ is not a singleton, the rule of breaking the tie is in favor of the coordinator, so that we can treat $\text{BR}_t$ as a deterministic function, resulting in
\ifonecol
       \begin{equation}
              \label{eq:BRs}
              g^d_t = \text{BR}^d_t(f_t,\beta),\quad g^a_t = \text{BR}^a_t(\beta), \quad g^m_t = \text{BR}^m_t(f_t,\beta).
       \end{equation}
\else
       \begin{align}
              g^d_t &= \text{BR}^d_t(f_t,\beta),\quad g^a_t = \text{BR}^a_t(\beta), \nonumber\\
              \label{eq:BRs}
              g^m_t &= \text{BR}^m_t(f_t,\beta).
       \end{align}
\fi

\subsection{Coordinator's Rationality}

A self-interested coordinator chooses a feasible mechanism $f=(f^e,f^x)$ to maximize his payoff, presuming that agents will perform their best responses to mechanism $f$. According to the analysis of agents' rationality, suppose $g$ is determined by~\eqref{eq:BRs}, a feasible $f$ needs to satisfy TT constraints \eqref{eq:truth-telling} and IR constraints \eqref{eq:IR}. Notice that the tax function $f^x_t$ only appears in IR constraints, as an object being upper bounded by a quantity determined by $f^e_t, g^a_t$, where $g^a=\text{BR}^a_t(\beta)$. To maximize the coordinator's profit, the tax function~$f^x$ must achieve the upper bound in~\eqref{eq:IR}:
\ifonecol
       \begin{equation}
              \label{eq:opt_tax}
              f^x_t(d^{t-1},a^{t-1}) = \sum_{y_t,w,m^{t-1}}  Q^y(y_t \vert w) \beta(w,m^{t-1} \vert h^d_t)  \left( f^e_t(w  \vert  d^{t-1},a^{t-1},m^t) - g^a_t(w  \vert  h^d_t,y_t) \right).
       \end{equation}
\else
       \begin{align}
              \label{eq:opt_tax}
              f^x_t&(d^{t-1},a^{t-1}) \nonumber \\
              &= \sum_{y_t,w,m^{t-1}}  Q^y(y_t \vert w) \beta(w,m^{t-1} \vert h^d_t)  \nonumber \\
              & \cdot \left( f^e_t(w  \vert  d^{t-1},a^{t-1},m^t) - g^a_t(w  \vert  h^d_t,y_t) \right).
       \end{align}
\fi
As a result, $f^x$ is a fixed function of $\beta$ once $f^e$ is fixed in the design.
We define the following class of mechanisms.
\begin{definition}
       A feasible mechanism $f=(f^e,f^x)$ with the tax form~\eqref{eq:opt_tax} is said to be a feasible and profit maximizing (FPM) mechanism.
\end{definition}
It should be clear from the above discussion that it is sufficient for a rational coordinator to restrict his attention to FPM mechanisms.

For a rational coordinator, an FPM mechanism $f$ maximizes the total expected revenue
\begin{equation}
       \label{eq:opt_obj}
       \E^{g} \left\{ \sum_{t=1}^T f^x_t(D^{t-1},A^{t-1}) \  \mid \  h_1=f  \right\},
\end{equation}
and if $T$ is taken to be infinite, we may instead consider an expected discounted revenue
\begin{equation}
       \E^{g}\left\{\sum_{t=1}^\infty \delta^{t-1} f^x_t(D^{t-1},A^{t-1})  \  \mid \  h_1=f  \right\}.
\end{equation}

With all these results, the coordinator is facing the optimization problem
\ifonecol
       \begin{subequations}
              \label{opt:offline}
              \begin{empheq}[box=\fbox]{align}
                     \underset{f^e}{\text{maximize}} \quad & \E^{g} \left\{ \sum_{t=1}^T f^x_t(D^{t-1},A^{t-1})  \  \mid \  h_1=f \right\}, \\
                     \text{subject to} \quad & \text{Truth-Telling: }\eqref{eq:truth-telling}, \quad \forall t, d^{t-1},  a^{t-1}, y_t, \\
                     \quad & \text{Profit Maximizing Tax: }\eqref{eq:opt_tax}, \quad \forall t, d^{t-1}, a^{t-1}, \\
                     \quad & g^a_t = \text{BR}^a_t(\beta), \quad \forall t,\\
                     \quad & \beta \text{ is consistent with } (f,g),
              \end{empheq}
       \end{subequations}
\else
\begin{subequations}
       \label{opt:offline}
        \begin{empheq}[box=\fbox]{align}
              \underset{f^e}{\max} \quad & \E^{g} \left\{ \sum_{t=1}^T f^x_t(D^{t-1},A^{t-1}) \mid h_1=f \right\} \\
              \text{s.t.} \quad & \text{TT: }\eqref{eq:truth-telling}, \quad \forall t, d^{t-1},  a^{t-1},\forall y_t, \\
              \quad & \text{PM Tax: }\eqref{eq:opt_tax}, \quad \forall t, d^{t-1}, a^{t-1}, \\
              \quad & g^a_t = \text{BR}^a_t(\beta), \quad \forall t,\\
              \quad & \beta \text{ is consistent with } (f,g),
        \end{empheq}
\end{subequations}
\fi
where the consistency of belief system~$\beta$ can be described by~\eqref{eq:belief_hd0}--\eqref{eq:belief_hm} on the equilibrium path, or by the iteratively update equations~\eqref{eq:belief_ha}, \eqref{eq:belief_hd_off} and~\eqref{eq:belief_hd_ignore} in Appendix~\ref{appx:belief_off_path} off the equilibrium path. The optimization problem~\eqref{opt:offline} is an offline optimization problem for information design. By solving~\eqref{opt:offline}, we can derive the equilibrium for this model.
Note however, that this is a very complex optimization problem due to the interdependence of the strategies on the beliefs and the beliefs on the strategies (due to the consistency constraint that renders beliefs strategy dependent).
In the next section we explore further the structure of the model in order to simplify this optimization problem.

\section{Summary-based Mechanism Design}
\label{sec:sb_md}

The recommendation mechanism $f^e_t$ at time~$t$ is a function of $d^{t-1},a^{t-1},m^t$. This type of design is problematic in practice as the space of dependencies grows linearly in time and therefore difficult to keep track on. In this section we will show that the coordinator can restrict attention to a summary-based mechanism instead, where the summary is a group of variables that summarize the history $d^{t-1},a^{t-1},m^t$ and whose space is time-invariant. We will see later that the summary-based design simplifies the coordinator's problem~\eqref{opt:offline}, revealing a systematic way of finding PBE rather than a brute force algorithm. It also provides a more concise framework for analysis.

In this section, we will first introduce the notion of \emph{partial strategy}, which is a function mapping the private part of the information to the final action. We will reformulate~\eqref{opt:offline} as a decision making problem with partial strategies as actions.
This decomposes every strategy, including $f^e$, into a public information part and a private information part.
%
Subsequently, utilizing the partial strategy representation as a tool, we show that the offline optimization problem~\eqref{eq:opt_obj} can be restated as a \emph{Markov decision process} (MDP). Based on this MDP, we will see that using summary-based strategies is without loss of generality for the original problem.

\subsection{Partial Strategies}

The idea of partial strategies (also known as prescriptions) originates from decentralized stochastic control~\cite{WaVa83}, and is widely used in dynamic teams and games with asymmetric information~\cite{nayyar2013, VaSiAn19, tang2022dynamic}. The use of a partial strategy decomposes a full strategy into a common-information-dependent part and a private-information-dependent part. By regarding partial strategies as actions, we end up with belief updates that are strategy independent, thus significantly simplifying the problem structure. In this part, we introduce the definitions of partial strategies for both the coordinator and the agents, and then reformulate~\eqref{opt:offline} accordingly.

Starting from the coordinator's strategy $f^e(\cdot \vert d^{t-1},a^{t-1},m^t)$, the introduction of a partial strategy decomposes the decision into two steps. The first step is to choose a partial strategy (conditional distribution) $\theta^e_t$ based on the public information $d^{t-1},a^{t-1}$ through a mapping $\phi_t$, i.e., $\theta^e_t=\phi_t[d^{t-1},a^{t-1}]$. Then, the second step is to generate the decision variable $E_t$ based on the private information $m^t$ (i.e., the information which is not public to all participants) through the conditional distribution $\theta^e_t$, i.e., $E_t \sim \theta^e_t(\cdot \vert m^{t})$. Formally we have
\ifonecol
       \begin{subequations}
       \begin{align}
              \theta^e_t : \cM^t \to \Delta(\cA)  &\quad \text{with } E_t \sim \theta^e_t(\cdot \vert m^{t}) \\
              \phi_t:  \cD^{t-1}\times \cA^{t-1} \to (\cM^t \to \Delta(\cA)) &\quad \text{with }\theta^e_t = \phi_t[d^{t-1},a^{t-1}], \\
              f^e_t(e_t \vert d^{t-1},a^{t-1},m^t) &= \phi_t[d^{t-1},a^{t-1}](e_t \vert m^t) = \theta^e_t(e_t \vert m^t).
       \end{align}
       \end{subequations}
\else
       \begin{subequations}
              \begin{align}
                     &\theta^e_t : \cM^t \to \Delta(\cA) \  \text{with } E_t \sim \theta^e_t(\cdot \vert m^{t}), \\
                     &\phi_t:  \cD^{t-1}\times \cA^{t-1} \to (\cM^t \to \Delta(\cA)),  \nonumber \\
                     &\qquad\qquad \text{with }\theta^e_t = \phi_t[d^{t-1},a^{t-1}], \\
                     &\splitfrac{f^e_t(e_t \vert d^{t-1},a^{t-1},m^t), }{= \phi_t[d^{t-1},a^{t-1}](e_t \vert m^t) = \theta^e_t(e_t \vert m^t).}
              \end{align}
       \end{subequations}
\fi
Note that for better readability we use square brackets for mappings that result in partial functions.
Observe that with the above decomposition, designing $f^e$ is equivalent to designing $\phi$.
Similarly, we can introduce partial functions for agents' strategies as follows.
\ifonecol
       \begin{subequations}
       \begin{align}
              \gamma^m_t : \cY \to \Delta(\cM)  &\quad \text{with } M_t \sim \gamma^m_t(\cdot \vert y_t) \\
              \psi^m_t:  \cF \times \cD^{t-1} \times \cA^{t-1} \to (\cY \to \Delta(\cM)) &\quad \text{with }\gamma^m_t = \psi^m_t[f,d^{t-1},a^{t-1}], \\
              g^m_t(m_t \vert f,d^{t-1},a^{t-1},y_t) &= \psi^m_t[f,d^{t-1},a^{t-1}](m_t \vert y_t) = \gamma^m_t(m_t \vert y_t).
       \end{align}
       \end{subequations}
       \begin{subequations}
              \label{eq:agent_partial}
       \begin{align}
              \gamma^a_t : \cY \to \Delta(\cA)  &\quad \text{with } A_t \sim \gamma^a_t(\cdot \vert y_t) \\
              \psi^a_t:  \cF \times \cD^{t-1} \times \cA^{t-1} \to (\cY \to \Delta(\cA)) &\quad \text{with }\gamma^a_t = \psi^a_t[f,d^{t-1},a^{t-1}], \\
              g^a_t(a_t \vert f,d^{t-1},a^{t-1},y_t) &= \psi^a_t[f,d^{t-1},a^{t-1}](a_t \vert y_t) = \gamma^a_t(a_t \vert y_t).
       \end{align}
       \end{subequations}
       \begin{subequations}
       \begin{align}
              \gamma^d_t \in \Delta(\cD)  &\quad \text{with } D_t \sim \gamma^d_t(\cdot) \\
              \psi^d_t:  \cF \times \cD^{t-1} \times \cA^{t-1} \to \Delta(\cD) &\quad \text{with }\gamma^d_t = \psi^d_t[f,d^{t-1},a^{t-1}], \\
              g^d_t(d_t \vert f,d^{t-1},a^{t-1}) &= \psi^d_t[f,d^{t-1},a^{t-1}](d_t) = \gamma^d_t(d_t),
       \end{align}
       \end{subequations}
\else
       \begin{subequations}
              \begin{align}
                     &\gamma^m_t : \cY \to \Delta(\cM) \ \text{with } M_t \sim \gamma^m_t(\cdot \vert y_t) \\
                     &\psi^m_t:  \cF \times \cD^{t-1} \times \cA^{t-1} \to (\cY \to \Delta(\cM))\nonumber \\
                      &\qquad \qquad \text{with }\gamma^m_t = \psi^m_t[f,d^{t-1},a^{t-1}], \\
                     &\splitfrac{g^m_t(m_t \vert f,d^{t-1},a^{t-1},y_t) }{= \psi^m_t[f,d^{t-1},a^{t-1}](m_t \vert y_t) = \gamma^m_t(m_t \vert y_t).}
              \end{align}
       \end{subequations}
       \begin{subequations}
              \label{eq:agent_partial}
              \begin{align}
                     &\gamma^a_t : \cY \to \Delta(\cA)  \quad \text{with } A_t \sim \gamma^a_t(\cdot \vert y_t) \\
                     &\splitfrac{\psi^a_t:  \cF \times \cD^{t-1} \times \cA^{t-1} \to (\cY \to \Delta(\cA))}{\text{with }\gamma^a_t = \psi^a_t[f,d^{t-1},a^{t-1}],} \\
                     &\splitfrac{g^a_t(a_t \vert f,d^{t-1},a^{t-1},y_t) }{= \psi^a_t[f,d^{t-1},a^{t-1}](a_t \vert y_t) = \gamma^a_t(a_t \vert y_t).}
              \end{align}
       \end{subequations}
       \begin{subequations}
              \begin{align}
                     &\gamma^d_t \in \Delta(\cD)  \ \text{with } D_t \sim \gamma^d_t(\cdot) \\
                     &\splitfrac{\psi^d_t:  \cF \times \cD^{t-1} \times \cA^{t-1} \to \Delta(\cD) }{\text{with }\gamma^d_t = \psi^d_t[f,d^{t-1},a^{t-1}],} \\
                     &\splitfrac{g^d_t(d_t \vert f,d^{t-1},a^{t-1}) }{= \psi^d_t[f,d^{t-1},a^{t-1}](d_t) = \gamma^d_t(d_t),}
              \end{align}
       \end{subequations}
\fi
where $\cF$ denotes the space of all strategies $f$.
Furthermore, we define the public belief~$\pi_t\in \Delta(\cM^{t-1})$ utilizing the belief system as follows:
\begin{equation}
       \pi_t(m^{t-1}) := \sum_w \beta(w,m^{t-1} \vert h^d_t).
\end{equation}
We now show that the belief $\pi_t$ can be evaluated in a recursive way independently of the policy~$\phi$.

\begin{lemma}[Belief Evolution]
       \label{lem:belief}
       The joint belief $\beta(w,m^{t-1} \vert h^d_t)$ can be evaluated from $\pi_t(m^{t-1})$ as follows
       \ifonecol
              \begin{align}
                     \label{eq:belief_decompose}
                     &\beta(w,m^{t-1} \vert h^d_t) = \pi_t(m^{t-1}) q(w \vert n(m^{t-1})),
              \end{align}
       \else
              \begin{align}
                     \label{eq:belief_decompose}
                     &\beta(w,m^{t-1} \vert h^d_t) = \pi_t(m^{t-1}) q(w \vert n(m^{t-1})),
              \end{align}
       \fi
       where the function $q(\cdot \vert \cdot)$ is defined as
       \begin{equation}
              q(1 \vert n_t) = \frac{(\bar{p}/p)^{n_t}}{(\bar{p}/p)^{n_t}+1}, \quad q(-1 \vert n_t) = \frac{1}{(\bar{p}/p)^{n_t}+1}.
       \end{equation}
       and $n(m^{t-1}):=\sum_{\tau=1}^{t-1} m_{\tau}$.
       Furthermore, the public belief~$\pi_t$, can be updated as
       \begin{equation}
              \label{eq:belief}
              \pi_{t+1} = T_t(\pi_t,d_t,a_t,\theta^e_t),
       \end{equation}
       which is $\phi$-policy independent. Consequently, $\pi_t=T_{1:t-1}(d^{t-1},a^{t-1},\theta^e_{1:t-1})$ is $\phi$-policy independent.
\end{lemma}
\begin{IEEEproof}
       See Appendix~\ref{appx:pf:lem:belief}.
\end{IEEEproof}

Now, according to agents' rationality~\eqref{eq:g^m_t}, \eqref{eq:g^a_t}, \eqref{eq:g^d_t}, and using a similar argument to the one we used to derive belief-based strategies~\eqref{eq:BRs}, we can show that at equilibrium a rational agent forms her partial strategies depending on history $h^d_t$ only through $\pi_t$, i.e.,
\ifonecol
       \begin{equation}
              \gamma^d_t = \hat{\psi}^d[\theta^e_t,\pi_t],
              \quad  \gamma^a_t = \hat{\psi}^a[\pi_t],
              \quad \text{ and } \gamma^m_t = \hat{\psi}^m[\theta^e_t,\pi_t].
       \end{equation}
\else
       \begin{align}
                     \gamma^d_t = \hat{\psi}^d[\theta^e_t,\pi_t],
              \quad  \gamma^a_t = \hat{\psi}^a[\pi_t],
              \nonumber \\
              \text{ and } \gamma^m_t = \hat{\psi}^m[\theta^e_t,\pi_t].
       \end{align}
\fi
The above equations are the analog of those in~\eqref{eq:BRs}.
With the notations of partial strategies, the profit maximizing tax can be rewritten as
\ifonecol
       \begin{align}
              \label{eq:tax_par}
              &f^x_t(d^{t-1},a^{t-1}) = \sum_{y_t,w,m^{t-1}} Q^y(y_t \vert w) q(w \vert n(m^{t-1}))\pi_t(m^{t-1}) \left(\theta^e_t(w \vert (m^{t-1}, y_t))-\gamma^a_t(w \vert y_t) \right) \nonumber \\
              &=: \hat{f}^x_t(\pi_t,\theta^e_t),
       \end{align}
\else
       \begin{align}
              &f^x_t(d^{t-1},a^{t-1}) \nonumber\\
              &= \sum_{y_t,w,m^{t-1}} Q^y(y_t \vert w) q(w \vert n(m^{t-1}))\pi_t(m^{t-1}) \nonumber\\
              &\quad \cdot \left(\theta^e_t(w \vert (m^{t-1}, y_t))-\gamma^a_t(w \vert y_t) \right) \nonumber\\
              \label{eq:tax_par}
              &=: \hat{f}^x_t(\pi_t,\theta^e_t),
       \end{align}
\fi

At time~$t$, suppose the history~$d^{t-1},a^{t-1}$ induces belief~$\pi_t$. Then, the truth-telling constraints~\eqref{eq:truth-telling} can be written as
\ifonecol
       \begin{equation}
              \label{eq:truth-telling_par}
              \sum_{w,m^{t-1}} Q^y(y_t \vert w)q(w \vert n(m^{t-1}))\pi_t(m^{t-1})  \left(\theta^e_t(w \vert (m^{t-1},y_t)) - \theta^e_t(w \vert (m^{t-1},-y_t)) \right) \geq 0, \ \forall y_t.
       \end{equation}
\else
       \begin{align}
              \label{eq:truth-telling_par}
              &\sum_{w,m^{t-1}} Q^y(y_t \vert w)q(w \vert n(m^{t-1}))\pi_t(m^{t-1}) \nonumber\\
              &\cdot \left(\theta^e_t(w \vert (m^{t-1},y_t)) - \theta^e_t(w \vert (m^{t-1},-y_t)) \right) \geq 0, \ \forall y_t.
       \end{align}
\fi

Substituting the above expressions to the problem~\eqref{opt:offline}, and substituting the optimization variable~$f^e$ by the policy~$\phi$, we obtain the new optimization problem
\ifonecol
       \begin{subequations}
              \label{opt:partial}
              \begin{empheq}[box=\fbox]{align}
                     \underset{\phi}{\text{maximize}} \quad& \E^\phi \left\{ \sum_{t=1}^{T}\hat{f}^x_t (\Pi_t,\Theta^e_t)\right\}, \\
                     \text{subject to}
                     \label{con:TT_par}
                     \quad& \text{Truth-Telling:~\eqref{eq:truth-telling_par}}, \quad \forall y_t, \\
                     \label{con:PMT_par}
                     \quad& \text{Profit Maximizing Tax:~\eqref{eq:tax_par}}, \\
                     \label{con:BR_par}
                     \quad& \theta^e_t = \phi_t[d^{t-1},a^{t-1}], \ \gamma^a_t = \hat{\psi}^a [\pi_t],\\
                     \label{con:belief_par}
                     \quad& \forall t, d^{t-1}, a^{t-1};\  \pi_{t+1} = T_t(\pi_t, d_t, a_t, \theta^e_t).
              \end{empheq}
       \end{subequations}
\else
       \begin{subequations}
              \label{opt:partial}
              \begin{empheq}[box=\fbox]{align}
                     \underset{\phi}{\max} \quad& \E^\phi \left\{ \sum_{t=1}^{T}\hat{f}^x_t (\Pi_t,\Theta^e_t)\right\} \\
                     \text{s.t.}
                     \label{con:TT_par}
                     \quad& \text{TT:~\eqref{eq:truth-telling_par}}, \quad \forall y_t, \\
                     \label{con:PMT_par}
                     \quad& \text{PM Tax:~\eqref{eq:tax_par}}, \\
                     \label{con:BR_par}
                     \quad& \theta^e_t = \phi_t[d^{t-1},a^{t-1}], \nonumber\\
                     &\qquad \gamma^a_t = \hat{\psi}^a [\pi_t], \\
                     \label{con:belief_par}
                     \quad& \forall t, d^{t-1}, a^{t-1}; \nonumber\\
                     &\qquad   \pi_{t+1} = T_t(\pi_t, d_t, a_t, \theta^e_t).
              \end{empheq}
       \end{subequations}
\fi
The solution of the new problem is equivalent to that of~\eqref{opt:offline}, because given a solution $\phi$ to the new problem, one can always set $f^e_t(e_t \vert d^{t-1},a^{t-1},m^t)=\phi_t[d^{t-1},a^{t-1}](e_t \vert m^t)$, and vice versa.

The partial strategy framework provides more than an alternate representation of the offline problem. With the introduction of the partial strategy $\theta^e$, the belief update~\eqref{con:belief_par} is disentangled from the strategy~$\phi$.
In the following we show that the state process~$\{\Pi_t\}_{t\geq 1}$
is an MDP with action $\theta^e_t$ and instantaneous reward $\hat{f}^x_t(\pi_t,\theta^e_t)$. Theorem~\ref{thm:MDP_suff} shows the formal connection between optimization problem~\eqref{opt:partial} and Decision Process~\ref{mdp:direct}.
       \begin{mdpmodel}
              \label{mdp:direct}
              \

              \begin{itemize}
                     \item \emph{State.} The public belief $\pi_t \in \Delta(\cM^{t-1})$.
                     \item \emph{Action.} A partial recommendation strategy~$\theta^e_t: \cM^t \rightarrow \Delta(\cA)$. 
                     \item \emph{Instantaneous reward.} $\hat{f}^x_t(\pi_t,\theta^e_t)$.
                     \item \emph{Feasible action set.} For each time~$t$, the action~$\theta^e_t$ satisfies the TT constraints~\eqref{eq:truth-telling_par}.
              \end{itemize}
       \end{mdpmodel}

       \begin{theorem}
              \label{thm:MDP_suff}
              The Decision Process~\ref{mdp:direct} is an MDP. Moreover, the optimal solution to this MDP is an optimal solution to the offline optimization problem~\eqref{opt:partial}.
       \end{theorem}
       \begin{IEEEproof}
              See Appendix~\ref{appx:pf:thm:MDP_suff}.
       \end{IEEEproof}

We will not further discuss the dynamic programming solution for Decision Process~\ref{mdp:direct}. The reason is that this MDP has a state space~$\Delta(\cM^{t-1})$ that increases with time, as well as a large (and increasing) action space $\cM^t \rightarrow \Delta(\cA)$,
resulting in a prohibitively complex dynamic program.
Instead, we proceed with a further simplification in the next subsection by summarizing the sequence $m^{t-1}$ into an integer sufficient statistic.

\subsection{Summary-Based Control}

In this subsection, we will show a simplification on the recommendation mechanism based on the summary $n_t$ defined by
\begin{equation}
       n_t := n(m^{t-1}) = \sum_{\tau=1}^{t-1} m_\tau,
\end{equation}
which is the difference between the number of $+1$'s and the number of $-1$'s in the sequences $m^{t-1}$. Let $m^{t-1}:n_t$ denote the sequences $m^{t-1}$ with $n(m^{t-1})=n_t$.
Define the belief $\eta_t\in \Delta(\Integer)$ on $n_t$ as
\begin{equation}
       \eta_t(n_t) :=  \sum_{m^{t-1} : n_t} \pi_t(m^{t-1}).
\end{equation}

The introduction of the belief $\eta_t(\cdot)$ together with~\eqref{eq:belief_decompose} indicates an alternative way (alternative to Lemma~\ref{lem:belief}) to form a belief on~$w$. To form a belief on~$w$ agents first form a belief on $n_t$, and then deduce a belief on $w$ from $n_t$. Therefore, if one wants to know the probability of a certain state~$w$, it is enough to know the belief on $n_t$ rather than learn the detailed joint distribution on $m^{t-1}$.
As a direct result, from~\eqref{eq:val_a} and~\eqref{eq:g^a_t}, since the agents' rationality at infosets $h^a_t$ depends on the belief through the marginal $\beta(w \vert h^d_t)$
\begin{equation}
       \sum_{m^{t-1}} \beta(w, m^{t-1} \vert h^d_t) = \sum_{n_t} \eta_t(n_t) q(w \vert n_t),
\end{equation}
the agents' strategy on infoset $h^a_t$ in~\eqref{eq:agent_partial} can be further expressed in terms of $\eta_t$:
\ifonecol
       \begin{subequations}
              \label{eq:agent_summary_based}
              \begin{align}
                     \gamma^a_t: \cY \to \Delta(\cA) &\quad \text{with } A_t \sim \gamma^a_t(\cdot \vert y_t), \\
                     \tilde{\psi}^a_t: \Delta(\mathbb{Z}) \to (\cY \to \Delta(\cA)) &\quad \text{with } \gamma^a_t=\tilde{\psi}^a_t[\eta_t], \\
                     g^a_t(a_t \vert f, d^{t-1}, a^{t-1}, y_t) = \tilde{\psi}[\eta_t](a_t \vert y_t) &= \gamma^a_t(a_t \vert y_t).
              \end{align}
       \end{subequations}
\else
       \begin{subequations}
              \label{eq:agent_summary_based}
              \begin{align}
                     &\gamma^a_t: \cY \to \Delta(\cA)\quad \text{with } A_t \sim \gamma^a_t(\cdot \vert y_t), \\
                     &\tilde{\psi}^a_t: \Delta(\mathbb{Z}) \to (\cY \to \Delta(\cA)) \quad \text{with } \gamma^a_t=\tilde{\psi}^a_t[\eta_t], \\
                     &g^a_t(a_t \vert f, d^{t-1}, a^{t-1}, y_t) = \tilde{\psi}[\eta_t](a_t \vert y_t) = \gamma^a_t(a_t \vert y_t).
              \end{align}
       \end{subequations}
\fi
The specific form is defined by~\eqref{eq:val_a} and~\eqref{eq:g^a_t}, but substituting $\sum_{m^{t-1}} \beta(w, m^{t-1} \vert h^d_t)$ with $\sum_{n_t} \eta_t(n_t) q(w \vert n_t)$.

The above discussion inspires us to think that the strategies implied by Theorem~\ref{thm:MDP_suff}
(which depend on $\pi_t\in\Delta(\cM^{t-1})$) can be further summarized so that they depend on $\eta_t\in\Delta(\Integer)$.
To show that, we will first construct a summary-based MDP in Lemma~\ref{lem:pi_hat_mdp}. Then, in Theorem~\ref{thm:belief_n}, we will show that solving this MDP provides a solution to the optimization problem~\eqref{opt:partial}.
Essentially we claim that the coordinator can without loss of optimality restrict attention to recommendation functions $f^e$ that have the following structure (utilizing again partial strategies):
%
\ifonecol
       \begin{subequations}
       \begin{align}
              \tilde{\theta}^e_t : \Integer \times \cM \to \Delta(\cA)  &\quad \text{with } E_t \sim \tilde{\theta}^e_t(\cdot \vert n_t,m_t) \\
              \tilde{\phi}_t:  \Delta(\Integer) \to (\Integer \times \cM \to \Delta(\cA)) &\quad \text{with }\tilde{\theta}^e_t = \tilde{\phi}_t[\eta_t], \\
              f^e_t(e_t \vert d^{t-1},a^{t-1},m^t) &= \tilde{\phi}_t[\eta_t](e_t \vert n_t,m_t) = \tilde{\theta}^e_t(e_t \vert n_t,m_t).
       \end{align}
       \end{subequations}
\else
       \begin{subequations}
              \begin{align}
                     &\tilde{\theta}^e_t : \Integer \times \cM \to \Delta(\cA),  \nonumber\\
                     &\qquad \text{with } E_t \sim \tilde{\theta}^e_t(\cdot \vert n_t,m_t) \\
                     &\tilde{\phi}_t:  \Delta(\Integer) \to (\Integer \times \cM \to \Delta(\cA)), \nonumber\\
                     & \qquad \text{with }\tilde{\theta}^e_t = \tilde{\phi}_t[\eta_t], \\
                     &f^e_t(e_t \vert d^{t-1},a^{t-1},m^t) = \tilde{\phi}_t[\eta_t](e_t \vert n_t,m_t) \nonumber\\
                     &= \tilde{\theta}^e_t(e_t \vert n_t,m_t).
              \end{align}
       \end{subequations}
\fi
Towards this goal we define the following process.

\begin{mdpmodel}[Summary-Based MDP]
       \label{mdp:summary-based}
       \
       \begin{itemize}
              \item State. The public belief on the summary: $\eta_t \in \Delta(\Integer)$.
              \item Action. A summary-based partial strategy $\tilde{\theta}^e_t: \Integer \times \cM \to \Delta(\cA)$.
              \item Instantaneous reward. $\tilde{f}^x(\eta_t, \tilde{\theta}^e_t)$ (to be determined in Lemma~\ref{lem:pi_hat_mdp}).
              \item Feasible action set. A subset $\tilde{S}_t(\eta_t)$ (to be determined in Lemma~\ref{lem:pi_hat_mdp}) that describes the TT constraints~\eqref{eq:truth-telling_par}.
       \end{itemize}
\end{mdpmodel}

\begin{lemma}
       \label{lem:pi_hat_mdp}
       The Decision Process~\ref{mdp:summary-based} is an MDP.
       Furthermore, the feasible set and the instantaneous reward have time-invariant forms $\tilde{S}(\eta_t)$ and $\tilde{f}^x(\eta_t, \tilde{\theta}^e_t)$ as follows:
       \ifonecol
              \begin{align}
                     \label{def:summary-based_feasible_set}
                     &\tilde{S}(\eta_t) = \left\{ \tilde{\theta}^e_t: \sum_{w,n_t} Q^y(y_t \vert w) \eta_t(n_t) q(w \vert n_t)
                     \left(\tilde{\theta}^e_t(w \vert n_t,y_t) - \tilde{\theta}^e_t(w \vert n_t,-y_t)\right) \geq 0, \forall y_t \right\}, \\
                     \label{def:summary-based_instantaneous_reward}
                     &\tilde{f}^x(\eta_t,\tilde{\theta}^e_t) = \sum_{y_t, w, n_t} Q^y(y_t \vert w) \eta_t(n_t) q(w \vert n_t)  \left(\tilde{\theta}^e_t(w \vert n_t,y_t)-\gamma^a_t(w \vert y_t)\right).
              \end{align}
       \else
              \begin{align}
                     \tilde{S}(\eta_t) = \left\{ \tilde{\theta}^e_t: \sum_{w,n_t} Q^y(y_t \vert w) \eta_t(n_t) q(w \vert n_t) \right. \nonumber\\
                     \label{def:summary-based_feasible_set}
                     \left. \cdot \left(\tilde{\theta}^e_t(w \vert n_t,y_t) - \tilde{\theta}^e_t(w \vert n_t,-y_t)\right) \geq 0, \forall y_t \right\},
              \end{align}
              \begin{align}
                     \tilde{f}^x&(\eta_t,\tilde{\theta}^e_t) = \sum_{y_t, w, n_t} Q^y(y_t \vert w) \eta_t(n_t) q(w \vert n_t)  \nonumber\\
                     \label{def:summary-based_instantaneous_reward}
                     &\cdot \left(\tilde{\theta}^e_t(w \vert n_t,y_t)-\gamma^a_t(w \vert y_t)\right).
              \end{align}
       \fi
\end{lemma}
\begin{IEEEproof}
       See Appendix~\ref{appx:pf:lem:pi_hat_mdp}.
\end{IEEEproof}

For Decision Process~\ref{mdp:summary-based}, Bellman equations characterize Markov optimal strategies:
\ifonecol
       \begin{subequations}
              \label{eq:bdp:finite}
              \begin{align}
                     \label{eq:DP_hat_T}
                     \tilde{V}_T(\eta_T) =& \max_{\tilde{\theta}^e \in \tilde{S}(\eta_T)} \tilde{f}^x_T(\eta_T,\tilde{\theta}^e_T), \\
                     \label{eq:DP_hat_T_strategy}
                     \tilde{\phi}^*_T(\eta_T) \in& \arg \max_{\tilde{\theta}^e_T \in \tilde{S}(\eta_T)} \tilde{f}^x(\eta_T,\tilde{\theta}^e_T), \\
                     \label{eq:DP_hat_t}
                     \tilde{V}_t(\eta_t) =& \max_{\tilde{\theta}^e_t \in \tilde{S}(\eta_t)} \tilde{f}^x(\eta_t,\tilde{\theta^e_t}) + \E \left\{ \tilde{V}_{t+1}(H_{t+1})  \vert  \eta_t, \tilde{\theta}^e_t \right\}, \quad t=1,\ldots,T-1, \\
                     \label{eq:DP_hat_t_strategy}
                     \tilde{\phi}^*_t(\eta_t) \in& \arg \max_{\tilde{\theta}^e_t \in \tilde{S}(\eta_t)} \tilde{f}^x(\eta_t,\tilde{\theta^e_t}) + \E \left\{ \tilde{V}_{t+1}(H_{t+1})  \vert  \eta_t, \tilde{\theta}^e_t \right\}, \quad t=1,\ldots,T-1.
              \end{align}
       \end{subequations}
\else
       \begin{subequations}
              \label{eq:bdp:finite}
              \begin{equation}
                     \label{eq:DP_hat_T}
                     \tilde{V}_T(\eta_T) = \max_{\tilde{\theta}^e \in \tilde{S}(\eta_T)} \tilde{f}^x_T(\eta_T,\tilde{\theta}^e_T),
              \end{equation}
              \begin{equation}
                     \label{eq:DP_hat_T_strategy}
                     \tilde{\phi}^*_T(\eta_T) \in \arg \max_{\tilde{\theta}^e_T \in \tilde{S}(\eta_T)} \tilde{f}^x(\eta_T,\tilde{\theta}^e_T),
              \end{equation}
              \begin{align}
                     &\tilde{V}_t(\eta_t) = \nonumber\\
                     \label{eq:DP_hat_t}
                     &\max_{\tilde{\theta}^e_t \in \tilde{S}(\eta_t)}  \tilde{f}^x(\eta_t,\tilde{\theta^e_t})
                     + \E \left\{ \tilde{V}_{t+1}(H_{t+1})  \vert  \eta_t, \tilde{\theta}^e_t \right\},
              \end{align}
              \begin{align}
                     \tilde{\phi}^*_t(\eta_t) \in \arg &\max_{\tilde{\theta}^e_t \in \tilde{S}(\eta_t)} \tilde{f}^x(\eta_t,\tilde{\theta^e_t}) \nonumber\\
                     \label{eq:DP_hat_t_strategy}
                     &+ \E \left\{ \tilde{V}_{t+1}(H_{t+1})  \vert  \eta_t, \tilde{\theta}^e_t \right\}.
              \end{align}
       \end{subequations}
\fi

Decision Process~\ref{mdp:summary-based} focuses seemingly on a subclass $\tilde{\phi}$ of all feasible mechanisms. Indeed, $\tilde{\phi}_t$ only utilizes $\eta_t$ (instead of $d^{t-1}a^{t-1}$, or $\pi_t$) to generate the partial strategy $\tilde{\theta}^e_t$, and $\tilde{\theta}^e_t$ only uses $n_t$ and $m_t$ (instead of $m^t$) to generate a recommendation. Since $\tilde{\phi}$ is a special case of a general mechanism in terms of the dependencies, one might wonder if there is a gap between the optimal solutions to the optimization problem~\eqref{opt:partial} and this MDP. Theorem~\ref{thm:belief_n} answers this question by showing the sufficiency of strategies $\tilde{\phi}$ for the optimization problem~\eqref{opt:partial}.

\begin{theorem}
       \label{thm:belief_n}
       Without loss of optimality in the optimization problem~\eqref{opt:partial}, it is sufficient for the coordinator to restrict attention to strategies $\tilde{\phi}$ of the form
       \begin{equation}
              \tilde{\phi}_t:\Delta({\Integer}) \to \left( \Integer \times \cM \to \Delta(\cA) \right).
       \end{equation}
       The optimal strategy $\tilde{\phi}^*$ can be found from the Bellman equations~\eqref{eq:bdp:finite}.
\end{theorem}
\begin{IEEEproof}
       We note here that the proof does not simply use an MDP argument on Decision Process~\ref{mdp:summary-based}. Unlike Theorem~\ref{thm:MDP_suff} which only narrows down the domain of the strategy $\phi$, Theorem~\ref{thm:belief_n} requires us to demonstrate two things: (i) at each time $t$, the change of actions from $\hat{\theta}^e$ to $\tilde{\theta}^e$ does not affect the result, (ii) it is sufficient to use a Markovian strategy depending only on $\eta_t$. See Appendix~\ref{appx:pf:thm_belief_n} for details.
\end{IEEEproof}

\subsection{Extension to Infinite Horizon}

In the finite horizon setting, we find it sufficient for the coordinator to use an optimal strategy derived from a simplified time-invariant MDP. Although with finite time horizon, the optimal strategy is not necessarily time-invariant, if this simplification can be applied to the infinite-horizon problem, the coordinator may be able to restrict attention to a time-invariant $\tilde{\phi}$, so that the analysis of the learning process over an infinite horizon is tractable.

We extend the information coordinator problem into infinite horizon by introducing a discount factor $\delta \in (0,1)$, so that the coordinator wants to maximize the following:
\begin{equation}
       \E^\phi \left\{ \sum_{t=1}^\infty \delta^{t-1} \hat{f}^x_t(\Pi_t,\Theta^e_t) \right\}.
\end{equation}

Similar to what we did in finite horizon, in the infinite-horizon setting, we can define a simplified MDP with state $\eta_t \in \Delta(\mathbb{Z})$, action $\tilde{\theta}^e_t: \mathbb{Z} \times \cM \to \Delta(\cA)$ and instantaneous reward $\tilde{f}^x_t(\eta_t,\tilde{\theta}^e_t)$. For this simplified MDP, the value function is in the space $\cV = \{ V:\Delta(\mathbb{Z}) \to \Real \}$. Define a norm for value function $V$: $\lVert V \rVert = \sup_{\eta \in \Delta(\mathbb{Z})}  \vert  V(\eta)  \vert $. Since $V(\eta) \in [0,1/(1-\delta)]$ is bounded, one can show that~$\cV$ is a Banach space in norm $\lVert \cdot \rVert$, so that in the simplified MDP it is sufficient to search for a time-invariant optimal strategy based on the basic result of infinite-horizon MDPs~\cite[Chap. 8]{kumar2015stochastic}.

The next theorem serves as the infinite-horizon version of Theorem~\ref{thm:belief_n}, which builds the connection between the optimal strategy of the original MDP and the simplified MDP.
\begin{theorem}
       \label{thm:belief_n_infinite}
       Without loss of optimality, in the coordinator's optimization problem with infinite horizon, it is sufficient for the coordinator to restrict the attention to stationary summary-based strategies $\tilde{\phi}$ with $\tilde{\phi}_t = \tilde{\phi}_0$ for all $t$, where
       \begin{equation}
              \tilde{\phi}_0: \Delta(\Integer) \to \left( \Integer \times \cM \to \Delta(\cA) \right).
       \end{equation}
       Furthermore, the optimal strategy $\tilde{\phi}_0^*$ satisfies the Bellman equation:
       \ifonecol
              \begin{subequations}
                     \label{eq:bdp:infinite}
                     \begin{align}
                            \tilde{V}(\eta) = \max_{\tilde{\theta}^e \in \tilde{S}(\eta)} \tilde{f}^x(\eta, \tilde{\theta}^e) + \E \left\{ \tilde{V}(H_{t+1}) \vert H_t = \eta, \tilde{\Theta}^e_t = \tilde{\theta}^e \right\}, \\
                            \tilde{\phi}_0^*(\eta) \in \arg \max_{\tilde{\theta}^e \in \tilde{S}(\eta)} \tilde{f}^x(\eta, \tilde{\theta}^e) + \E \left\{ \tilde{V}(H_{t+1}) \vert H_t = \eta, \tilde{\Theta}^e_t = \tilde{\theta}^e \right\}.
                     \end{align}
              \end{subequations}
       \else
              \begin{subequations}
                     \label{eq:bdp:infinite}
                     \begin{align}
                            &\tilde{V}(\eta) = \nonumber\\
                            & \max_{\tilde{\theta}^e \in \tilde{S}(\eta)} \tilde{f}^x(\eta, \tilde{\theta}^e) + \E \left\{ \tilde{V}(H_{t+1}) \vert H_t = \eta, \tilde{\Theta}^e_t = \tilde{\theta}^e \right\},
                     \end{align}
                     \begin{align}
                            \tilde{\phi}_0^*(\eta) \in& \arg \max_{\tilde{\theta}^e \in \tilde{S}(\eta)} \tilde{f}^x(\eta, \tilde{\theta}^e) \nonumber\\
                            &+ \E \left\{ \tilde{V}(H_{t+1}) \vert H_t = \eta, \tilde{\Theta}^e_t = \tilde{\theta}^e \right\}.
                     \end{align}
              \end{subequations}
       \fi
\end{theorem}
\begin{IEEEproof}
       See Appendix~\ref{appx:pf:thm:belief_n_infinite}.
\end{IEEEproof}

\section{Suboptimal Solutions}
\label{sec:suboptimal}

We showed in Theorem~\ref{thm:belief_n} that a PBE can be found by solving the MDP in Decision Process~\ref{mdp:summary-based}. However, solving the dynamic program in~\eqref{eq:bdp:finite} is still a difficult task since the domain of the value functions is $\Delta(\Integer)$ and the action space is the set of functions $\Integer \times \cM \to \Delta(\cA)$. For a finite horizon $T$, since the summary $ \vert n_t \vert \leq T$ we can easily deduce that the value functions have domain $[0,1]^{2T+1}$ and the action space is $[0,1]^{2(2T+1)}$. Although one may attempt to solve such dynamic program for small values of $T$ using standard approximation techniques, for moderate to large values of $T$ the complexity becomes prohibitive.

In this section, we study suboptimal solutions for the coordinator that possess certain desirable properties.
To motivate this study consider the sequential Bayesian learning system with a coordinator, but now the coordinator is not the one who designs the mechanism. Instead, there is a (non-for-profit) third party that wants to improve the social welfare by designing a mechanism, but would like to find someone else to administer it. To incentivize a self-interested coordinator to take up the administration task, the mechanism need not maximize the coordinator's profit, but should at least provide a positive profit (this can be regarded as the individual rationality constraint for the coordinator).
Motivated by the above scenario, in this section, we define a subclass of FPM mechanisms, and construct a profitable FPM mechanism in that subclass that also improves the social welfare.


We consider a subclass of feasible mechanisms called nonnegative taxation and feasible profit maximizing (NT-FPM) mechanisms, defined as follows.
\begin{definition}[NT-FPM Mechanism]
       A nonnegative taxation and profit maximization feasible mechanism is an FPM mechanism with nonnegative taxation, i.e., $f^x_t(\cdot)$ is a nonnegative function for all~$t$.
\end{definition}
As mentioned above, NT-FPM mechanisms can be desirable when a third party wants to improve social welfare by introducing an information coordinator and wants to ensure the coordinator has no loss (may even have a positive profit) from administering this task, so that the coordinator is willing to take this responsibility.

In the remaining part of this section, we will present the construction of a special NT-FPM mechansim and show that it can strictly improve the net social welfare in comparison to the baseline non-coordinator sequential learning in~\cite{bikhchandani1992}, known as the BHW mechanism.
The gross social welfare (GSW) and the net social welfare (NSW) are defined by the discounted gross income of the agents before (after) tax:
\begin{subequations}
\begin{align}
       \text{GSW} &= \sum_{t=1}^{\infty} \delta^{t-1}u_t(w,a_t),\\
       \text{NSW} &= \sum_{t=1}^{\infty} \delta^{t-1}
       \left( u_t(w,a_t) - f^x_t(d^{t-1},a^{t-1}) \right).
\end{align}
\end{subequations}

\subsection{The BHW Baseline Mechanism}
In the BHW setting~\cite{bikhchandani1992}, agents observe previous actions $a^{t-1}$, receive a private signal $y_t$, and take an action $a_t$. Although in that setting there are no decisions $d_t$, no reports $m_t$, and no suggestions $e_t$, we can think of the BHW mechanism as a special case of our model.

Let us restate the BHW setting under the framework of summary-based mechanisms.
Although not used in the original BHW, the belief $\eta_t$ on summary $n_t=n(m^{t-1})$ with the update rule~\eqref{eq:pi_hat_update} will be used in the definition here, because it sufficiently implies the public belief on state $w$ at time~$t$.
From Lemma~\ref{lem:belief}, the belief toward the state $W$ conditioned on public information (denoted by $\hat{\pi}_t$ in order to distinguish it from $\pi_t \in \Delta(\cM^{t-1})$) can be written as
\begin{equation}
       \hat{\pi}_t(w) := \sum_{n_t} \eta_t(n_t) q(w \vert n_t).
\end{equation}
Then, agent~$t$ forms her private likelihood ratio (LR) on the state, and compares it with $1$
\begin{align}
              \text{LR} &= \frac{\hat{\pi}_t(+1)Q^y(y_t \vert +1)}{\hat{\pi}_t(-1)Q^y(y_t \vert -1)} \nonumber\\
              &= \frac{\sum_{n_t} \eta_t(n_t) q(+1 \vert n_t)Q^y(y_t \vert +1)}{\sum_{n_t} \eta_t(n_t) q(-1 \vert n_t)Q^y(y_t \vert -1)}.
\end{align}
Notice that $Q^y(y_t \vert +1)/Q^y(y_t \vert -1)$ can only be $\bar{p}/p$ or $p/\bar{p}$, so if the public LR is large or small enough, agent~$t$ doesn't have to know $y_t$ in the comparison between private LR and $1$; otherwise, $y_t$ is needed.
This idea leads to the definition of a belief set $\cL$:
\begin{equation}
       \cL = \left\{\eta_t: \frac{p}{\bar{p}} \leq \frac{\sum_{n_t} \eta_t(n_t) q(+1 \vert n_t)}{\sum_{n_t} \eta_t(n_t) q(-1 \vert n_t) } \leq \frac{\bar{p}}{p}\right\},
\end{equation}
where the letter ``$\cL$'' represents ``learning''. The reason for the name ``learning set'' is that the agent community is able to learn agent~$t$'s private signal from her action, if and only if $\eta_t \in \cL$.

The original BHW paper shows that there are two phases in sequential social learning, which we call them ``learning phase'' and ``cascade phase''. Specifically, $\eta_t$ starts from learning phase with $\eta_1=1_0$. In this phase, $\eta_t \in \cL$, a rational strategy for an agent is simply playing according to her private signal\footnote{In the original analysis of BHW\cite{bikhchandani1992}, an agent would play randomly when she knows $n_t=+1$ or $-1$ for sure and her private signal $y_t=-n_t$. This randomized strategy speeds up the occurrence of the cascade and decreases the expected social welfare. Thus, in our comparison, we do not introduce this randomness.}. In this specific setting, one can verify that there are only three possible beliefs in this phase: $1_{-1}, 1_{0}, 1_{+1}$.

The cascade phase occurs right after the belief $\eta_t = 1_{+1}$ and agent~$t$'s action $a_t=+1$ (or $\eta_t=1_{-1}, a_t=-1$, respectively). In the cascade phase, agents blindly follow the action of the agent who triggered the cascade, and therefore, the public belief $\hat{\pi}_t(w)$ stops evolving (even if $\eta_t$ evolves). Note that the occurence of cascade in BHW is irreversible, i.e., once $\eta_t$ jumps out of $\cL$, it will never come back.

In the setting of summary-based mechanisms, BHW can be ``simulated'' as follows. In the learning phase, it uses the partial strategy
\begin{equation}
       \tilde{\theta}^L(\cdot  \vert  n_t, m_t) = 1_{m_t}(\cdot).
\end{equation}
In a cascade phase, depending on what type of cascade agents are facing (the superscript ``$C+$'' means ``cascade on action~$+1$''. ``$C-$'' is similar), it uses the partial strategy
\begin{subequations}
       \begin{align}
              \tilde{\theta}^{C+} (\cdot  \vert n_t, m_t) = 1_{+1}(\cdot),\\
              \tilde{\theta}^{C-} (\cdot  \vert  n_t, m_t) = 1_{-1} (\cdot).
       \end{align}
\end{subequations}
The mechanism $\tilde{\phi}^{B}$ (superscript ``B'' stands for BHW) can then be expressed in the summary-based form:
\ifonecol
       \begin{equation}
              \label{def:fe_bhw}
              \tilde{\phi}^{B}_t[\eta_t] =
              \begin{cases}
                     \tilde{\theta}^{L}, & \eta_t \in \mathcal{L}, \\
                     \tilde{\theta}^{C+}, & \eta_t \notin \mathcal{L} \text{ and }
                     \frac{\sum_{n_t} \eta_t(n_t) q(+1 \vert n_t)}{\sum_{n_t} \eta_t(n_t) q(-1 \vert n_t) } > \frac{\bar{p}}{p} , \\
                     \tilde{\theta}^{C-}, & \text{otherwise}.
              \end{cases}
       \end{equation}
\else
       \begin{align}
              &\tilde{\phi}^{B}_t[\eta_t] = \nonumber \\
              \label{def:fe_bhw}
              &\begin{cases}
                     \tilde{\theta}^{L}, \ \eta_t \in \mathcal{L}, \\
                     \tilde{\theta}^{C+}, \ \eta_t \notin \mathcal{L} \text{ and }
                     \frac{\sum_{n_t} \eta_t(n_t) q(+1 \vert n_t)}{\sum_{n_t} \eta_t(n_t) q(-1 \vert n_t) } > \frac{\bar{p}}{p} , \\
                     \tilde{\theta}^{C-}, \ \text{otherwise}.
              \end{cases}
       \end{align}
\fi

It can easily be shown that $\tilde{\phi}^B$ is an NT-FPM mechanism because all recommendations are simulating agents' rationality, which means it is both harmless and profitless to join the mechanism and meaningless to lie to the mechanism.

From the definition of profit maximizing tax, the tax function of BHW is
\ifonecol
       \begin{equation}
              \tilde{f}^{B,x}(\eta_t,\tilde{\theta}^e_t) = \sum_{y_t, w, n_t} Q^y(y_t \vert w) \eta_t(n_t) q(w \vert n_t)  \left(\tilde{\theta}^e_t(w \vert n_t,y_t)-\tilde{\psi}^a_t[\eta_t](w \vert y_t)\right).
       \end{equation}
\else
       \begin{align}
                     \tilde{f}^{B,x}&(\eta_t,\tilde{\theta}^e_t) = \sum_{y_t, w, n_t} Q^y(y_t \vert w) \eta_t(n_t) q(w \vert n_t)  \nonumber\\
                     &\cdot \left(\tilde{\theta}^e_t(w \vert n_t,y_t)-\tilde{\psi}^a_t[\eta_t](w \vert y_t)\right).
       \end{align}
\fi
Since all the recommendations match agents' rationality, $\tilde{\theta}^e_t(w \vert n_t,y_t)=\tilde{\psi}^a_t[\eta_t](w \vert y_t)$, so $\tilde{f}^{B,x} \equiv 0$.

\subsection{The NSII Mechanism}

We now construct a new NT-FPM mechanism by modifying the BHW mechanism as follows.
\begin{equation}
       \label{def:fes}
       \tilde{\phi}^N_t[\eta_t] = \begin{cases}
              \tilde{\theta}^L, & \eta_t \in \mathcal{L}, \\
              \tilde{\theta}^N, & \text{otherwise},
       \end{cases}
\end{equation}
where $\tilde{\theta}^L$ is defined above, and
\ifonecol
       \begin{equation}
              \tilde{\theta}^N(\cdot \vert n_t, m_t) = \begin{cases}
                     1_{\text{sign}(n_t+m_t)}(\cdot), & n_t + m_t \neq 0, \\
                     1_{\text{sign}(n_t)}(\cdot), & n_t + m_t = 0.
              \end{cases}
       \end{equation}
\else
       \begin{align}
              &\tilde{\theta}^N(\cdot \vert n_t, m_t) = \nonumber\\
              &\begin{cases}
                     1_{\text{sign}(n_t+m_t)}(\cdot), \ n_t + m_t \neq 0, \\
                     1_{\text{sign}(n_t)}(\cdot), \ n_t + m_t = 0.
              \end{cases}
       \end{align}
\fi
%
The intuition of how the partial strategy $\tilde{\theta}^N$ works goes as follows. Note that if the mechanism is truth-telling (will be proved in Theorem~\ref{thm:nsii_benefits}), $n_t$ represents the difference between the number of $+1$'s and $-1$'s among sequence $y_{1:t-1}$, so given $m_t=y_t$, the likelihood ratio of the state is $(\bar{p}/p)^{n_t+m_t}$.
There are three cases we need to consider: (i) $n_t + m_t > 0$, (ii) $n_t + m_t < 0$, (iii) $n_t + m_t = 0$.
In case~(i), the likelihood of the state being $W=+1$ is strictly greater than the likelihood of $W=-1$. Consequently, the mechanism recommends action~$+1$ in~\eqref{def:fes}.
In case~(ii), the likelihood of~$W=-1$ is strictly greater than that of $W=-1$. Since $m_t\in\{\pm 1\}$, $n_t+m_t<0$ implies that $n_t < -m_t \leq 1$. Therefore, when $n_t+m_t < 0$, it follows that $n_t \leq 0$, which means the mechanism recommends action~$-1$. Consequently, in the first two cases, the mechanism always recommends the best action to take according to the likelihood of the state.
The remaining question is what to recommend in case (iii). In case (iii), $n_t+m_t=0$, so the likelihood ratio is not biased towards either of the states. Given the range of $m_t$, there are two subcases: either $n_t=+1, m_t=-1$ or $n_t=-1, m_t=+1$. If $n_t=+1$ the mechanism recommends action $+1$, and otherwise it recommends $-1$. Notice that $n_t=n_{t-1} + m_{t-1}$, so one may infer that $n_t=+1$ implies the recommendation at $t-1$ was $+1$, and $n_t=-1$ implies the previous recommendation was $-1$. Thus, what the mechanism recommends at time~$t$ in case (iii) is indeed the same as the recommendation at $t-1$, i.e., the recommendation does not switch from one to another if the current likelihood ratio is indifferent between two states.
Due to this reason, $\tilde{\theta}^N$ is called ``no switch if indifference'' (NSII) partial strategy, with a superscript ``$N$''. The mechanism $\tilde{\phi}^N$ is therefore called an NSII mechanism.  Note that $\tilde{\phi}^N$ does not alter the learning process when $\eta_t \in \cL$. During a cascade in the BHW sense, the mechanism keeps monitoring the summary $n_t$, and breaks the cascade once the summary suggests a different action from the current cascade.

From the definition of profit maximizing profit, the tax function is also in the form of
\ifonecol
       \begin{equation}
              \tilde{f}^{N,x}(\eta_t,\tilde{\theta}^e_t) = \sum_{y_t, w, n_t} Q^y(y_t \vert w) \eta_t(n_t) q(w \vert n_t)  \left(\tilde{\theta}^e_t(w \vert n_t,y_t)-\tilde{\psi}^a_t[\eta_t](w \vert y_t)\right).
       \end{equation}
\else
       \begin{align}
              \tilde{f}^{N,x}&(\eta_t,\tilde{\theta}^e_t) = \sum_{y_t, w, n_t} Q^y(y_t \vert w) \eta_t(n_t) q(w \vert n_t) \nonumber  \\
              &\cdot \left(\tilde{\theta}^e_t(w \vert n_t,y_t)-\tilde{\psi}^a_t[\eta_t](w \vert y_t)\right).
       \end{align}
\fi

The following theorem establishes a number of properties of NSII and provides motivation for accepting the NSII mechanism for the three parties: individual agents, the agent community as a whole, and the coordinator.
\begin{theorem}
       \label{thm:nsii_benefits}
       The following are true for NSII:
       \begin{enumerate}
              \item NSII is an NT-FPM mechanism.
              \item NSII improves expected individual welfares, i.e., for all $t$,
              \ifonecol
                     \begin{equation}
                            \label{eq:individual_imp}
                            \E^{\tilde{\phi}^N}[u_t(W, A_t) - \tilde{f}^{N,x}(H_t, \tilde{\Theta}^e_t)] \geq \E^{\tilde{\phi}^B}[u_t(W, A_t)- \tilde{f}^{B,x}(H_t, \tilde{\Theta}^e_t)].
                     \end{equation}
              \else
                     \begin{align}
                            &\E^{f^N}[u_t(W, A_t) - f^{N,x}(D^{t-1},A^{t-1})] \nonumber\\
                            \label{eq:individual_imp}
                            &\geq \E^{f^B}[u_t(W, A_t)- f^{B,x}(D^{t-1},A^{t-1})].
                     \end{align}
              \fi
              \item NSII strictly improves the expected net social welfare, i.e.,
              \begin{equation}
                     \E^{\tilde{\phi}^N}[NSW] > \E^{\tilde{\phi}^B}[NSW].
              \end{equation}
              \item For the coordinator, NSII mechanism brings a positive expected revenue, i.e.,
              \begin{equation}
                     \E^{\tilde{\phi}^N}\left[\sum_{t=1}^\infty \tilde{f}^{N,x}(H_t,\tilde{\Theta}^e_t)\right] > 0.
              \end{equation}
       \end{enumerate}
       In the statements above, $H_t$ is a random variable corresponding to realization $\eta_t$.
\end{theorem}
\begin{IEEEproof}
       See Appendix~\ref{appx:pf:thm:nsii_benefits}.
\end{IEEEproof}

The coordinator-related statement in Theorem~\ref{thm:nsii_benefits} (statement 4) shows strict profitability, but it is not clear whether the profit is infinitesimally above 0.
In Appendix~\ref{appx:exact_analysis_nsii} we complete an exact analysis of the NSII mechanism and propose approaches to evaluate the coordinator's revenue and the gross social welfare under this mechanism. We state the main results of this analysis in the following theorem.
\begin{theorem}
       \label{thm:welfare_char}
       NSII has the following payoff characterizations:
       \begin{enumerate}
              \item The coordinator's expected revenue can be characterized by a system of difference equations~\eqref{eq:val_nsii_REV}, in which $R_0^C$ corresponds to the coordinator's expected revenue.
              \item The expected gross social welfare has a closed form expression~\eqref{eq:val_nsii_gsw}.
       \end{enumerate}
\end{theorem}
\begin{IEEEproof}
       See Appendix~\ref{appx:exact_analysis_nsii} for detailed analysis.
\end{IEEEproof}

In the next section we provide numerical results based on the analysis of Theorem~\ref{thm:welfare_char}.

\section{Numerical Analysis}
\label{sec:num}

\begin{figure}
       \centering
       \includegraphics[width=0.95\columnwidth]{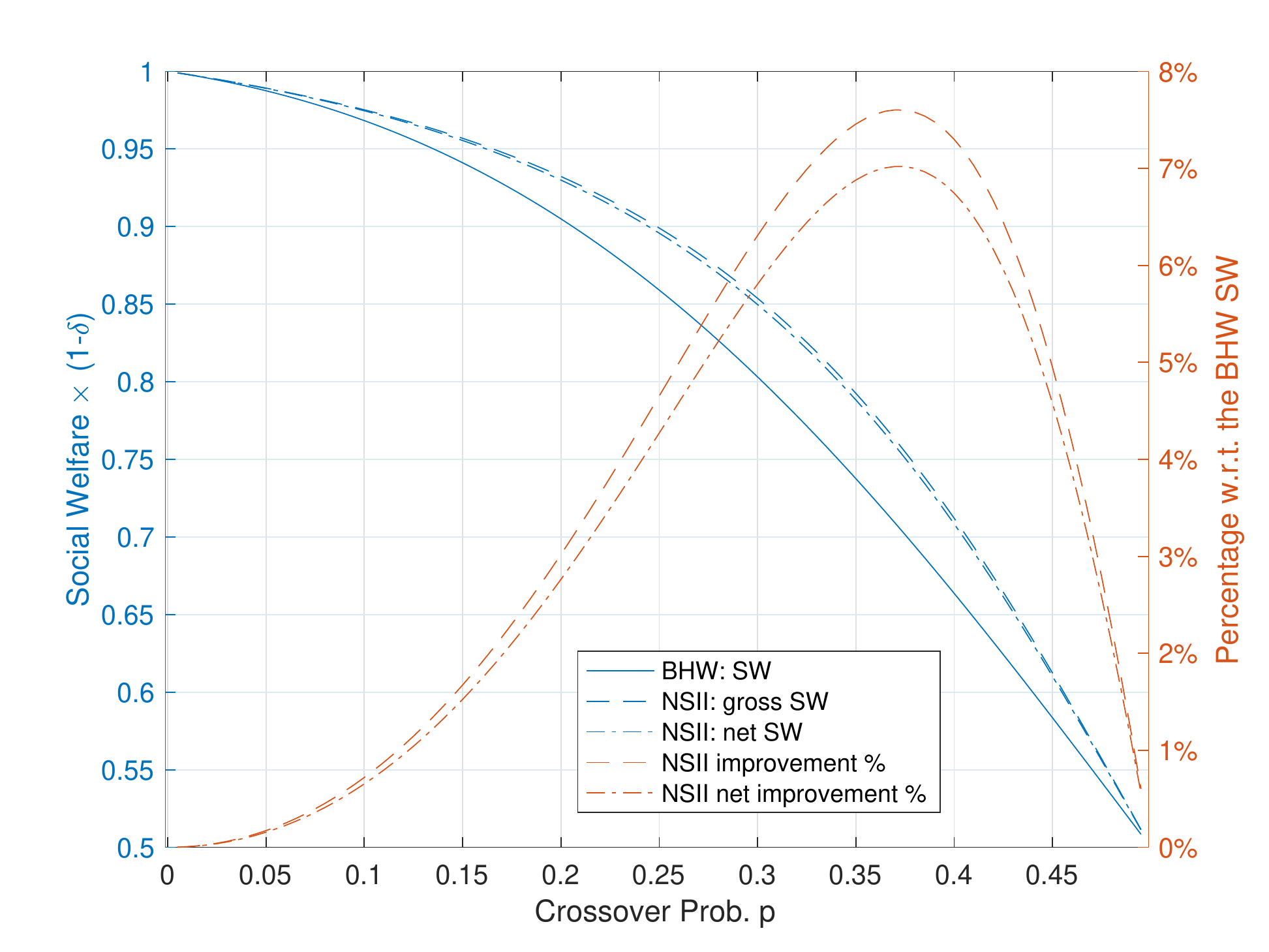}
       \vspace*{-0.25cm}
       \caption{Social welfares (SW) comparison between~\cite{bikhchandani1992} and NSII mechanism.\label{fig:SW_comp}}
\end{figure}

In this section, we present numerical results for the performance of the NSII mechanism. The numerical analysis is done with discount factor $\delta=0.9$ and crossover probabilities $p$ ranging from 0.005 to 0.495, normalized by multiplying $(1-\delta)$. For all values of $p$, the expected revenue of the coordinator and the  expected GSW under the NSII mechanism are evaluated by~\eqref{eq:val_nsii_REV} (using value iteration) and \eqref{eq:val_nsii_gsw} in Appendix~\ref{appx:exact_analysis_nsii}. The expected social welfare of BHW is calculated by~\eqref{eq:val_BHW_SW} in Appendix~\ref{appx:exact_analysis_nsii}. All the percentage numbers are calculated with respect to the social welfare of BHW.
Fig.~\ref{fig:SW_comp} shows the social welfare for BHW and NSII, and the percentages of the social welfare improvements (gross and net).
The results corroborate the social welfare improvement stated in Theorem~\ref{thm:nsii_benefits}. Both the gross and net social welfares of NSII are better than that of BHW. The improvement is substantial under crossover probabilities away from $0$ (fully-informative) and $0.5$ (uninformative). At crossover probability $p=0.37$, the percentage improvement reaches $7.60\%$ for the gross, and $7.02\%$ for the net.

\begin{figure}
       \centering
       \vspace*{-0.5cm}
       \includegraphics[width=0.95\columnwidth]{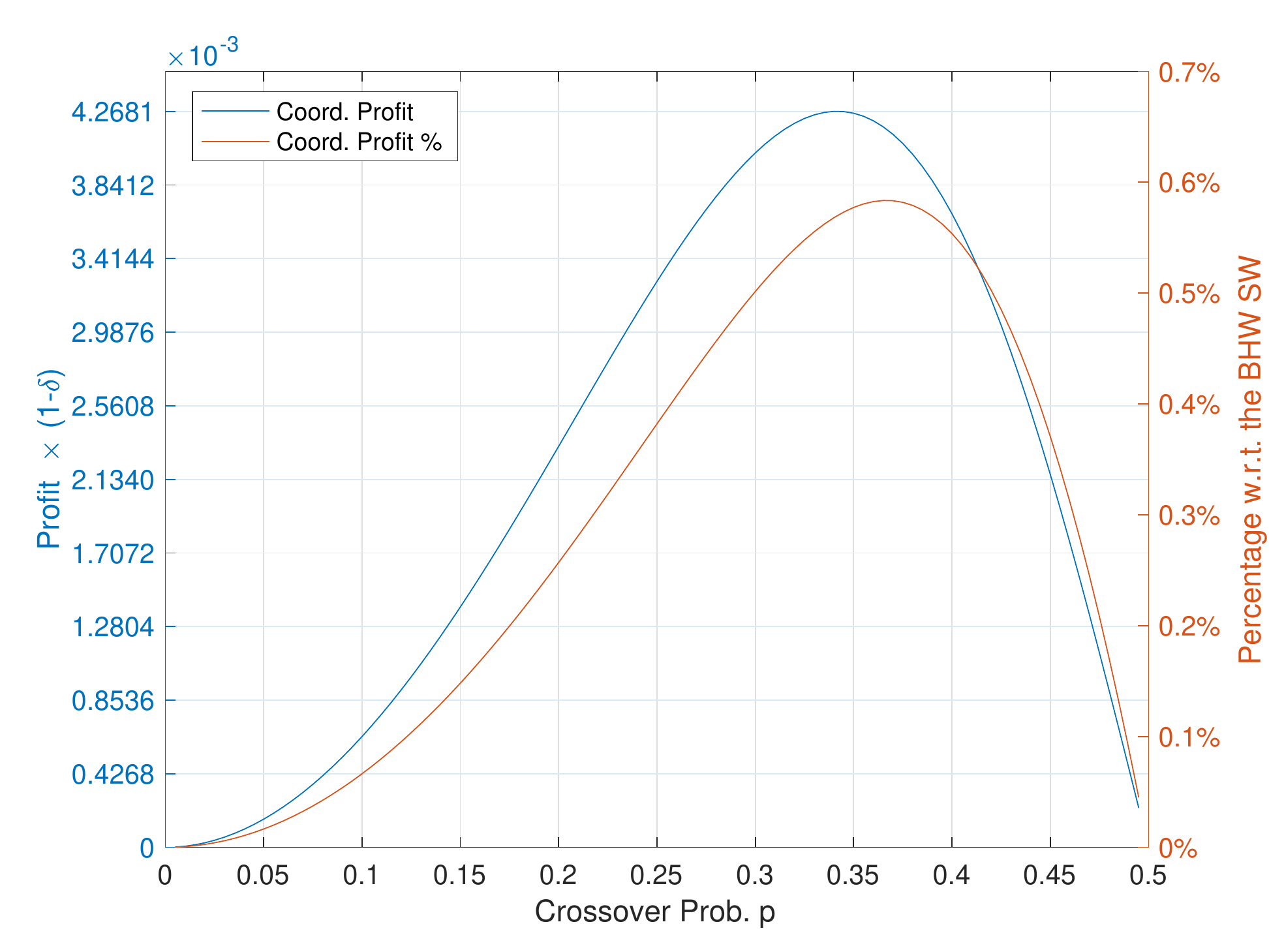}
       \vspace*{-0.25cm}
       \caption{The coordinator's profit and the percentile of the SW improvement w.r.t.~\cite{bikhchandani1992}.\label{fig:Impr_perc}}
\end{figure}

Fig.~\ref{fig:Impr_perc} presents the coordinator's profit (in absolute and relative terms).
The results  show that the NSII coordinator gains a nontrivial proportion of profit under the crossover probabilities away from $p=0$ and $p=0.5$ as well. At $p=0.37$, the percentage of the profit comparing to BHW social welfare is $0.58\%$.

By further comparison on figures \ref{fig:SW_comp} and \ref{fig:Impr_perc}, we find that the percentages of net social welfare improvement and the coordinator's profit share a similar shape, in which both obtain a larger value when private information is neither too informative nor uninformative. This somehow reveals the essence of the recommendation mechanism. By utilizing the mechanism, the coordinator assembles previous private information to build an information asymmetry against future agents. The motivation for agents to join is that this asymmetry improves utilities. Hence, if private information is too informative, agents' private information is close to full information, which leaves less room for asymmetry. On the other side, if private information is close to uninformative case, the coordinator's cumulative information won't grow to an informative one in a short term. As a result, the improvement brought by NSII is more significant under intermediate informative cases rather than the extreme cases.

\section{Conclusion}
\label{sec:conclusion}

We introduced a class of recommendation mechanisms to a sequential Bayesian learning problem. By modeling the coordinator's problem as an MDP, we showed that it is sufficient for a self-interested coordinator to use a summary-based mechanism for the purpose of achieving the optimal profit. We also investigated NT-FPM mechanisms, and specifically, a mechanism named NSII was constructed. The NSII mechanism improves the social welfare, so the community of agents has motivation to join, and it is strictly profitable, so the coordinator is willing to deploy it. The numerical analysis part demonstrated a nontrivial benefit when this mechanism is used.

An interesting future research direction is to treat the recommendation mechanism as a special coordination device in game theory, much like a randomization/recommendation device is introduced in the context of correlated equilibrium.
The sequential Bayesian learning system is indeed a Markov game with a static state, $W$, where players have only an one-shot interaction with the environment.
A more general scenario would include a dynamic system state and multiple users acting at each stage over the entire time horizon. To facilitate coordination among users, a coordinator might provide suggestions of actions/strategies to the agents at each stage of the game.

\appendix

\subsection{Belief Update Off the Equilibrium Path}
\label{appx:belief_off_path}

Consider the deviation made by the coordinator first. Since the coordinator announces $f$ in the beginning, $f$ is public no matter if it is a deviation or not. As long as a feasible $f$ is specified, the recursive update equations on equilibrium still work for the case where the coordinator deviates.

We then consider the deviation made by agents. An infoset $h_t$ is said to be ``off the equilibrium path'' if $\P^{g}(h_t) = 0$, i.e., it can never be achieved (with probability 1) if no one deviates. Nevertheless, the beliefs on some of them still need to be defined because these beliefs help justify the rationality of agents' off-equilibrium strategies (so that they won't be treated as ``empty threat''). The off-equilibrium infosets also belong to one of the three types: $h^m, h^a$ or $h^d$. We now consider the belief update on these three types of off-equilibrium infosets.

The first case is that $h^m_t$ is an off-equilibrium infoset. Since $h^m_t=(h^d_t, d_t=1, y_t)$, if we know $\beta(w, m^{t-1} \vert h^d_t)$, we may continue adopting the update equation~\eqref{eq:belief_hm}.

The second possible type of off-equilibrium infoset is $h^a_t$. $h^a_t=(h^d_t, d_t=0, y_t)$, so it is almost the same as $h^m_t$ except for the value of $d_t$. Suppose we've formed $\beta(w, m^{t-1} \vert h^d_t)$, we can obtain a belief on $h^a_t$ by Bayes' rule
\ifonecol
       \begin{align}
              \beta&(w,m^{t-1} \vert h^a_t) = \frac{\P^{\tilde{g}}(w,m^{t-1},y_t, d_t=0 \vert h^d_t)}{\P^{\tilde{g}}(y_t, d_t=0 \vert h^d_t)} \nonumber \\
              \label{eq:belief_ha}
              &= \frac{\P^{\tilde{g}}(w,m^{t-1},y_t \vert h^d_t, d_t=0)\tilde{g}^d_t(0 \vert h^d_t)}{\P^{\tilde{g}}(y_t \vert h^d_t, d_t=0)\tilde{g}^d_t(0 \vert h^d_t)} = \frac{Q^y(y_t \vert w)\beta(w,m^{t-1} \vert h^d_t)}{\sum_{\bar{w},\bar{m}^{t-1}} Q^y(y_t \vert \bar{w})\beta(\bar{w},\bar{m}^{t-1} \vert h^d_t) },
       \end{align}
\else
       \begin{align}
              \beta&(w,m^{t-1} \vert h^a_t) = \frac{\P^{\tilde{g}}(w,m^{t-1},y_t, d_t=0 \vert h^d_t)}{\P^{\tilde{g}}(y_t, d_t=0 \vert h^d_t)} \nonumber\\
              &= \frac{\P^{\tilde{g}}(w,m^{t-1},y_t \vert h^d_t, d_t=0)\tilde{g}^d_t(0 \vert h^d_t)}{\P^{\tilde{g}}(y_t \vert h^d_t, d_t=0)\tilde{g}^d_t(0 \vert h^d_t)} \nonumber\\
              \label{eq:belief_ha}
              &= \frac{Q^y(y_t \vert w)\beta(w,m^{t-1} \vert h^d_t)}{\sum_{\bar{w},\bar{m}^{t-1}} Q^y(y_t \vert \bar{w})\beta(\bar{w},\bar{m}^{t-1} \vert h^d_t) },
       \end{align}
\fi
where $\tilde{g}$ is the strategy that led to the off-equilibrium infoset, and $\tilde{g}^d_t \neq g^d_t$, because we know agent~$t$ is not playing $g^d_t$.
Observe that since $\tilde{g}^d_t(0 \vert h^d_t)> 0$ it is cancelled out.
Also, note that the second equality holds because agent~$t$'s decision on $d_t$ fully depends on $h^d_t$ and won't influence $w, m^{t-1}$ or $y_t$, so it can be dropped from the condition part.

Lastly, we describe the belief update at an off-equilibrium infoset $h^d_{t+1}$.
If $d_t=1$, $\beta(w,m^t \vert h^d_{t+1})$ can still be updated by~\eqref{eq:belief_hd} using $\beta(w, m^{t-1} \vert h^d_t)$, even if $h^d_t$ is off equilibrium.
If $d_t=0$, write down the recursive relation:
\ifonecol
       \begin{align}
              \beta&(w,m^t \vert h^d_{t+1}) = 1\{m_t=\emptyset\} \frac{\P^{\tilde{g}}(w, m^{t-1}, d_t=0, a_t  \vert  h^d_t)}{\P^{\tilde{g}}(d_t=0, a_t  \vert  h^d_t)} \nonumber \\
              &= 1\{m_t=\emptyset\} \frac{\sum_{y_t} \beta(w,m^{t-1} \vert h^d_t) \tilde{g}^d_t(0 \vert h^d_t) Q^y(y_t \vert w) g^a_t(a_t \vert h^d_t, d_t=0,y_t) }
              {\sum_{w, m^{t-1}, y_t} \beta(w,m^{t-1} \vert h^d_t) \tilde{g}^d_t(0 \vert h^d_t) Q^y(y_t \vert w) g^a_t(a_t \vert h^d_t, d_t=0,y_t)} \nonumber\\
              \label{eq:belief_hd_off}
              &= 1\{m_t=\emptyset\} \frac{\beta(w,m^{t-1} \vert h^d_t) \sum_{y_t} Q^y(y_t \vert w) g^a_t(a_t \vert h^d_t, d_t=0,y_t) }{\sum_{w, m^{t-1}} \beta(w,m^{t-1} \vert h^d_t)  \sum_{y_t} Q^y(y_t \vert w) g^a_t(a_t \vert h^d_t, d_t=0,y_t)}.
       \end{align}
\else
       \begin{align}
              \beta&(w,m^t \vert h^d_{t+1}) \nonumber\\
              &= 1\{m_t=\emptyset\} \frac{\P^{\tilde{g}}(w, m^{t-1}, d_t=0, a_t  \vert  h^d_t)}{\P^{\tilde{g}}(d_t=0, a_t  \vert  h^d_t)} \nonumber\\
              &= 1\{m_t=\emptyset\} \nonumber\\
              &\cdot \frac{\splitfrac{\sum_{y_t} \beta(w,m^{t-1} \vert h^d_t) }
              {\cdot \tilde{g}^d_t(0 \vert h^d_t) Q^y(y_t \vert w) g^a_t(a_t \vert h^d_t, d_t=0,y_t)} }
              {\splitfrac{\sum_{w, m^{t-1}, y_t} \beta(w,m^{t-1} \vert h^d_t) }
              {\cdot \tilde{g}^d_t(0 \vert h^d_t) Q^y(y_t \vert w) g^a_t(a_t \vert h^d_t, d_t=0,y_t)}} \nonumber\\
              &= 1\{m_t=\emptyset\} \nonumber\\
              \label{eq:belief_hd_off}
              &\cdot  \frac{\splitfrac{\beta(w,m^{t-1} \vert h^d_t) }{ \cdot \sum_{y_t} Q^y(y_t \vert w) g^a_t(a_t \vert h^d_t, d_t=0,y_t) }
              }
              {\splitfrac{\sum_{w, m^{t-1}} \beta(w,m^{t-1} \vert h^d_t)  }{\cdot \sum_{y_t} Q^y(y_t \vert w) g^a_t(a_t \vert h^d_t, d_t=0,y_t)}}.
       \end{align}
\fi
There are two cases: (i) $g^a_t(a_t \vert h^d_t, d_t=0, y_t) > 0$ for some $y_t$, (ii) $g^a_t(a_t \vert h^d_t, d_t=0, y_t)=0$ for all $y_t$. Equation~\eqref{eq:belief_hd_off} can be directly used for belief update in case (i), but it is problematic in case (ii) because both numerator and denominator become zero. To resolve case (ii), assume $\tilde{g}^a_t$ is the actual strategy being used. Since agent~$t+1$ knows nothing about $\tilde{g}^a_t$ except that $\tilde{g}^a_t(a_t \vert h^d_t, d_t=0, y_t)>0$ for some $y_t$,
we can only assume agent~$t$ considers any possible $\tilde{g}^a_t$ with equal probability. Denote by $G^a$ the uniform prior over all such strategies $\tilde{g}^a_t$. Then, in~\eqref{eq:belief_hd_off}, the $g^a_t$-term has to be replaced by
\begin{equation}
       \label{eq:replace_offpath_gat}
       \frac{\int_{\tilde{g}^a_t \in \cD_{a_t}} \tilde{g}^a_t(a_t \vert h^d_t, d_t=0,y_t) \text{d}G^a }{ \int_{\tilde{g}^a_t \in \cD_{a_t}} \text{d}G^a},
\end{equation}
where
\begin{align}
       \cD_{a_t} = \left\{ \tilde{g}^a_t:  \tilde{g}^a_t(a_t \vert h^d_t, d_t=0, \tilde{y}_t)>0 \text{ for some $\tilde{y}_t$} \right\}.
\end{align}
Similarly, $\cD_{a_t}^c$ is the subset of strategies $\tilde{g}^a_t$ that assign probability 0 to $a_t$ for all $\tilde{y}_t$. $\int_{\tilde{g}^a_t \in \cD_{a_t}^c} \text{d}G^a=0$ because $\cD_{a_t}^c$ is only a subspace of the strategy space. Since agent~$t+1$ assumes uniform distribution $G^a$ as prior, there is no reason for her to believe the numerator is larger under one $a_t$-value than the other $a_t$-value. Therefore,
\ifonecol
       \begin{align}
              1 &= \int_{\tilde{g}^a_t} \text{d}G^a = \int_{\tilde{g}^a_t} \left(\tilde{g}^a_t(a_t \vert h^d_t, d_t=0,y_t) + \tilde{g}^a_t(1-a_t \vert h^d_t, d_t=0,y_t)\right) \text{d}G^a \nonumber \\
              &= 2 \int_{\tilde{g}^a_t} \tilde{g}^a_t(a_t \vert h^d_t, d_t=0,y_t) \text{d}G^a = 2 \int_{\tilde{g}^a_t \in \cD_{a_t}} \tilde{g}^a_t(a_t \vert h^d_t, d_t=0,y_t) \text{d}G^a,
       \end{align}
\else
       \begin{align}
              1 &= \int_{\tilde{g}^a_t} \text{d}G^a = \int_{\tilde{g}^a_t} \left(\tilde{g}^a_t(a_t \vert h^d_t, d_t=0,y_t)\right. \nonumber \\
              &\qquad \left. + \tilde{g}^a_t(1-a_t \vert h^d_t, d_t=0,y_t)\right) \text{d}G^a \nonumber\\
              &= 2 \int_{\tilde{g}^a_t} \tilde{g}^a_t(a_t \vert h^d_t, d_t=0,y_t) \text{d}G^a \nonumber\\
              &= 2 \int_{\tilde{g}^a_t \in \cD_{a_t}} \tilde{g}^a_t(a_t \vert h^d_t, d_t=0,y_t) \text{d}G^a,
       \end{align}
\fi
Therefore, the value of~\eqref{eq:replace_offpath_gat} is $1/2$. Plug it in the places of $g^a_t$ in~\eqref{eq:belief_hd_off}, we have the update rule
\ifonecol
       \begin{align}
              \beta(w,m^t \vert h^d_{t+1}) & = 1\{m_t=\emptyset\}\frac{\beta(w,m^{t-1} \vert h^d_t) \sum_{y_t} Q^y(y_t \vert w)}{\sum_{w, m^{t-1}} \beta(w,m^{t-1} \vert h^d_t)  \sum_{y_t} Q^y(y_t \vert w)} \nonumber\\
              \label{eq:belief_hd_ignore}
              & = 1\{m_t=\emptyset\} \beta(w, m^{t-1} \vert h^d_t).
       \end{align}
\else
       \begin{align}
              &\beta(w,m^t \vert h^d_{t+1}) \nonumber\\
              &= 1\{m_t=\emptyset\}\frac{\beta(w,m^{t-1} \vert h^d_t) \sum_{y_t} Q^y(y_t \vert w)}{\sum_{w, m^{t-1}} \beta(w,m^{t-1} \vert h^d_t)  \sum_{y_t} Q^y(y_t \vert w)} \nonumber\\
              \label{eq:belief_hd_ignore}
              &= 1\{m_t=\emptyset\} \beta(w, m^{t-1} \vert h^d_t).
       \end{align}
\fi

\subsection{Proof of Lemma~\ref{lem:belief}}
\label{appx:pf:lem:belief}
\begin{IEEEproof}
       We have $\beta(w,m^{t-1} \vert h^d_t)=\beta(w \vert m^{t-1},h^d_t)\pi_t(m^{t-1})$. Now one can evaluate the likelihood ratio:
       \begin{align}
              &\frac{\beta(+1 \vert m^{t-1},h^d_t)}{\beta(-1 \vert m^{t-1},h^d_t)} = \frac{\P^\phi (w=+1,m^{t-1},d^{t-1},a^{t-1})}{\P^\phi (w=-1,m^{t-1},d^{t-1},a^{t-1})} \nonumber \\
              &= \frac{Q^w(+1)\prod_{\tau=1}^{t-1} Q^y(m_\tau \vert +1)}{Q^w(-1)\prod_{\tau=1}^{t-1} Q^y(m_\tau \vert -1)} \nonumber \\
              &= (\bar{p}/p)^{\text{\# of $+1$'s in }m^{t-1}} (p/\bar{p})^{\text{\# of $-1$'s in }m^{t-1}} \nonumber  \\
              &= (\bar{p}/p)^{n_t}.
       \end{align}
       This proves~\eqref{eq:belief_decompose}.
       Since $\pi_t$ is defined through the belief system~$\beta$, it inherits the update equations~\eqref{eq:belief_hd}, \eqref{eq:belief_hd_off}, and~\eqref{eq:belief_hd_ignore}.

       On the paths with $d_t=1$,
       \ifonecol
              \begin{align*}
                     \pi_{t+1}(m^t) = \sum_w \beta(w,m^t \vert h^d_{t+1})
                     = \frac{\sum_w q(w \vert n(m^{t-1}))\pi_t(m^{t-1})Q^y(m_t \vert w)\theta^e_t(a_t \vert m^t)}
                            {\sum_{\bar{w},\bar{m}^t}q(\bar{w} \vert n(\bar{m}^{t-1}))\pi_t(\bar{m}^{t-1})Q^y(\bar{m}_t \vert \bar{w})\theta^e_t(a_t \vert \bar{m}^t)},
              \end{align*}
       \else
              \begin{align*}
                     &\pi_{t+1}(m^t) = \sum_w \beta(w,m^t \vert h^d_{t+1})   \\
                     &= \frac{\sum_w q(w \vert n(m^{t-1})) \pi_t(m^{t-1}) Q^y(m_t \vert w)\theta^e_t(a_t \vert m^t)}
                            {\sum_{\bar{w},\bar{m}^t}q(\bar{w} \vert n(\bar{m}^{t-1}))\pi_t(\bar{m}^{t-1})Q^y(\bar{m}_t \vert \bar{w})\theta^e_t(a_t \vert \bar{m}^t)},
              \end{align*}
       \fi
       so it is sufficient to use $\pi_t, d_t=1, a_t, \theta^e_t$ to evaluate the belief~$\pi_{t+1}$ on the path with $d_t=1$.

       For the off-equilibrium path with $d_t=0$, $\pi_t$ is updated according to equations~\eqref{eq:belief_hd_off} and~\eqref{eq:belief_hd_ignore}.
       \ifonecol
              \begin{align}
                     &\pi_{t+1}(m^t) = 1\{m_t=\emptyset\} \nonumber \\
                     & \cdot \begin{cases}
                            \frac{\sum_w q(w \vert n(m^{t-1}))\pi_t(m^{t-1}) \sum_{y_t} Q^y(y_t \vert w) g^a_t(a_t \vert h^d_t, d_t=0,y_t) }
                            {\sum_{w, m^{t-1}} q(w \vert n(m^{t-1}))\pi_t(m^{t-1})  \sum_{y_t} Q^y(y_t \vert w) g^a_t(a_t \vert h^d_t, d_t=0,y_t)}, & g^a_t(a_t \vert h^d_t, d_t=0,y_t) > 0\text{ for some }t, \\
                            \pi_t( m^{t-1}), & \text{otherwise.}
                     \end{cases}
              \end{align}
       \else
              \begin{align}
                     &\pi_{t+1}(m^t) = 1\{m_t=\emptyset\} \nonumber\\
                     & \cdot \begin{cases}
                            \frac{\splitfrac{\sum_w q(w \vert n(m^{t-1}))\pi_t(m^{t-1})  }{\cdot \sum_{y_t} Q^y(y_t \vert w) g^a_t(a_t \vert h^d_t, d_t=0,y_t)} }
                            {\splitfrac{\sum_{w, m^{t-1}} q(w \vert n(m^{t-1}))\pi_t(m^{t-1}) }{\cdot \sum_{y_t} Q^y(y_t \vert w) g^a_t(a_t \vert h^d_t, d_t=0,y_t)}} \\
                            \qquad \text{, }g^a_t(a_t \vert h^d_t, d_t=0,y_t) > 0\text{ for some }t, \\
                            \pi_t(m^{t-1}), \qquad \text{otherwise.}
                     \end{cases}
              \end{align}
       \fi
       The expression of the update rule contains the on-equilibrium $g^a_t$, but recall that from~\eqref{eq:val_a}, \eqref{eq:g^a_t}, $g^a_t$ on the equilibrium path is only determined by $\beta(\cdot,\cdot \vert h^d_t)$ which is determined by $\pi_t(\cdot)$. Therefore, the update rule of $\pi_t$ can be expressed as a recursion depending on $\pi_t, d_t, a_t$ off the equilibrium path.
\end{IEEEproof}

\subsection{Proof of Theorem~\ref{thm:MDP_suff}}
\label{appx:pf:thm:MDP_suff}

\begin{IEEEproof}
       First, we show that the Decision Process~\ref{mdp:direct} is an MDP.

       \emph{State transition.}
       Since individual rationality constraint is satisfied automatically by the choice of the profit maximizing tax, it is sufficient to consider~$d_t=1$ for all~$t$. Consider the joint distribution of the public observation~$d_t,a_t$ conditioned on the public history~$h^d_t$:
       \ifonecol
              \begin{equation}
                     \label{eq:prob_da}
                     \P^\phi(d_t,a_t \vert h^d_t) = \sum_{w,m^t} 1\{d_t=1\}q(w \vert n(m^{t-1}))\pi_t(m^{t-1}) Q^y(m_t \vert w) \theta^e_t(a_t \vert m^t),
              \end{equation}
       \else
              \begin{align}
                     &\P^\phi(d_t,a_t \vert h^d_t) = 1\{d_t=1\} \sum_{w,m^t} q(w \vert n(m^{t-1})) \nonumber\\
                     \label{eq:prob_da}
                     &\cdot  \pi_t(m^{t-1}) Q^y(m_t \vert w) \theta^e_t(a_t \vert m^t),
              \end{align}
       \fi
       where we also utilize the truth-telling property of~$\phi$. Notice that given $h^d_t=(\phi,d^{t-1},a^{t-1})$, we can calculate~$\theta^e_{1:t}$, and then~$\pi^t$ can be determined by~$d^{t-1},a^{t-1},\theta^e_{1:t}$. Therefore,
       \ifonecol
              \begin{equation}
                     \P^\phi(d_t,a_t \vert h^d_t) = \P^\phi(d_t,a_t \vert h^d_t,\theta^e_{1:t},\pi^t) = \P(d_t,a_t \vert \pi_t,\theta^e_t).
              \end{equation}
       \else
              \begin{align}
                     \P^\phi(d_t,a_t \vert h^d_t) &= \P^\phi(d_t,a_t \vert h^d_t,\theta^e_{1:t},\pi^t) \nonumber \\
                     &= \P(d_t,a_t \vert \pi_t,\theta^e_t).
              \end{align}
       \fi
       Then we have
       \begin{align}
              \P^\phi&(\pi_{t+1} \mid \pi^t,\theta^e_{1:t}) \nonumber \\
              &= \sum_{d_t,a_t} \P^\phi(\pi_{t+1},d_t,a_t \mid \pi^t,\theta^e_{1:t}) \nonumber\\
              &= \sum_{d_t,a_t} 1_{T_t(\pi_t,d_t,a_t,\theta^e_t)}(\pi_{t+1})\P(d_t,a_t \mid \pi_t,\theta^e_t) \nonumber\\
              &= \P(\pi_{t+1}\mid \pi_t, \theta^e_t).
       \end{align}

       \emph{Reward.} The instantaneous reward at time~$t$ is exactly the profit maximizing tax~$\hat{f}^x_t(\pi_t,\theta_t)$, which is a function depending on state~$\pi_t$ and action~$\theta_t$.

       \emph{Feasible action set.} In the optimization problem~\eqref{opt:partial}, the feasible action set is defined by~\eqref{con:TT_par}, which depends on~$d^{t-1},a^{t-1}$ through~$\pi_t$.

       Therefore, the process described above is an MDP.
       Furthermore, the objective function and constraints in the offline optimization problem~\eqref{opt:partial} coincide with the MDP objective and constraints, thus, the optimal MDP solution is a solution to~\eqref{opt:partial}.
       From standard results on MDP, the optimal Markov deterministic policy for Decision Process~\ref{mdp:direct} is also an optimal solution to the original problem~\eqref{opt:partial}.
\end{IEEEproof}

\subsection{Proof of Lemma~\ref{lem:pi_hat_mdp}}
\label{appx:pf:lem:pi_hat_mdp}
\begin{IEEEproof}
       \textit{Feasible Action Set.}
       Truth-telling constraints in~\eqref{con:TT_par} need to be checked. Given a summary-based partial strategy $\tilde{\theta}^e_t$, the TT constraints~\eqref{eq:truth-telling_par} can be written as
       \ifonecol
              \begin{equation}
                     \label{eq:truth-telling_hat}
                     \sum_{w,n_t} Q^y(y_t \vert w) \eta_t(n_t) q(w \vert n_t) \cdot (\tilde{\theta}^e_t(w \vert n_t,y_t) - \tilde{\theta}^e_t(w \vert n_t,-y_t)) \geq 0.
              \end{equation}
       \else
              \begin{align}
                     \sum_{w,n_t} & Q^y(y_t \vert w) \eta_t(n_t) q(w \vert n_t) \nonumber\\
                     \label{eq:truth-telling_hat}
                     &\cdot (\tilde{\theta}^e_t(w \vert n_t,y_t) - \tilde{\theta}^e_t(w \vert n_t,-y_t)) \geq 0.
              \end{align}
       \fi
       Therefore, in Decision Process~\ref{mdp:summary-based}, if the coordinator restricts attention to partial strategies of the form $\tilde{\theta}^e_t: \Integer \times \cM \to \Delta(\cA)$, the feasible action set can be described by a set $\tilde{S}(\eta_t)$ defined by~\eqref{eq:truth-telling_hat}. Note that this feasible set is time-invariant given $\eta_t$.

       \textit{State Transition.} Since the coordinator only chooses an action from the feasible strategies satisfying~\eqref{eq:truth-telling_hat}, he should expect that $d_t=1$. Thus, in Decision Process~\ref{mdp:summary-based}, we only consider the history with $d_t=1$. For arbitrary $n_{t+1} \leq t$, since $\tilde{\theta}^e_t$ can be regarded as a special case of the general partial strategy $\theta^e_t$, the update equation~\eqref{eq:belief} still works for $\pi_t$, by marginalizing the result of~$T_t(\pi_t,d_t=1,a_t,\tilde{\theta}^e_t)$, we have
       \ifonecol
              \begin{align}
                     \eta_{t+1}&(n_{t+1}) = \sum_{m^t:n_{t+1}} \pi_{t+1}(m^t)  \nonumber\\
                     &= \frac{\sum_{w, m^{t}:{n_{t+1}}} q(w \vert n(m^{t-1}))\pi_t(m^{t-1}) Q^y(m_t \vert w)\tilde{\theta}^e_t(a_t \vert n(m^{t-1}),m_t)}{\sum_{\bar{w},\bar{m}^t} q(\bar{w} \vert n(\bar{m}^{t-1}))\pi_t(\bar{m}^{t-1})Q^y(\bar{m}_t \vert \bar{w})\tilde{\theta}^e_t(a_t \vert n(\bar{m}^{t-1}),\bar{m}_t)} \nonumber\\
                     &= \frac{\sum_w \left( \splitfrac{\eta_t(n_{t+1}-1) q(w \vert n_{t+1}-1) Q^y(1 \vert w) \tilde{\theta}^e_t(a_t \vert n_{t+1}-1,1)}{+\eta_t(n_{t+1}+1) q(w \vert n_{t+1}+1) Q^y(-1 \vert w) \tilde{\theta}^e_t(a_t \vert n_{t+1}+1,-1)}   \right)}{\sum_{\bar{w},n_t,\bar{m}_t} \eta_t(n_t) q(\bar{w} \vert n_t) Q^y(\bar{m}_t \vert \bar{w})\tilde{\theta}^e_t(a_t \vert n_t,\bar{m}_t)} \nonumber\\
                     \label{eq:pi_hat_update}
                     &=: \tilde{T}_t(\eta_t,d_t=1,a_t,\tilde{\theta}^e_t)(n_{t+1}).
              \end{align}
       \else
              \begin{align}
                     &\eta_{t+1}(n_{t+1}) = \sum_{ m^t:n_{t+1}} \pi_{t+1}(m^t)  \nonumber \\
                     \label{eq:pi_hat_update}
                     &= \frac{NUM}{DEN} =: \tilde{T}_t(\eta_t,d_t=1,a_t,\tilde{\theta}^e_t)(n_{t+1}),
              \end{align}
              where
              \begin{align*}
                            NUM =& \sum_{w, m^{t}:{n_{t+1}}} q(w \vert n(m^{t-1}))\pi_t(m^{t-1}) \\
                            &\qquad \cdot Q^y(m_t \vert w)\tilde{\theta}^e_t(a_t \vert n(m^{t-1}),m_t) \\
                            =&\sum_w \left( \eta_t(n_{t+1}-1) q(w \vert n_{t+1}-1) \right.\\
                            &\qquad \cdot Q^y(1 \vert w) \tilde{\theta}^e_t(a_t \vert n_{t+1}-1,1) \\
                            +& \eta_t(n_{t+1}+1) q(w \vert n_{t+1}+1) \\
                            &\qquad \left. \cdot Q^y(-1 \vert w) \tilde{\theta}^e_t(a_t \vert n_{t+1}+1,-1)   \right)
              \end{align*}
              \begin{align*}
                     DEN =& \sum_{\bar{w},\bar{m}^t} q(\bar{w} \vert n(\bar{m}^{t-1}))\pi_t(\bar{m}^{t-1})\\
                     &\qquad \cdot Q^y(\bar{m}_t \vert \bar{w})\tilde{\theta}^e_t(a_t \vert n(\bar{m}^{t-1}),\bar{m}_t)\\
                     =& \sum_{\bar{w},n_t,\bar{m}_t} \eta_t(n_t) q(\bar{w} \vert n_t) \\
                     &\qquad \cdot Q^y(\bar{m}_t \vert \bar{w})\tilde{\theta}^e_t(a_t \vert n_t,\bar{m}_t)
              \end{align*}
       \fi
       Equation~\eqref{eq:pi_hat_update} is a recursive update equation for~$\eta_t$. The updated $\eta_{t+1}$ depends on $\pi_t$ only through $\eta_t$.

       The state transition probability is
       \ifonecol
              \begin{align}
                     \P^{\tilde{\phi}}(\eta_{t+1} \vert \eta_{1:t},\tilde{\theta}^e_{1:t}) =& \sum_{a_t}\P^{\tilde{\phi}} (\eta_{t+1},d_t,a_t  \vert  \eta_{1:t},\tilde{\theta}^e_{1:t}) \nonumber \\
                     =& \sum_{a_t} 1_{\tilde{T}_t(\eta_t,d_t,a_t,\tilde{\theta^e_t})} (\eta_{t+1}) \sum_{w,n_t, y_t} \tilde{\theta}^e_t(a_t \vert n_t,y_t) Q^y(y_t \vert w) q(w \vert n_t) \eta_t(n_t) \nonumber\\
                     =:& \P(\eta_{t+1} \vert \eta_t,\tilde{\theta}^e_t).
              \end{align}
       \else
              \begin{align}
                     &\P^{\tilde{\phi}}(\eta_{t+1} \vert \eta_{1:t},\tilde{\theta}^e_{1:t}) \nonumber\\
                     &= \sum_{a_t}\P^{\tilde{\phi}} (\eta_{t+1},d_t,a_t  \vert  \eta_{1:t},\tilde{\theta}^e_{1:t}) \nonumber\\
                     &= \sum_{a_t} 1_{\tilde{T}_t(\eta_t,d_t,a_t,\tilde{\theta^e_t})} (\eta_{t+1}) \nonumber\\
                     &\cdot \sum_{w,n_t, y_t} \tilde{\theta}^e_t(a_t \vert n_t,y_t) Q^y(y_t \vert w) q(w \vert n_t) \eta_t(n_t) \nonumber\\
                     &=: \P(\eta_{t+1} \vert \eta_t,\tilde{\theta}^e_t).
              \end{align}
       \fi

       \textit{Reward.}
       Notice that
       \ifonecol
              \begin{align}
                     \hat{f}^x_t(\pi_t,\tilde{\theta}^e_t) =& \sum_{y_t, w, m^{t-1}} Q^y(y_t \vert w) q(w \vert n(m^{t-1}))\pi_t(m^{t-1}) \cdot (\tilde{\theta}^e_t(w \vert n(m^{t-1}),y_t)-\tilde{\psi}^a[\eta_t](w \vert y_t))\nonumber \\
                     =& \sum_{y_t, w, n_t} Q^y(y_t \vert w) q(w \vert n_t) \sum_{m^{t-1} :{n_t}} \pi_t(m^{t-1}) \cdot (\tilde{\theta}^e_t(w \vert n(m^{t-1}),y_t)-\tilde{\psi}^a[\eta_t](w \vert y_t)) \nonumber\\
                     =& \sum_{y_t, w, n_t} Q^y(y_t \vert w) \eta_t(n_t) q(w \vert n_t) \cdot (\tilde{\theta}^e_t(w \vert n_t,y_t)-\tilde{\psi}^a[\eta_t](w \vert y_t)) \nonumber\\
                     =:& \tilde{f}^x(\eta_t,\tilde{\theta}^e_t).
              \end{align}
       \else
              \begin{align}
                     &\hat{f}^x_t(\pi_t,\tilde{\theta}^e_t) \nonumber\\
                     &= \sum_{y_t, w, m^{t-1}} Q^y(y_t \vert w) q(w \vert n(m^{t-1}))\pi_t(m^{t-1}) \nonumber\\
                     &\cdot (\tilde{\theta}^e_t(w \vert n(m^{t-1}),y_t)-\tilde{\psi}^a[\eta_t](w \vert y_t)) \nonumber \\
                     &= \sum_{y_t, w, n_t, m^{t-1} :{n_t}} Q^y(y_t \vert w) q(w \vert n_t)\pi_t(m^{t-1}) \nonumber\\
                     &\cdot (\tilde{\theta}^e_t(w \vert n(m^{t-1}),y_t)-\tilde{\psi}^a[\eta_t](w \vert y_t)) \nonumber\\
                     &= \sum_{y_t, w, n_t} Q^y(y_t \vert w) \eta_t(n_t) q(w \vert n_t) \nonumber\\
                     &\cdot (\tilde{\theta}^e_t(w \vert n_t,y_t)-\tilde{\psi}^a[\eta_t](w \vert y_t))\nonumber\\
                     &=: \tilde{f}^x(\eta_t,\tilde{\theta}^e_t).
              \end{align}
       \fi
       Therefore, for any summary-based partial strategy $\tilde{\theta}^e_t$,
       \begin{equation}
              \E^\phi\{\hat{f}^x_t(\Pi_t,\Theta^e_t) \mid \eta_{1:t}, \tilde{\theta}^e_{1:t}\} = \tilde{f}^x(\eta_t,\tilde{\theta}^e_t).
       \end{equation}
\end{IEEEproof}

\subsection{Proof of Theorem~\ref{thm:belief_n}}
\label{appx:pf:thm_belief_n}
\begin{IEEEproof}
       We will show a stronger result in Lemma~\ref{lem:essence_thm_belief_n}, which implies Theorem~\ref{thm:belief_n}.
       \begin{lemma}
              \label{lem:essence_thm_belief_n}
              Suppose $\hat{\phi}$ is optimal, and $\hat{V}_t(\pi_t)$ for $t=1,\ldots, T$ is its value function. Then, for each time $t$:
              \begin{enumerate}
                     \item If $\pi_t, \pi'_t$ induce the same $\eta_t$, $\hat{V}_t(\pi_t)=\hat{V}_t(\pi'_t)$.
                     \item There exists a summary-based $\tilde{\phi}_t:\Delta(\Integer) \to (\Integer \to \Delta(\cA))$, such that $\hat{\phi}_{1:t-1}, \tilde{\phi}_t, \hat{\phi}_{t+1:T}$ is also an optimal strategy.
              \end{enumerate}
       \end{lemma}
       By the second statement in Lemma~\ref{lem:essence_thm_belief_n}, for an optimal belief-based strategy $\hat{\phi}$, one may obtain an optimal summary-based strategy $\tilde{\phi}$ by iteratively substituting each $\hat{\phi}_t$ with some $\tilde{\phi}$, while maintaining the optimality. Therefore, it is sufficient for the coordinator to focus on the summary-based strategies.

       Now we want to prove Lemma~\ref{lem:essence_thm_belief_n}. Since strategies $\hat{\phi}_t$ and $\tilde{\phi}_t$ have different range, when constructing the output of $\tilde{\phi}_t$, we will ``smooth'' $m^t$-based partial strategies into $(n_t, m_t)$-based ones. Lemma~\ref{lem:smoothing} shows several good properties of the smoothing operation.
       \begin{lemma}
              \label{lem:smoothing}
              For a belief-based strategy $\hat{\phi}$, suppose $\pi_t$ induces $\eta_t$, and $\hat{\theta}^e_t = \hat{\phi}_t[\pi_t]$. Define a summary-based strategy $\tilde{\theta}^e_t$ by
              \begin{equation}
                     \label{eq:smoothing}
                     \tilde{\theta}^e_t(\cdot \vert n_t, m_t) := \frac{\sum_{m^{t-1}:n_t}\pi_t(m^{t-1})\hat{\theta}^e_t(\cdot \vert m^t)}{\sum_{m^{t-1}:n_t}\pi_t(m^{t-1})}.
              \end{equation}
              The following statements are true for any $\pi'_t$ that also induces $\eta_t$:
              \begin{enumerate}
                     \item $\tilde{\theta}^e_t$ is a feasible partial strategy for $\pi'_t$.
                     \item $\hat{f}^x(\pi'_t, \tilde{\theta}^e_t) = \hat{f}^x(\pi_t, \hat{\theta}^e_t) = \tilde{f}^x(\eta_t, \tilde{\theta}^e_t)$.
                     \ifonecol
                     \item $\P(\eta_{t+1} \vert \pi'_t, \tilde{\theta}^e_t) = \P(\eta_{t+1} \vert \pi_t, \hat{\theta}^e_t) = \P(\eta_{t+1} \vert \eta_t, \tilde{\theta}^e_t)$.
                     \else
                     \item
                     \begin{align*}
                            \P(\eta_{t+1} \vert \pi'_t, \tilde{\theta}^e_t) = \P(\eta_{t+1} \vert \pi_t, \hat{\theta}^e_t) = \P(\eta_{t+1} \vert \eta_t, \tilde{\theta}^e_t).
                     \end{align*}
                     \fi
              \end{enumerate}
       \end{lemma}
       The proof of Lemma~\ref{lem:smoothing} is basically ``plugging-in-and-verifying'', so we postpone the tedious calculations to the end.
       Next, we show how Lemma~\ref{lem:smoothing} implies Lemma~\ref{lem:essence_thm_belief_n}.

       \emph{Proof of Lemma~\ref{lem:essence_thm_belief_n}.} The proof is done by induction. Let's first assume at time $t+1$, $\hat{V}_{t+1}(\pi_{t+1})=\hat{V}_{t+1}(\pi'_{t+1}) = \tilde{V}_{t+1}(\eta_{t+1})$ for any $\pi_{t+1}, \pi'_{t+1}$ that induce the same $\eta_{t+1}$. We are going to show statements 1) and 2) in Lemma~\ref{lem:essence_thm_belief_n} are true under this assumption. For statement 1), suppose $\pi^{(1)}_t, \pi^{(2)}_t$ induce the same $\eta_t$ but $\hat{V}_t(\pi^{(1)}_t) > \hat{V}_t(\pi^{(2)}_t)$. For any $\pi_t$ that induces $\eta_t$, $\hat{V}_t(\pi_t)$ can be written as
       \ifonecol
       \begin{equation}
              \hat{V}_t(\pi_t) =
                     \hat{f}^x(\pi_t, \hat{\phi}_t[\pi_t]) + \sum_{\eta_{t+1}} \tilde{V}_{t+1}(\eta_{t+1}) \P(\eta_{t+1} \vert \pi_t, \hat{\phi}_t[\pi_t]).
       \end{equation}
       \else
       \begin{align}
                     &\hat{V}_t(\pi_t) =  \hat{f}^x(\pi_t, \hat{\phi}_t[\pi_t]) \nonumber\\
                     &+ \sum_{\eta_{t+1}} \tilde{V}_{t+1}(\eta_{t+1}) \P(\eta_{t+1} \vert \pi_t, \hat{\phi}_t[\pi_t]).
       \end{align}
       \fi
       From Lemma~\ref{lem:smoothing}, we may construct a $\tilde{\theta}^{e(1)}_t$ from $\hat{\theta}^{e(1)}_t=\hat{\phi}_t[\pi^{(1)}_t]$. Then, for $\pi^{(2)}_t$, if we use this $\tilde{\theta}^{e(1)}_t$ instead of $\hat{\phi}_t[\pi^{(2)}_t]$, from Lemma~\ref{lem:smoothing}, the tax $\hat{f}^x(\pi^{(2)}_t, \tilde{\theta}^{e(1)}_t)$ and transition probability $\P(\eta_{t+1} \vert \pi^{(2)}_t, \tilde{\theta}^{e(1)}_t)$ will be the same as those of $\pi^{(1)}_t$. This implies the value function at $\pi^{(2)}_t$ should be able to achieve a value of at least $\hat{V}_t(\pi^{(1)}_t)$, which produces the contradiction. Therefore, $\hat{V}_t(\pi_t) = \hat{V}_t(\pi'_t)$ for any $\pi_t, \pi'_t$ that induce the same $\eta_t$.

       To show statement 2), for each possible $\eta_t$, we arbitrarily choose one $\pi_t$ that induces $\eta_t$, and let $\tilde{\phi}_t[\eta_t]$ be the smooth partial strategy $\tilde{\theta}^e_t$ derived from $\hat{\phi}_t[\pi_t]$ by Lemma~\ref{lem:smoothing}. As a result, all beliefs $\pi'_t$ that induce $\eta_t$ will have value function equal to $\hat{V}_t(\pi_t)$. However, recall that $\hat{V}_t(\pi_t) = \hat{V}_t(\pi'_t)$ for any $\pi_t, \pi'_t$ that induce the same $\eta_t$, and so, the use of the constructed $\tilde{\phi}$ indeed does not change the value function. Thus, $\hat{\phi}_{1:t-1}, \tilde{\phi}_t, \hat{\phi}_{t+1:T}$ is also an optimal strategy.

       Lastly, to complete the backward inductive proof, we need to show is $\hat{V}_T(\pi_T) = \hat{V}_T(\pi'_T)$ for any $\pi_T, \pi'_T$ that induce the same $\eta_T$. However, notice that at time $T$ the value function is a special case of the above, since $\hat{V}_{T+1}(\cdot) \equiv 0$, and so no extra work is needed.

       \emph{Proof of Lemma~\ref{lem:smoothing}.}

       \begin{enumerate}
              \item \emph{Feasibility.} For the truth-telling constraint~\eqref{eq:truth-telling_par}, we compare the expected payoff of truth-telling (denoted by ``TT'') with that of misreporting (denoted by ``MT'') for~$\tilde{\theta}^e_t$ constructed by $\hat{\theta}^e_t=\hat{\phi}_t[\pi_t]$, but under some $\pi'_t$ that induces the same $\eta_t$ as $\pi_t$ does. For any $y_t$,
              \ifonecol
                     \begin{align*}
                                   \text{TT} &= \sum_{n_t, w, m^{t-1}:{n_t}} Q^y(y_t \vert w) \pi'_t (m^{t-1}) q(w \vert n_t) \tilde{\theta}^e_t(w \vert n_t,y_t) \\
                                   &= \sum_{w} Q^y(y_t \vert w) \sum_{n_t} q(w \vert n_t) \sum_{m^{t-1}:{n_t}} \pi'_t (m^{t-1}) \tilde{\theta}^e_t(w \vert n_t,y_t).
                     \end{align*}
              \else
                     \begin{align*}
                                   \text{TT} &= \sum_{n_t, w, m^{t-1}:{n_t}} Q^y(y_t \vert w) \pi'_t (m^{t-1}) \\
                                   &\qquad \cdot q(w \vert n_t) \tilde{\theta}^e_t(w \vert n_t,y_t) \\
                                   &= \sum_{w} Q^y(y_t \vert w) \sum_{n_t} q(w \vert n_t) \\
                                   &\qquad \cdot \sum_{m^{t-1}:{n_t}} \pi'_t (m^{t-1}) \tilde{\theta}^e_t(w \vert n_t,y_t).
                     \end{align*}
              \fi
              Note that for a given $n_t$, $\sum_{m^{t-1}:n_t} \pi_t(m^{t-1}) = \sum_{m^{t-1}:n_t} \pi'_t(m^{t-1}) $, so if we substitute the definition of $\tilde{\theta}^e_t$, after some algebraic steps we get
              \ifonecol
                     \begin{equation*}
                            \text{TT} = \sum_{w, m^{t-1}} \pi_t (m^{t-1}) q(w \vert n(m^{t-1})) Q^y(y_t \vert w) \hat{\theta}^e_t(w \vert (m^{t-1}, y_t)).
                     \end{equation*}
              \else
                     \begin{align*}
                            \text{TT} =& \sum_{w, m^{t-1}} \pi_t (m^{t-1}) q(w \vert n(m^{t-1})) \\
                            &\qquad \cdot Q^y(y_t \vert w) \hat{\theta}^e_t(w \vert (m^{t-1}, y_t)).
                     \end{align*}
              \fi
              Similarly,
              \ifonecol
                     \begin{align*}
                                   \text{MT} =& \sum_{w} Q^y(y_t \vert w) \sum_{n_t} q(w \vert n_t) \sum_{m^{t-1}:{n_t}} \pi'_t (m^{t-1}) \tilde{\theta}^e_t(w \vert n_t,-y_t) \\
                                   =& \sum_{w, m^{t-1}} \pi_t (m^{t-1}) q(w \vert n(m^{t-1})) Q^y(y_t \vert w) \hat{\theta}^e_t(w \vert (m^{t-1}, -y_t)).
                     \end{align*}
              \else
                     \begin{align*}
                                   \text{MT} =& \sum_{w} Q^y(y_t \vert w) \sum_{n_t} q(w \vert n_t) \\
                                   &\qquad \cdot \sum_{m^{t-1}:{n_t}} \pi'_t (m^{t-1}) \tilde{\theta}^e_t(w \vert n_t,-y_t) \\
                                   =& \sum_{w, m^{t-1}} \pi_t (m^{t-1}) q(w \vert n(m^{t-1})) \\
                                   &\qquad \cdot Q^y(y_t \vert w) \hat{\theta}^e_t(w \vert (m^{t-1}, -y_t)).
                     \end{align*}
              \fi
              With the results of TT and MT, we can evaluate the difference as
              \ifonecol
                     \begin{equation*}
                            \text{TT} - \text{MT} = \sum_{w, m^{t-1}} \pi_t (m^{t-1}) q(w \vert n(m^{t-1})) Q^y(y_t \vert w) (\hat{\theta}^e_t(w \vert (m^{t-1},y_t)) - \hat{\theta}^e_t(w \vert (m^{t-1},-y_t))),
                     \end{equation*}
              \else
                     \begin{align*}
                            &\text{TT} - \text{MT}\\
                            &= \sum_{w, m^{t-1}} \pi_t (m^{t-1}) q(w \vert n(m^{t-1})) Q^y(y_t \vert w) \\
                            &\cdot \left(\theta^e_t(w \vert (m^{t-1},y_t)) - \theta^e_t(w \vert (m^{t-1},-y_t))\right),
                     \end{align*}
              \fi
              which is nonnegative due to the truth-telling~\eqref{eq:truth-telling_par} of the original policy~$\hat{\phi}$, so the newly constructed~$\tilde{\theta}^e_t$ is truth-telling for any $\pi'_t$ that induces $\eta_t$.

              \item \emph{The Tax Property.} In the proof of Lemma~\ref{lem:pi_hat_mdp}, we've shown that if a summary-based partial strategy $\tilde{\theta}^e_t$ is used, then $\hat{f}^x(\pi, \tilde{\theta}^e_t)=\tilde{f}^x(\eta, \tilde{\theta}^e_t)$, i.e., tax depends on $\pi$ through $\eta$ in this case. Therefore, the tax property in Lemma~\ref{lem:smoothing} is true.

              \item \emph{The Transition Probability Property.} We first show $\P(a_t \vert \pi'_t, \tilde{\theta}^e_t) = \P(a_t \vert \pi_t, \hat{\theta}^e_t) = \P(a_t \vert \eta_t, \tilde{\theta}^e_t)$.
              \ifonecol
                     \begin{align*}
                            \P(a_t \vert \pi'_t, \tilde{\theta}^e_t) =& \sum_{w, m^{t-1}, y_t} \pi'_t(m^{t-1}) q(w \vert n(m^{t-1})) Q^y(y_t \vert w) \tilde{\theta}^e_t(a_t \vert n(m^{t-1}), y_t) \\
                            =& \sum_{w, n_t, y_t} q(w \vert n_t) Q^y(y_t \vert w) \left( \sum_{m^{t-1}:n_t} \pi'_t(m^{t-1}) \right) \cdot \tilde{\theta}^e_t(a_t \vert n_t, y_t) \\
                            =& \sum_{w, m^{t-1}, y_t} \pi_t(m^{t-1}) q(w \vert n(m^{t-1})) Q^y(y_t \vert w) \hat{\theta}^e_t(a_t \vert (m^{t-1}, y_t)) \\
                            =& \P(a_t \vert \pi_t, \hat{\theta}^e_t),
                     \end{align*}
              \else
                     \begin{align*}
                            &\P(a_t \vert \pi'_t, \tilde{\theta}^e_t) \\
                            &= \sum_{w, m^{t-1}, y_t} \pi'_t(m^{t-1}) q(w \vert n(m^{t-1})) \\
                            &\qquad \cdot Q^y(y_t \vert w) \tilde{\theta}^e_t(a_t \vert n(m^{t-1}), y_t) \\
                            &= \sum_{w, n_t, y_t} q(w \vert n_t) Q^y(y_t \vert w) \\
                            &\qquad \cdot \left( \sum_{m^{t-1}:n_t} \pi'_t(m^{t-1}) \right) \cdot \tilde{\theta}^e_t(a_t \vert n_t, y_t) \\
                            &= \sum_{w, m^{t-1}, y_t} \pi_t(m^{t-1}) q(w \vert n(m^{t-1})) \\
                            &\qquad \cdot Q^y(y_t \vert w) \hat{\theta}^e_t(a_t \vert (m^{t-1}, y_t)) \\
                            &= \P(a_t \vert \pi_t, \hat{\theta}^e_t),
                     \end{align*}
              \fi
              where right after the second equality we have $\sum_{m^{t-1}:n_t} \pi'_t(m^{t-1})=\eta_t(n_t)$, and if we substitute that in, we will obtain the result $\P(a_t \vert \pi'_t, \tilde{\theta}^e_t) = \P(a_t \vert \eta_t, \tilde{\theta}^e_t)$.

              Then we show the belief $\eta_{t+1}$ derived from $\pi'_t, a_t, \tilde{\theta}^e_t$ is consistent with the belief derived from $\pi_t, a_t, \hat{\theta}^e_t$:
              \ifonecol
                     \begin{align*}
                            &\eta_{t+1}(n_{t+1}) = \sum_{m^t:{n_{t+1}}} [T_t(\pi'_t,d_t=1,a_t,\tilde{\theta}^e_t)](m^t)\\
                            &= \frac{\sum_{w,m_t} q(w \vert n_{t+1}-m_t) Q^y(m_t \vert w) \left( \sum_{m^{t-1}:{n_{t+1}-m_t}} \pi'_t(m^{t-1})\right) \tilde{\theta}^e_t(a_t \vert n_{t+1}-m_t,m_t)}
                            {\sum_{\bar{w},\bar{n}_t,\bar{m}_t} q(\bar{w} \vert \bar{n}_{t}) Q^y(\bar{m}_t \vert \bar{w})\left(\sum_{\bar{m}^{t-1}:{\bar{n}_t}} \pi'_t(\bar{m}^{t-1})\right) \tilde{\theta}^e_t(a_t \vert \bar{n}_t,\bar{m}_t)} \\
                            &= \frac{\sum_{w,m^t:n_{t+1}} q(w \vert n_{t+1}-m_t) Q^y(m_t \vert w)  \pi_t(m^{t-1}) \hat{\theta}^e_t(a_t \vert m^t)}
                            {\sum_{\bar{w},\bar{m}^t} q(\bar{w} \vert \bar{n}_{t}) Q^y(\bar{m}_t \vert \bar{w}) \pi_t(\bar{m}^{t-1}) \hat{\theta}^e_t(a_t \vert \bar{m}^t)} \\
                            &= \sum_{m^{t}:{n_{t+1}}} T_t(\pi_t,d_t=1,a_t,\hat{\theta}^e_t)(m^t),
                     \end{align*}
              \else
                     \begin{align*}
                            &\eta_{t+1}(n_{t+1}) = \sum_{m^t:{n_{t+1}}} [T_t(\pi'_t,d_t=1,a_t,\tilde{\theta}^e_t)](m^t)\\
                            &= \frac{NUM}{DEN} \\
                            &= \sum_{m^{t}:{n_{t+1}}} T_t(\pi_t,d_t=1,a_t,\hat{\theta}^e_t)(m^{t}),
                     \end{align*}
                     where
                     \begin{align*}
                            &NUM = \sum_{w,m_t} q(w \vert n_{t+1}-m_t) Q^y(m_t \vert w) \\
                            &\cdot \left( \sum_{m^{t-1}:{n_{t+1}-m_t}} \pi'_t(m^{t-1})\right) \\
                            &\qquad \cdot \tilde{\theta}^e_t(a_t \vert n_{t+1}-m_t,m_t) \\
                            &= \sum_{w,m^t:n_{t+1}} q(w \vert n_{t+1}-m_t) Q^y(m_t \vert w) \\
                            &\qquad \cdot \pi_t(m^{t-1}) \hat{\theta}^e_t(a_t \vert m^t),
                     \end{align*}
                     \begin{align*}
                            &DEN \\
                            &= \sum_{\bar{w},\bar{n}_t,\bar{m}_t} q(\bar{w} \vert \bar{n}_{t}) Q^y(\bar{m}_t \vert \bar{w})\\
                            &\cdot \left(\sum_{\bar{m}^{t-1}:{\bar{n}_t}} \pi'_t(\bar{m}^{t-1})\right) \tilde{\theta}^e_t(a_t \vert \bar{n}_t,\bar{m}_t) \\
                            &= \sum_{\bar{w},\bar{m}^t} q(\bar{w} \vert \bar{n}_{t}) Q^y(\bar{m}_t \vert \bar{w}) \pi_t(\bar{m}^{t-1}) \hat{\theta}^e_t(a_t \vert \bar{m}^t)
                     \end{align*}
              \fi
              Moreover, the belief $\eta_{t+1}$ updated from $\pi'_t, a_t, \tilde{\theta}^e_t$ depends on $\pi'_t$ through $\eta_t$, because in the intermediate step, $\sum_{m^{t-1}:n_t} \pi'_t(m^{t-1})$ can be substituted by $\eta_t(n_t)$.

              Since (i) $\P(a_t \vert \pi'_t, \tilde{\theta}^e_t) = \P(a_t \vert \pi_t, \hat{\theta}^e_t) = \P(a_t \vert \eta_t, \tilde{\theta}^e_t)$, and (ii) $(\pi'_t, \tilde{\theta}^e_t, a_t)$,  $(\pi_t, \hat{\theta}^e_t, a_t)$, $(\eta_t, \tilde{\theta}^e_t, a_t)$ lead to the same $\eta_{t+1}$, we conclude that $\P(\eta_{t+1} \vert \pi'_t, \tilde{\theta}^e_t) = \P(\eta_{t+1} \vert \pi_t, \hat{\theta}^e_t) = \P(\eta_{t+1} \vert \eta_t, \tilde{\theta}^e_t)$.

       \end{enumerate}

\end{IEEEproof}

\subsection{Proof of Theorem~\ref{thm:belief_n_infinite}}
\label{appx:pf:thm:belief_n_infinite}
\begin{IEEEproof}
       We first show that in the infinite horizon case, it is sufficient to use $\tilde{\phi}$, which implies the sufficiency of summary-based MDP in Decision Process~\ref{mdp:summary-based} for the optimization.
       Note that the fact used in the induction step of the proof of Theorem~\ref{thm:belief_n} also works for the infinite time horizon: for a general strategy $\phi$, if $\phi_{t+1:\infty}$ are of type $\tilde{\phi}$, then (i) $V^{\phi}_{t+1}(\cdot)$ is a function of $\eta_{t+1}$, (ii) one can find a $\tilde{\phi}_t$, such that the expected reward to go is weakly improved, and $V^{\tilde{\phi}_t, \phi_{t+1:\infty}}_t(\cdot)$ is a function of $\eta_t$.

       Suppose $\tilde{\phi}^*$ is an optimal strategy to Decision Process~\ref{mdp:summary-based} with infinite time horizon, but not an optimal strategy to the optimization problem~\eqref{opt:partial}. Then, by the one-shot deviation principle, there is a profitable one-shot deviation~$\hat{\phi}_t$ with $\theta^e_t$ for some state~$\pi_t$ at some time~$t$. From the fact stated above, one can substitute the~$\hat{\phi}_t$ with $\tilde{\phi}_t$ that weakly improve the reward. However, the newly constructed mechanism is a valid solution to Decision Process~\ref{mdp:summary-based} but with a better value than $\tilde{\phi}^*$, which contradicts to the optimality of $\tilde{\phi}^*$. Thus, the optimal strategy $\tilde{\phi}^*$ for Decision Process~\ref{mdp:summary-based} has to be an optimal strategy to the optimization problem~\eqref{opt:partial} in the infinite horizon scenario. Now from basic results of MDPs, one can find a stationary optimal $\tilde{\phi}^*$ that satisfies the Bellman equation~\eqref{eq:bdp:infinite}.
\end{IEEEproof}


\subsection{Proof of Theorem~\ref{thm:nsii_benefits}}
\label{appx:pf:thm:nsii_benefits}
\begin{IEEEproof}
       \begin{enumerate}
              \item First check the TT constraint, i.e., for any $\eta_t$, $\tilde{\phi}^N[\eta_t]$ is in the feasible set $\tilde{S}(\eta_t)$ defined in~\eqref{def:summary-based_feasible_set}. Suppose $\eta_t \in \cL$, then $\tilde{\phi}^N[\eta_t] = \tilde{\theta}^L = 1_{m_t}$. Denote the left hand side of the inequality in~\eqref{def:summary-based_feasible_set} by $COL$ (cost of lying):
              \ifonecol
                     \begin{align*}
                            &COL = \sum_{w, n_t} Q^y(y_t \vert w)\eta_t(n_t)q(w \vert n_t) \cdot (1_{y_t}(w) - 1_{-y_t}(w)) \\
                            &= \sum_{n_t} Q^y(y_t \vert y_t) \eta_t(n_t)q(y_t \vert n_t)  - \sum_{n_t} Q^y(-y_t \vert y_t) \eta_t(n_t)q(-y_t \vert n_t) \\
                            &= \bar{p} \sum_{n_t} \eta_t(n_t)q(y_t \vert n_t) \cdot \left( 1 - \frac{p \sum_{n_t} \eta_t(n_t)q(-y_t \vert n_t)}{\bar{p} \sum_{n_t} \eta_t(n_t)q(y_t \vert n_t)} \right)\\
                            &\geq \bar{p} \sum_{n_t} \eta_t(n_t)q(y_t \vert n_t)  \cdot \left( 1 - \frac{p}{\bar{p}} \cdot \frac{\bar{p}}{p} \right)\\
                            &\geq 0,
                     \end{align*}
              \else
                     \begin{align*}
                            &COL = \sum_{w, n_t} Q^y(y_t \vert w)\eta_t(n_t)q(w \vert n_t) \\
                            &\quad \cdot (1_{y_t}(w) - 1_{-y_t}(w)) \\
                            &= \sum_{n_t} Q^y(y_t \vert y_t) \eta_t(n_t)q(y_t \vert n_t) \\
                            &\quad - \sum_{n_t} Q^y(-y_t \vert y_t) \eta_t(n_t)q(-y_t \vert n_t) \\
                            &= \bar{p} \sum_{n_t} \eta_t(n_t)q(y_t \vert n_t) \\
                            &\quad \cdot \left( 1 - \frac{p \sum_{n_t} \eta_t(n_t)q(-y_t \vert n_t)}{\bar{p} \sum_{n_t} \eta_t(n_t)q(y_t \vert n_t)} \right)\\
                            &\geq \bar{p} \sum_{n_t} \eta_t(n_t)q(y_t \vert n_t)  \cdot \left( 1 - \frac{p}{\bar{p}} \cdot \frac{\bar{p}}{p} \right)\\
                            &\geq 0,
                     \end{align*}
              \fi
              where the first inequality is implied by the definition of $\cL$. Then we check $COL$ for $\eta_t \notin \cL$. $\tilde{\phi}^N[\eta_t]=\tilde{\theta}^N$ in this case, which recommends $\text{sign}(n_t)$ when $n_t \neq 0$, but $m_t$ if $n_t=0$, so lying makes a difference only when $n_t=0$.
              \ifonecol
                     \begin{align*}
                            &COL = \sum_{w, n_t} Q^y(y_t \vert w)\eta_t(n_t)q(w \vert n_t) \cdot (\tilde{\theta}^N(w \vert n_t, y_t) - \tilde{\theta}^N(w \vert n_t, -y_t)) \\
                            &= \sum_w Q^y(y_t \vert w)\eta_t(0)q(w \vert 0)  \cdot (1_{y_t}(w) - 1_{-y_t}(w)) \\
                            &= Q^y(y_t \vert y_t)\eta_t(0)q(y_t \vert 0) - Q^y(y_t \vert -y_t)\eta_t(0)q(-y_t \vert 0) \\\
                            &= \frac{1}{2} \eta_t(0) (\bar{p} - p) \geq 0.
                     \end{align*}
              \else
                     \begin{align*}
                            &COL = \sum_{w, n_t} Q^y(y_t \vert w)\eta_t(n_t)q(w \vert n_t) \\
                            &\quad \cdot (\tilde{\theta}^N(w \vert n_t, y_t) - \tilde{\theta}^N(w \vert n_t, -y_t)) \\
                            &= \sum_w Q^y(y_t \vert w)\eta_t(0)q(w \vert 0) \\
                            &\quad \cdot (1_{y_t}(w) - 1_{-y_t}(w)) \\
                            &= Q^y(y_t \vert y_t)\eta_t(0)q(y_t \vert 0) \\
                            & \quad - Q^y(y_t \vert -y_t)\eta_t(0)q(-y_t \vert 0) \\\
                            &= \frac{1}{2} \eta_t(0) (\bar{p} - p) \geq 0.
                     \end{align*}
              \fi
              Hence, TT constraints hold. The tax function follows the profit maximizing form~\eqref{def:summary-based_instantaneous_reward}. Thus, NSII is FPM. The remaining thing is to check the NT constraint. Define a function $G$:
              \ifonecol
                     \begin{align}
                            &G(\eta_t, \tilde{\theta}, n_t, y_t) = \sum_{w} Q^y(y_t \vert w) \eta_t(n_t) q(w \vert n_t) \tilde{\theta}(w \vert n_t,y_t),
                     \end{align}
              \else
                     \begin{align}
                            &G(\eta_t, \tilde{\theta}, n_t, y_t) = \sum_{w} Q^y(y_t \vert w) \eta_t(n_t) \nonumber\\
                            &\quad \cdot q(w \vert n_t) \tilde{\theta}(w \vert n_t,y_t),
                     \end{align}
              \fi
              so that
              \ifonecol
                     \begin{align}
                            \tilde{f}^{N,x}&(\eta_t, \tilde{\theta}^e_t) = \sum_{y_t, n_t} G(\eta_t, \tilde{\theta}^e_t, n_t, y_t) - G(\eta_t, \tilde{\psi}^a_t[\eta_t], n_t, y_t).
                     \end{align}
              \else
                     \begin{align}
                            \tilde{f}^{N,x}&(\eta_t, \tilde{\theta}^e_t) = \sum_{y_t, n_t} G(\eta_t, \tilde{\theta}^e_t, n_t, y_t) \nonumber\\
                            &- G(\eta_t, \tilde{\psi}^a_t[\eta_t], n_t, y_t).
                     \end{align}
              \fi
              To show $\tilde{f}^{N,x}(\eta_t, \tilde{\phi}^N[\eta_t]) \geq 0$, it is sufficient to show that $\tilde{\phi}^N[\eta_t]$ maximizes $G(\eta_t, \tilde{\theta}^e_t, n_t, y_t)$ with respect to $\tilde{\theta}^e_t$. If we rewrite $G$:
              \ifonecol
                     \begin{align*}
                            &G(\eta_t, \tilde{\theta}, n_t, y_t) \\
                            &= Q^y(y_t \vert +1) \eta_t(n_t) q(+1 \vert n_t) \tilde{\theta}(+1 \vert n_t, y_t) + Q^y(y_t \vert -1) \eta_t(n_t) q(-1 \vert n_t) \tilde{\theta}(-1 \vert n_t, y_t) \\
                            &= Q^y(y_t \vert -1) \eta_t(n_t) q(-1 \vert n_t) \left( \frac{Q^y(y_t \vert +1) q(+1 \vert n_t)}{Q^y(y_t \vert -1) q(-1 \vert n_t)} \cdot \tilde{\theta}(+1 \vert n_t, y_t) + \tilde{\theta}(-1 \vert n_t, y_t) \right),
                     \end{align*}
              \else
                     \begin{align*}
                            &G(\eta_t, \tilde{\theta}, n_t, y_t) \\
                            &= Q^y(y_t \vert +1) \eta_t(n_t) q(+1 \vert n_t) \tilde{\theta}(+1 \vert n_t, y_t) \\
                            & \quad + Q^y(y_t \vert -1) \eta_t(n_t) q(-1 \vert n_t) \tilde{\theta}(-1 \vert n_t, y_t) \\
                            &= Q^y(y_t \vert -1) \eta_t(n_t) q(-1 \vert n_t) \\
                            & \ \cdot \left( \frac{Q^y(y_t \vert +1) q(+1 \vert n_t)}{Q^y(y_t \vert -1) q(-1 \vert n_t)} \cdot \tilde{\theta}(+1 \vert n_t, y_t) \right.\\
                            & \quad \qquad \left. + \tilde{\theta}(-1 \vert n_t, y_t) \right),
                     \end{align*}
              \fi
              where
              \ifonecol
                     \begin{align}
                            \label{eq:temp_frac_lem4}
                            &\frac{Q^y(y_t \vert +1) q(+1 \vert n_t)}{Q^y(y_t \vert -1) q(-1 \vert n_t)} =\frac{Q^y(y_t \vert +1)}{Q^y(y_t \vert -1)} \cdot \left(\frac{\bar{p}}{p}\right)^{n_t} = \left(\frac{\bar{p}}{p}\right)^{n_t + y_t}.
                     \end{align}
              \else
                     \begin{align}
                            &\frac{Q^y(y_t \vert +1) q(+1 \vert n_t)}{Q^y(y_t \vert -1) q(-1 \vert n_t)} \nonumber\\
                            \label{eq:temp_frac_lem4}
                            &=\frac{Q^y(y_t \vert +1)}{Q^y(y_t \vert -1)} \cdot \left(\frac{\bar{p}}{p}\right)^{n_t} = \left(\frac{\bar{p}}{p}\right)^{n_t + y_t}.
                     \end{align}
              \fi
              Therefore, if~$n_t+y_t>0$, the $\tilde{\theta}$ that maximizes $G$ has to have $\tilde{\theta}(+1 \vert n_t, y_t)=1$; if~$n_t+y_t<1$, $\tilde{\theta}(-1 \vert n_t, y_t)=1$; otherwise, $\tilde{\theta}$ can be anything. If $\eta_t \notin \cL$, one can verify that $\tilde{\theta}^N$ satisfies the three conditions above, so $\tilde{\phi}^N[\eta_t]$ maximizes the value of $G$. On the other hand, if $\eta_t \in \cL$, $\tilde{\phi}^N[\eta_t] = \tilde{\theta}^L = \tilde{\psi}^a[\eta_t]$. Thus, $\tilde{f}^{N,x}$ is always nonnegative.

              \item This proof will be done through random variables $H^N_{1:t}, H^B_{1:t}$ induced by $Y_{1:t-1}$, each $H^i_t$ can be viewed as what $\eta_t$ is supposed to be given $Y_{1:t-1}$ under mechanism $\tilde{\phi}^i$, $i=N, B$.

              We start with the motivation and formal definitions of $H^N_{1:t}, H^B_{1:t}$. We notice that both $\tilde{\phi}^N$ and $\tilde{\phi}^B$ produce deterministic partial strategies for given $\eta_t$. This implies the following fact: suppose $Y_{1:t-1}=y_{1:t-1}$ is fixed, under either $\tilde{\phi}^i$ ($i=N, B$), at every time step $\tau$, if we know $n_\tau, \eta_\tau, d_\tau=1$, and $a_\tau$ is determined by $\tilde{\phi}^i[\eta_t](\cdot  \vert  n_t, y_t)$ in a deterministic way, then the next $\eta_{\tau+1} = \tilde{T}(\eta_\tau, d_\tau, a_\tau, \tilde{\phi}^i[\eta_\tau])$ and $n_{\tau+1} = n_\tau + y_\tau$ are determined without any uncertainty. Consequently, since we know $n_1=0, \eta_1 = 1_0$, if $Y_{1:t-1}=y_{1:t-1}$ and $\tilde{\phi}^i$ ($i=N, B$) are known, $\eta_{1:t}$ can be determined, without any uncertainty. This fact indicates a deterministic mapping $\kappa^i$ from $Y_{1:t-1}$ to $H_{1:t}$ under the mechanism $\tilde{\phi}^i, i=N, B$. We define
              \begin{equation}
                     H^i_{1:t} := \kappa^i_t (Y_{1:t-1}), \ i=N, B.
              \end{equation}
              The measurability of $H^i_{1:t}$ is guaranteed under any $\P^{\tilde{\phi}^i}$, because $H^i_{1:t}$ is a function of $Y_{1:t-1}$, and $Y_{1:t-1}$ is measurable and takes finitely many values.

              Before proceeding to the main body of this proof, we provide three important facts: (i) $\P^{\tilde{\phi}}(W, Y_{1:t}) = \P(W, Y_{1:t})$ for any mechanism; (ii) $\P^{\tilde{\phi}^i} (H^i_{1:t} = H_{1:t})=1, i=N, B$; (iii) there is a function $\varphi$, $H^B_{1:t} = \varphi_t(H^N_{1:t})$. Facts (i), (ii) are straightforward. Here we explain why fact (iii) is correct. Suppose there is an underlying $Y_{1:t-1}$, and $H^N_{1:t} = \kappa^N_t(Y_{1:t-1})$, $H^B_{1:t} = \kappa^B_t(Y_{1:t-1})$. $H^B_{1:t}, H^N_{1:t}$ both start with $H^B_1=H^N_1=1_0 \in \cL$. If $H^B_\tau = H^N_\tau \in \cL$, both BHW and NSII play the same $\tilde{\theta}^L$, so $H^B_{\tau+1} = H^N_{\tau+1}$. If $H^B_\tau \notin \cL$, $H^B_\tau$ may be in a positive or negative cascade. Depending on the specific type,
              \begin{align*}
                     \eta^B_{\tau+1} = \tilde{T}(\eta^B_\tau, d_\tau=1, a_\tau=+1, \tilde{\theta}^{C+}), \\
                     \text{or } \eta^B_{\tau+1} = \tilde{T}(\eta^B_\tau, d_\tau=1, a_\tau=-1, \tilde{\theta}^{C-}).
              \end{align*}
              Therefore, one can always infer $H^B_{1:t}$ sequence from $H^N_{1:t}$ by the recursion described above, which means $H^B_{1:t}$ is a function of $H^N_{1:t}$.

              With the definition of $H^N_t, H^B_t$ and the important facts established above, we now prove the statement. Let $A^{\text{out}}_t$ denote the action agent~$t$ would take if she did not join in the mechanism. By individual rationality, given observation $H_{1:t}=\eta^N_{1:t}$,
              \begin{align}
                     &\E^{\tilde{\phi}^N} \left[ u_t(W, A_t) - \tilde{f}^{N,x}(H_t,\tilde{\Theta}^e_t)  \vert  H^N_{1:t} = \eta^N_{1:t} \right] \nonumber \\
                     &= \E^{\tilde{\phi}^N} \left[ u_t(W, A_t) - \tilde{f}^{N,x}(H_t,\tilde{\Theta}^e_t)  \vert  H_{1:t} = \eta^N_{1:t} \right] \nonumber \\
                     &\geq \E^{\tilde{\phi}^N} \left[ u_t(W, A^{\text{out}}_t)  \vert  H_{1:t} = \eta^N_{1:t} \right] \nonumber \\
                     &= \E^{\tilde{\phi}^N} \left[ u_t(W, A^{\text{out}}_t)  \vert  H^N_{1:t} = \eta^N_{1:t} \right],
              \end{align}
              where we exchange $H_t$ and $H^N_t$ based on fact (ii).

              On the other hand, notice that under BHW mechanism, the action $a_t$ can be determined by $\eta_t, y_t$, so $A_t$ under $\tilde{\phi}^B$ can be written as a function $g^B(\eta^B_t, y_t)$. Then, for all $\eta^N_{1:t}, y_t$:
              \ifonecol
                     \begin{align}
                            \label{ineq:BHW_vs_outNSII}
                            &\E^{\tilde{\phi}^B} \left[ u_t(W, A_t)  \vert  H^N_{1:t} = \eta^N_{1:t}, Y_t = y_t \right] \nonumber \\
                            &= \E^{\tilde{\phi}^B} \left[ u_t(W, A_t)  \vert  H^N_{1:t} = \eta^N_{1:t}, Y_t=y_t, H^B_{1:t} = \varphi_t(\eta^N_{1:t}) \right] \nonumber \\
                            &= \P^{\tilde{\phi}^B} \left(W = g^B((\varphi_t(\eta^N_{1:t}))_t, y_t)  \vert  H^N_{1:t} = \eta^N_{1:t}, Y_t=y_t, H^B_{1:t} = \varphi_t(\eta^N_{1:t}) \right) \nonumber \\
                            &= \P \left(W = g^B((\varphi_t(\eta^N_{1:t}))_t, y_t)  \vert  H^N_{1:t} = \eta^N_{1:t}, Y_t=y_t, H^B_{1:t} = \varphi(\eta^N_{1:t}) \right) \nonumber \\
                            &= \P^{\tilde{\phi}^N} \left(W = g^B((\varphi_t(\eta^N_{1:t}))_t, y_t)  \vert  H^N_{1:t} = \eta^N_{1:t}, Y_t=y_t, H^B_{1:t} = \varphi(\eta^N_{1:t}) \right) \nonumber \\
                            &= \P^{\tilde{\phi}^N} \left(W = g^B((\varphi_t(\eta^N_{1:t}))_t, y_t)  \vert  H_{1:t} = \eta^N_{1:t}, Y_t=y_t \right) \nonumber \\
                            &\leq \P^{\tilde{\phi}^N} \left(W = A^{\text{out}}_t  \vert  H_{1:t} = \eta^N_{1:t}, Y_t=y_t\right) \nonumber \\
                            &= \E^{\tilde{\phi}^N} \left[ u_t(W, A^{\text{out}}_t)  \vert  H^N_{1:t} = \eta^N_{1:t}, Y_t=y_t \right],
                     \end{align}
              \else
                     \begin{align}
                            \label{ineq:BHW_vs_outNSII}
                            &\E^{\tilde{\phi}^B} \left[ u_t(W, A_t)  \vert  H^N_{1:t} = \eta^N_{1:t}, Y_t = y_t \right] \nonumber \\
                            &= \E^{\tilde{\phi}^B} \left[ u_t(W, A_t)  \vert  H^N_{1:t} = \eta^N_{1:t}, \right. \nonumber \\
                            & \qquad \qquad \left. Y_t=y_t, H^B_{1:t} = \varphi_t(\eta^N_{1:t}) \right] \nonumber \\
                            &= \P^{\tilde{\phi}^B} \left(W = g^B((\varphi_t(\eta^N_{1:t}))_t, y_t)  \vert  H^N_{1:t} = \eta^N_{1:t}, \right. \nonumber \\
                            & \qquad \qquad \left. Y_t=y_t, H^B_{1:t} = \varphi_t(\eta^N_{1:t}) \right) \nonumber \\
                            &= \P \left(W = g^B((\varphi_t(\eta^N_{1:t}))_t, y_t)  \vert  H^N_{1:t} = \eta^N_{1:t}, \right. \nonumber \\
                            & \qquad \qquad \left. Y_t=y_t, H^B_{1:t} = \varphi(\eta^N_{1:t}) \right) \nonumber \\
                            &= \P^{\tilde{\phi}^N} \left(W = g^B((\varphi_t(\eta^N_{1:t}))_t, y_t)  \vert  H^N_{1:t} = \eta^N_{1:t}, \right. \nonumber \\
                            &\qquad \qquad \left.Y_t=y_t, H^B_{1:t} = \varphi(\eta^N_{1:t}) \right) \nonumber \\
                            &= \P^{\tilde{\phi}^N} \left(W = g^B((\varphi_t(\eta^N_{1:t}))_t, y_t)  \vert  H_{1:t} = \eta^N_{1:t}, \right. \nonumber \\
                            & \qquad \qquad \left.Y_t=y_t \right) \nonumber \\
                            &\leq \P^{\tilde{\phi}^N} \left(W = A^{\text{out}}_t  \vert  H_{1:t} = \eta^N_{1:t}, Y_t=y_t\right) \nonumber \\
                            &= \E^{\tilde{\phi}^N} \left[ u_t(W, A^{\text{out}}_t)  \vert  H^N_{1:t} = \eta^N_{1:t}, Y_t=y_t \right],
                     \end{align}
              \fi
              where for the third equality, $\tilde{\phi}^B$ can be dropped because all the random variables involved are functions of $Y_{1:t}$, and for the same reason, we add $\tilde{\phi}^N$ to the superscript for the fourth equality. The inequality comes from agent~$t$'s rationality: given all of agent~$t$'s observation including $\eta^N_{1:t}, y_t$, agent~$t$ generates $A^{\text{out}}_t$ from an optimal estimator of $W$, depending only on $\eta_t, y_t$.

              If we evaluate the weighted sum of both sides of~\eqref{ineq:BHW_vs_outNSII} with weight $\P(Y_t=y_t \vert H^N_{1:t} = \eta^N_{1:t})$,
              \begin{align}
                     \label{eq:pf:individual_imp}
                     &\E^{\tilde{\phi}^B} \left[ u_t(W, A_t)  \vert  H^N_{1:t} = \eta^N_{1:t} \right] \nonumber \\
                     & \leq \E^{\tilde{\phi}^N} \left[ u_t(W, A^{\text{out}}_t)  \vert  H^N_{1:t} = \eta^N_{1:t},  \right] \nonumber \\
                     & \quad \leq \E^{\tilde{\phi}^N} \left[ u_t(W, A_t) - \tilde{f}^{N,x}(H_t,\tilde{\Theta}^e_t)  \vert  H^N_{1:t} = \eta^N_{1:t} \right].
              \end{align}
              Finally, we evaluate the weighted sum of both sides with weight~$\P(H^N_{1:t} = \eta^N_{1:t})$ resulting in
              \ifonecol
                     \begin{align}
                            &\E^{\tilde{\phi}^B} \left[ u_t(W, A_t) \right] \leq \E^{\tilde{\phi}^N} \left[ u_t(W, A_t) - \tilde{f}^{N,x}(H_t,\tilde{\Theta}^e_t) \right].
                     \end{align}
              \else
                     \begin{align}
                            &\E^{\tilde{\phi}^B} \left[ u_t(W, A_t) \right] \nonumber \\
                            & \quad \leq \E^{\tilde{\phi}^N} \left[ u_t(W, A_t) - \tilde{f}^{N,x}(H_t,\tilde{\Theta}^e_t) \right].
                     \end{align}
              \fi

              \item The inequality~\eqref{eq:pf:individual_imp} shows a weak improvment on agent~$t$'s welfare conditioned on any $\eta^N_{1:t}$, so it is sufficient to find a sample path that strictly improves the welfare.
              Consider a sample path $\eta^N_{1:6}$ generated by
              \begin{equation}
                     y_{1:5} = (+1, +1, -1, -1, -1).
              \end{equation}
              By checking NSII step by step given the underlying $y_{1:5}$, one will find the following action sequence:
              \begin{equation}
                     a_{1:5} = (+1, +1, +1, +1, -1)
              \end{equation}
              and the following state sequence:
              \begin{equation}
                     \eta^N_{1:6} = (1_0, 1_1, 1_2, \eta^N_4, \eta^N_5, 1_{-1}),
              \end{equation}
              where
              \ifonecol
                     \begin{align}
                            \eta^N_4 &= \tilde{T}(1_2, d_t=1, a_t=1,\tilde{\theta}^N) \nonumber\\
                            &= \frac{1}{\bar{p}^2 + p^2} \left( (\bar{p}^2 p + \bar{p}p^2) 1_1 + (\bar{p}^3 + p^3) 1_3\right) \\
                            \eta^N_5 &= \tilde{T}(\eta^N_4, d_t=1, a_t=1,\tilde{\theta}^N) \nonumber\\
                            &= \frac{1}{\bar{p}^2 + p^2} \left(  2\bar{p}^2 p^2 \cdot 1_0 + (\bar{p}^3p + \bar{p}p^3) \cdot 1_2 + (\bar{p}^4 + p^4) \cdot 1_4\right)
                     \end{align}
              \else
                     \begin{align}
                            \eta^N_4 &= \tilde{T}(1_2, d_t=1, a_t=1,\tilde{\theta}^N) \nonumber\\
                            &= \frac{1}{\bar{p}^2 + p^2} \left( (\bar{p}^2 p + \bar{p}p^2) 1_1 + (\bar{p}^3 + p^3) 1_3\right) \\
                            \eta^N_5 &= \tilde{T}(\eta^N_4, d_t=1, a_t=1,\tilde{\theta}^N) \nonumber\\
                            &= \frac{1}{\bar{p}^2 + p^2} \left(  2\bar{p}^2 p^2 \cdot 1_0 + (\bar{p}^3p + \bar{p}p^3) \cdot 1_2\right. \nonumber\\
                            &\qquad \left.+ (\bar{p}^4 + p^4) \cdot 1_4\right)
                     \end{align}
              \fi
              The explanation for this sequence of $\eta$'s is as follows: $\eta_1, \eta_2$ are in the learning phase, so agents~$1$ and $2$ play actions that reveal their signals. $\eta_3, \eta_4$ are in the cascade phase, and the underlying $n_t+y_t \geq 0$, so two ``$+1$'' follow as actions. At time 5, the underlying $n_5=0$, $y_5=-1$, so action switches to $-1$. From $\eta^N_5$, agent~$6$ knows $n_5=0, 2$ or $4$, but among all possible $(n_5, y_5)$ pairs, only $n_5=0, y_5=-1$ does not contradict to $a_5=-1$, so agent~$6$ forms a belief $\eta^N_6 = \tilde{T}(\eta^N_5, d_5=1, a_5=-1, \tilde{\theta}^N) = 1_{-1}$.

              For agent~6 under NSII, since $\eta_6=1_{-1} \in \cL$, she will adopt partial strategy $\tilde{\phi}^N[\eta^N_6] = \tilde{\theta}^L$ and pay zero tax, while under BHW, she plays $\tilde{\theta}^{C+}$.
              These two partial strategies provide different actions when $y_6=-1$. So
              \ifonecol
                     \begin{align}
                            &\E^{\tilde{\phi}^N} [u_6(W, A_6)  \vert  \eta^N_{1:6}] - \E^{\tilde{\phi}^B} [u_6(W, A_6)  \vert  \eta^N_{1:6}] \nonumber\\
                            &= \sum_{y_6, n_6, w}  \P(y_6, n_6, w \vert \eta^N_{1:6}) \left( \E^{\tilde{\phi}^N} [u_6(w, A_6)  \vert  \eta^N_{1:6}, y_6, n_6, w] - \E^{\tilde{\phi}^B} [u_6(w, A_6)  \vert  \eta^N_{1:6}, y_6, n_6, w] \right) \nonumber\\
                            &=  \sum_{y_6, n_6, w}  1_{-1}(n_6) q(w \vert n_6) Q^y(y_6 \vert w) \nonumber \\
                            &\qquad \cdot \left( \E^{\tilde{\phi}^N} [u_6(w, A_6)  \vert  \eta^N_{1:6}, y_6, n_6,w]  - \E^{\tilde{\phi}^B} [u_6(w, A_6)  \vert  \eta^N_{1:6}, y_6, n_6,w] \right) \nonumber\\
                            &= \sum_w q(w \vert -1) Q^y(-1 \vert w)  \cdot (u_6(w, -1) - u_6(w,+1)) \nonumber\\
                            &= q(-1 \vert -1) Q^y(-1 \vert -1) - q(+1 \vert -1) Q^y(-1 \vert +1) \nonumber \\
                            &= \frac{\bar{p}}{\bar{p} + p} \bar{p} - \frac{p}{\bar{p} + p} p = \frac{\bar{p}^2 - p^2}{\bar{p} + p} > 0.
                     \end{align}
              \else
                     \begin{align}
                            &\E^{\tilde{\phi}^N} [u_6(W, A_6)  \vert  \eta^N_{1:6}] - \E^{\tilde{\phi}^B} [u_6(W, A_6)  \vert  \eta^N_{1:6}] \nonumber\\
                            &= \sum_{y_6, n_6, w}  \P(y_6, n_6, w \vert \eta^N_{1:6}) \nonumber \\
                            &\qquad \cdot \left( \E^{\tilde{\phi}^N} [u_6(W, A_6)  \vert  \eta^N_{1:6}, y_6, n_6, w]\right. \nonumber\\
                            & \quad \qquad \left. - \E^{\tilde{\phi}^B} [u_6(W, A_6)  \vert  \eta^N_{1:6}, y_6, n_6, w] \right) \nonumber\\
                            &=  \sum_{y_6, n_6, w}  1_{-1}(n_6) q(w \vert n_6) Q^y(y_6 \vert w) \nonumber \\
                            & \qquad \cdot \left( \E^{\tilde{\phi}^N} [u_6(W, A_6)  \vert  \eta^N_{1:6}, y_6, n_6, w]\right. \nonumber\\
                            & \qquad \left. - \E^{\tilde{\phi}^N} [u_6(W, A_6)  \vert  \eta^N_{1:6}, y_6, n_6, w] \right) \nonumber\\
                            &= \sum_w q(w \vert -1) Q^y(-1 \vert w) \nonumber \\
                            &\qquad \cdot (u_6(w, -1) - u_6(w,+1)) \nonumber\\
                            &= q(-1 \vert -1) Q^y(-1 \vert -1) \nonumber\\
                            & \qquad - q(+1 \vert -1) Q^y(-1 \vert +1) \nonumber \\
                            &= \frac{\bar{p}}{\bar{p} + p} \bar{p} - \frac{p}{\bar{p} + p} p \nonumber \\
                            &= \frac{\bar{p}^2 - p^2}{\bar{p} + p} > 0.
                     \end{align}
              \fi

              Since individual welfare is weakly improved by NSII, and there exists agent~$t$, such that there is at least one sample path~$\eta^N_{1:t}$ leading to strict improvements, NSII strictly improves the expected net social welfare.

              \item Since NSII is an NT-FPM mechanism it has non-negative taxes. We only need to show that there exists a positive probability history that results in strictly positive taxes. Under the realization $y_{1:4}=(+1,+1,-1,-1)$, $a_{1:3}=(+1,+1,+1)$ is true for both BHW and NSII. Then, BHW gets into an information cascade and switches to $\tilde{\theta}^{C+}$, and NSII system switches to $\tilde{\theta}^N$. From the definition, $a_4=+1$ in both systems. For agent~5, by the update rule of $\eta_t$ results in
              \ifonecol
                     \begin{equation}
                            \eta_5 = \frac{1}{\bar{p}^2+p^2} (2\bar{p}^2 p^2 \cdot 1_0 + (2\bar{p}^3 p  + 2\bar{p} p^3 ) \cdot 1_{+2} + (\bar{p}^4 + p^4) \cdot 1_{+4}),
                     \end{equation}
              \else
                     \begin{align}
                            \eta_5 &= \frac{1}{\bar{p}^2+p^2} (2\bar{p}^2 p^2 \cdot 1_0 + (2\bar{p}^3 p  + 2\bar{p} p^3 ) \nonumber\\
                            &\qquad \cdot 1_{+2} + (\bar{p}^4 + p^4) \cdot 1_{+4})
                     \end{align}
              \fi
              which induces $\hat{\pi}_5(+1) = \bar{p}^2 / (\bar{p}^2 + p^2) > 1 / 2$. Therefore, $\tilde{\psi}^a_t[\eta_5] = 1_{+1}$. Agent~$5$ will take action $-1$ only if $n_5=0, y_5 = -1$. As a result, the tax simplifies to
              \ifonecol
                     \begin{align}
                            &\tilde{f}^{N, x}(\eta_5, \tilde{\theta}^N) \nonumber \\
                            &= \sum_{w} Q^y(-1 \vert w)\eta_5(0) q(w \vert 0) (1_{-1}(w) - 1_{+1}(w)) \nonumber\\
                            &= \bar{p} \frac{2\bar{p}^2p^2}{\bar{p}^2+p^2} \frac{1}{(\bar{p}/p)^0+1} - p \frac{2\bar{p}^2p^2}{\bar{p}^2+p^2} \frac{(\bar{p}/p)^0}{(\bar{p}/p)^0+1} \nonumber\\
                            &= \frac{\bar{p}^2 p^2(\bar{p}-p)}{\bar{p}^2 + p^2} > 0.
                     \end{align}
              \else
                     \begin{align}
                                   &\tilde{f}^{N, x}(\eta_5, \tilde{\theta}^N) \nonumber\\
                                   &= \sum_{w} Q^y(-1 \vert w)\eta_5(0) q(w \vert 0) \nonumber\\
                                   &\qquad \cdot (1_{-1}(w) - 1_{+1}(w)) \nonumber\\
                                   &= \bar{p} \frac{2\bar{p}^2p^2}{\bar{p}^2+p^2} \frac{1}{(\bar{p}/p)^0+1} \nonumber\\
                                   &\qquad - p \frac{2\bar{p}^2p^2}{\bar{p}^2+p^2} \frac{(\bar{p}/p)^0}{(\bar{p}/p)^0+1} \nonumber\\
                                   &= \frac{\bar{p}^2 p^2(\bar{p}-p)}{\bar{p}^2 + p^2} > 0.
                     \end{align}
              \fi
       \end{enumerate}
\end{IEEEproof}

\subsection{Exact Analysis of NSII Mechanism}
\label{appx:exact_analysis_nsii}

In this part, we investigate NSII mechanism in order to characterize the expected coordinator's revenue and social welfare. Lemma~\ref{lem:NSII_chain} characterizes the Markov Reward Process (MRP) $\{\eta_t\}_{t}$ starting from $\eta_1=1_0$ under NSII.

\begin{lemma}
\label{lem:NSII_chain}
       For NSII mechanism, starting from $\eta_1=1_0$, the following are true:
       \begin{enumerate}
              \item The belief $\eta_t$ takes values in the set $\Xi = \{\xi_k\}_{k \in \Integer} \subset \Delta(\Integer)$, where
              \ifonecol
                     \begin{subequations}
                            \label{def:chi_set}
                            \begin{align}
                                   &\xi_k = 1_{k}, \quad k \in \{0, \pm 1, \pm 2\}, \\
                                   &\xi_{+k} = \tilde{T}(\xi_{+(k-1)},d_t=1, a_t=+1,  \tilde{\theta}^N), \quad k > 2, \\
                                   &\xi_{-k} = \tilde{T}(\xi_{-(k-1)},d_t=1, a_t=-1,  \tilde{\theta}^N), \quad -k < -2,
                            \end{align}
                     \end{subequations}
              \else
                     \begin{subequations}
                            \label{def:chi_set}
                            \begin{align}
                                   \xi_k = 1_{k}, \quad k \in \{0, \pm 1, \pm 2\}, \\
                                   \xi_{+k} = \tilde{T}(\xi_{+(k-1)},d_t=1, a_t=+1,  \tilde{\theta}^N), \\
                                   \quad k > 2, \nonumber\\
                                   \xi_{-k} = \tilde{T}(\xi_{-(k-1)},d_t=1, a_t=-1,  \tilde{\theta}^N),\\
                                    \quad -k < -2,\nonumber
                            \end{align}
                     \end{subequations}
              \fi
              where $\tilde{T}$ is defined in~\eqref{eq:pi_hat_update}. Moreover, $\Xi \cap \cL = \{\xi_{-1}, \xi_{0}, \xi_{+1}\}$.

              \item Let $\chi(\eta_t)$ denote the set of possible values of $\eta_{t+1}$ given the current $\eta_t$, then
              \ifonecol
                     \begin{subequations}
                            \begin{align}
                                   \label{eq:reachable_k01}
                                   &\chi(\xi_k) = \{\xi_{k-1}, \xi_{k+1}\},\quad k=0, \pm 1, \\
                                   \label{eq:reachable_k2}
                                   &\chi(\xi_{+2}) = \{\xi_{+3}\}, \quad  \chi(\xi_{-2}) = \{\xi_{-3}\},\\
                                   \label{eq:reachable_odd}
                                   &\chi(\xi_{i \cdot (2k-1)}) = \{\xi_{i \cdot (2k)}\}, \quad k = 2, 3, \ldots, \ i=\pm 1, \\
                                   \label{eq:reachable_even}
                                   &\chi(\xi_{i \cdot (2k)}) = \{\xi_{i \cdot (2k+1)}, \xi_{-i}\},\quad k=2,3, \ldots, \ i=\pm 1.
                            \end{align}
                     \end{subequations}
              \else
                     \begin{subequations}
                            \begin{align}
                                   \label{eq:reachable_k01}
                                   \chi(\xi_k) = \{\xi_{k-1}, \xi_{k+1}\},\quad k=0, \pm 1, \\
                                   \label{eq:reachable_k2}
                                   \chi(\xi_{+2}) = \{\xi_{+3}\}, \quad  \chi(\xi_{-2}) = \{\xi_{-3}\},\\
                                   \label{eq:reachable_odd}
                                   \chi(\xi_{i \cdot (2k-1)}) = \{\xi_{i \cdot (2k)}\}, \\
                                   \quad k = 2, 3, \ldots, \ i=\pm 1, \nonumber \\
                                   \label{eq:reachable_even}
                                   \chi(\xi_{i \cdot (2k)}) = \{\xi_{i \cdot (2k+1)}, \xi_{-i}\},\\
                                   \quad k=2,3, \ldots, \ i=\pm 1 \nonumber.
                            \end{align}
                     \end{subequations}
              \fi
              \item Let $q_{j,k}$ denote the transition probability from $\xi_j$ to $\xi_k$. We have
\begin{subequations}
       \begin{align}
              &q_{0, +1} = 1/2, \\
              &q_{+1, +2} = \bar{p}^2 + p^2 , \\
              &q_{+2, +3} = 1 , \\
              &q_{+k, +(k+1)} = 1, \quad k>2, \text{odd}, \\
              &q_{+k, +(k+1)} = 1 - \frac{1}{2}\xi_{+k}(0), \quad k>2, \text{even}, \\
              &q_{+k, -1} = \frac{1}{2}\xi_{+k}(0), \quad k>2, \text{even}, \\
              &q_{-k, -(k+1)}= q_{+k, +(k+1)}, \quad k=0,1,\ldots
       \end{align}
\end{subequations}

              \item Let $r_{k}$ denote the tax $\tilde{f}^{N,x}(\xi_k, \tilde{\phi}^N[\xi_k])$. The following are true:
              \begin{subequations}
                     \begin{align}
                            &r_k = r_{-k}, \quad \forall k,\\
                            &r_0 = 0 \\
                            &r_k = \frac{1}{2} (\bar{p} - p) \xi_k(0), \quad k \neq 0.
                     \end{align}
              \end{subequations}
       \end{enumerate}
\end{lemma}
\begin{IEEEproof}
       Here we prove statements 1) and 2) together. Statements 3) and 4) can be derived by the definition of $\tilde{T}$ and $\tilde{f}^{N,x}$ if 1) and 2) are true.

       The sequence $\{\eta_t\}$ starts from $ \xi_0=1_0$. Since $\tilde{\phi}^N[1_0] = \tilde{\theta}^L$, and with $\tilde{\theta}^L$ agent~$t$ plays $y_t$, there are two possible outcomes with $y_t=\pm 1$. By the update rule $\tilde{T}$, $\eta_{t+1}=1_{+1}=\xi_{+1}$ if $y_t=+1$ and $\eta_{t+1}=1_{-1}=\xi_{-1}$ if $y_t=-1$. Similarly, we may show~\eqref{eq:reachable_k01} for $k=\pm 1$, and $\xi_k$ for $k=0, \pm 1, \pm 2$ are possible beliefs during the process. When $\eta_t$ equals $1_{+2}$ or $1_{-2}$, NSII partial strategy $\tilde{\theta}^N$ is triggered, regardless of $y_t$, the next action is $+1$ or $-1$ respectively, and correspondingly $\eta_{t+1}$ equals $\xi_{+3}$ or $\xi_{-3}$. This proves~\eqref{eq:reachable_k2}.

       Then, we consider the evolution of $\eta_t$ starting from $\xi_{+3}$ (the case starting from $\xi_{-3}$ is similar). For each $\eta_t$, there are at most two reachable states, caused by $a_t=1$ and $-1$. We claim the following: \emph{if the support of $\eta_t\notin \cL$ contains only nonnegative numbers, $\eta_{t+1}$ can be either $1_{-1}$, or another belief not in $\cL$ with a support containing only nonnegative numbers.} Because $\tilde{\theta}^L$ recommends $-1$ only if $n_t+y_t<0$, and the only way that this happens is when $n_t=0, y_t=-1$, then, given $\eta_t \notin \cL$ with nonnegative support, once agent~$t+1$ sees action $a_t=-1$, an immediate inference of $n_t=-1$ can be made.
       If $a_t=+1$, this implies $n_{t+1}=n_t+y_t<0$ does not happen, so the support of $\eta_{t+1}$ is still nonnegative. To show $\eta_{t+1} \notin \cL$ under the assumptions of $\eta_t$ above and $a_t=+1$, it is more cumbersome to go through the update rule $\tilde{T}$, so here we evaluate $\hat{\pi}_{t+1}(w)$ instead:
       \ifonecol
              \begin{equation}
                     \hat{\pi}_{t+1}(w) = \P^{\tilde{\phi}^N} (W=w \vert  \eta_t, A_t=+1) = \frac{\P^{\tilde{\phi}^N} (W=w, A_t=+1 \vert  \eta_t)}{\P^{\tilde{\phi}^N} (A_t=+1 \vert  \eta_t)},
              \end{equation}
       \else
              \begin{align}
                     &\hat{\pi}_{t+1}(w) = \P^{\tilde{\phi}^N} (W=w \vert  \eta_t, A_t=+1) \nonumber \\
                     &= \frac{\P^{\tilde{\phi}^N} (W=w, A_t=+1 \vert  \eta_t)}{\P^{\tilde{\phi}^N} (A_t=+1 \vert  \eta_t)},
              \end{align}
       \fi
       where
       \begin{align}
              &\P^{\tilde{\phi}^N} (W=w, A_t=+1 \vert  \eta_t) \nonumber \\
              &= \P^{\tilde{\phi}^N} (W=w \vert  \eta_t) - \P^{\tilde{\phi}^N} (W=w, A_t=-1 \vert  \eta_t) \nonumber\\
              &= \hat{\pi}_t(w)  - \P^{\tilde{\phi}^N} (W=w, N_t=0, Y_t=-1 \vert  \eta_t)   \nonumber \\
              &= \hat{\pi}_t(w)  - \eta_t(0)q(w \vert 0) Q^y(-1 \vert w) \nonumber\\
              &= \hat{\pi}_t(w) - \frac{1}{2} \eta_t(0) Q^y(-1 \vert w).
       \end{align}
       Note that the assumption on $\eta_t$ implies that $\pi_t(+1)/\pi_t(-1) > \bar{p}/p$, and as a result we have
       \begin{equation}
                     \frac{\hat{\pi}_{t+1}(+1)}{\hat{\pi}_{t+1}(-1)} = \frac{\hat{\pi}_{t}(+1) - \frac{1}{2}\eta_t(0)p}{\hat{\pi}_{t}(-1) - \frac{1}{2}\eta_t(0)\bar{p}} \geq \frac{\hat{\pi}_{t}(+1)}{\hat{\pi}_{t}(-1)} > \frac{\bar{p}}{p},
       \end{equation}
       where we use the fact that $\hat{\pi}_t(w) \geq \eta_t(0) q(w \vert 0) = \eta_t(0) / 2$ and if $a > c > 0, b > d > 0$, $a/b \geq c/d$, then $(a-c)/(b-d) \geq a/b$. This result proves $\eta_{t+1} \notin \cL$. Thus, our claim is true.

       Based on the claim and the fact that $\xi_{+2} = 1_{+2} \notin \cL$ and with a nonnegative support, starting from $\xi_{+2}$, if $a_t=+1$, $\eta_{t+1} = \tilde{T}(\eta_t, d_t=1, a_t=+1, \tilde{\theta}^L)$, and it will maintain the properties described in the claim; if $a_t=-1$, $\eta_{t+1}$ collapses to $1_{-1}$. A similar statement applies to the trajectories starting from $\xi_{-3}$. This finishes the proof of~\eqref{def:chi_set} and~\eqref{eq:reachable_even}. For the remaining~\eqref{eq:reachable_odd}, note that $\xi_{i\cdot (2k-1)}$ starts from $\xi_{+3}$ or $\xi_{-3}$, which implies $n_t$ starts from $+3$ or $-3$. After even numbers of steps, $n_t$ should still be an odd number and therefore $n_t \neq 0$, which means  $\xi_{i\cdot (2k-1)}$ will never induce action $-i$.
\end{IEEEproof}

Next we evaluate the expected revenue of the coordinator.

Use $R_{k}^C$ to denote the expected reward-to-go at $\xi_{k}$ for the coordinator (the superscript ``c'' for coordinator). By symmetry, $R_{k}^C=R_{-k}^C$, so it suffices to focus on $R_k^C$ with $k=0,1,\ldots$. These $R$'s can be characterized by recursive equations:

\begin{subequations}
       \label{eq:val_nsii_REV}
       \begin{align}
              &R_0^C = \delta R_1^C, \\
              &R_1^C = \delta \left( (1-\bar{p}^2-p^2)R_0^C + (\bar{p}^2 + p^2) R_2^C \right), \\
              &R_2^C = \delta R_3^C, \\
              &R_{2k-1}^C = \delta R_{2k}^C, \quad k=2,3,\ldots, \\
              &R_{2k}^C = \delta \left( \frac{1}{2} \xi_{+(2k)}(0) R_1^C + (1 - \frac{1}{2} \xi_{+(2k)}(0))\right. \nonumber\\
              &\left.  \qquad\cdot R_{2k+1}^C \right) \quad k=2,3,\ldots.
       \end{align}
\end{subequations}
where $R_0^C$ is the expected income of the coordinator. The expected net social welfare can then be evaluated by
\begin{equation}
       \label{eq:nsii_expected_NSW}
       \E^{\tilde{\phi}^N}[NSW] = \E^{\tilde{\phi}^N} [GSW] - R_0^C.
\end{equation}

The gross social welfare under NSII can be evaluated through another MRP. We point out two straightforward facts without proof: given the trajectory $Y_{1:t}$, (i) $\tilde{\theta}^N$ is always an optimal partial strategy to play, (ii) $\tilde{\theta}^L$ is also an optimal partial strategy to play if $\eta_t \in \cL$. Since agents under NSII keeps playing optimal strategies, their gross social welfare should equal the one induced by any other optimal estimator based on $\{Y_t\}$. To simplify the calculation, we assume $\tilde{\theta}^N$ is adopted all the time.
By symmetry, conditioning on $w=\pm 1$ yields the same expected social welfare, i.e., $\E^{\tilde{\phi}^N}[GSW \vert W=w]=\E^{\tilde{\phi}^N}[GSW]$ for any $w$. Thus, we consider $w=1$ only.
Conditioned on $w=1$, $n_t$ is a Markov chain with the state transition: $n_{t+1}=n_t+y_t$, and $Y_{t+1}=1,-1$ with probability $\bar{p},p$ respectively. Let $R_k^S$ be the expected reward-to-go for $n_t=k$ (``$S$'' represents ``society''). Then,
\begin{subequations}
       \label{eq:val_nsii_SW}
       \begin{align}
              R_k^S &= 1 + \delta (\bar{p} R_{k+1}^S + p R_{k-1}^S),\quad k > 0,\\
              R_k^S &= \delta (\bar{p} R_{k+1}^S + p R_{k-1}^S),\quad k<0,\\
              R_0^S &= \bar{p} + \delta (\bar{p} R_1^S + p R_{-1}^S),
       \end{align}
\end{subequations}
and $\E^{\tilde{\phi}^N}[GSW]=\E^{\tilde{\phi}^N}[GSW \vert W=1]=R_0^S$. The difference equations~\eqref{eq:val_nsii_SW} can be solved employing the Z-transform. Using the notation $r[k]:=R_k^S$, we have the following system equation
\begin{equation}
       r[k] = (\bar{p} - 1) 1_0[k] + u[k] + \delta \bar{p} r[k+1] + \delta p r[k-1], \ \forall k.
\end{equation}
The two-sided Z-transform, $R(z)$, of the sequence $r[n]$ can be found to be
\begin{equation}
       R(z) = \frac{\bar{p} z^2 + p z}{(z - 1)(-\delta  \bar{p}  z^2 + z - \delta p)}
\end{equation}
with three poles
\ifonecol
       \begin{equation}
              p_1 = 1,\quad  p_2 = \frac{1 + \sqrt{1 - 4 \delta^2 p \bar{p}}}{2 \delta \bar{p}},\quad  p_3 =  \frac{1 - \sqrt{1 - 4 \delta^2 p \bar{p}}}{2 \delta \bar{p}},
       \end{equation}
\else
       \begin{align}
              p_1 = 1,\quad  p_2 = \frac{1 + \sqrt{1 - 4 \delta^2 p q}}{2 \delta q}, \nonumber\\
              \quad  p_3 =  \frac{1 - \sqrt{1 - 4 \delta^2 p q}}{2 \delta q},
       \end{align}
\fi
and one can prove that $(p_2 - 1)(p_3 - 1) < 0$, so $0 < p_3 < p_1 < p_2$. Depending on the region of convergence, the inverse Z-transform could be different. However, in our problem, we have the following conditions for $r[k]$:
\begin{itemize}
       \item Nonnegativity: $r[k] \geq 0$;
       \item As $k$ becomes sufficiently large, there could be little probability for state $n$ coming back to non-positive part, so agents keep acting 1, leading to $\lim_{k\to +\infty} r[k] = \frac{1}{1-\delta}$;
       \item As $k$ becomes sufficiently small, agents keep acting $-1$, leading to $\lim_{k \to -\infty} r[k] = 0$.
\end{itemize}
Then, the only region of convergence having such an inverse~$r[k]$ is $\{z: 1 = p_1 < \lvert z \rvert < p_2\}$. As a result,
\begin{subequations}
       \begin{align}
              &\E^{\tilde{\phi}^N}[GSW] = r[0] = -A_2 p_2^{-1}, \\
              &A_2 = \lim_{z \to p_2} (z - p_2) R(z).
       \end{align}
\end{subequations}
Therefore,
\ifonecol
       \begin{equation}
              \label{eq:val_nsii_gsw}
              \E^{\tilde{\phi}^N}[GSW] = \frac{\bar{p}}{(1 - \delta) (1 + \sqrt{1 - 4 \delta^2 p \bar{p}})} \left( 1 + \frac{1 - 2 \delta^2 p}{\sqrt{1 - 4 \delta^2 p \bar{p} }} \right).
       \end{equation}
\else
       \begin{align}
              \E^{\tilde{\phi}^N}[GSW] =& \frac{\bar{p}}{(1 - \delta) (1 + \sqrt{1 - 4 \delta^2 p \bar{p}})} \nonumber \\
              \label{eq:val_nsii_gsw}
              &\cdot \left( 1 + \frac{1 - 2 \delta^2 p}{\sqrt{1 - 4 \delta^2 p \bar{p} }} \right).
       \end{align}
\fi

The expected gross social welfare $\E^{\tilde{\phi}^N}[GSW]$ under NSII will be compared with that under $\tilde{\phi}^{B}$,
which turns out to be
\begin{equation}
       \label{eq:val_BHW_SW}
       \E^{\tilde{\phi}^B}[GSW]= \frac{\bar{p}(1-p\delta^2)}{(1-\delta)(1-2p\bar{p}\delta^2)}.
\end{equation}

\bibliographystyle{IEEEtran}


\end{document}